\documentclass[twocolumn,pra,amssymb,amsmath,superscriptaddress]{revtex4-1}
\usepackage{braket}
\usepackage{tikz}
\usepackage{multirow}
\usepackage{booktabs}
\usepackage[export]{adjustbox}
\usepackage{graphicx}
\usepackage{afterpage}
\usepackage[page,toc]{appendix}
\usepackage{amsthm}
\usepackage{dsfont}
\usepackage{xcolor}
\usepackage[normalem]{ulem}
\usepackage{soul}
\newcommand{\outprod}[1]{\ket{#1}\!\!\bra{#1}}
\newcommand{\proj}[2]{\ket{#1}\!\!\bra{#2}}
\usepackage{dcolumn}
\usepackage{bm}
\usepackage{blkarray}
\usepackage{xparse}
\usepackage{mathtools}
\usepackage{subfiles}
\graphicspath{{images/}}
\DeclareMathOperator{\Tr}{Tr}

\usepackage{float}

\begin{document}

\preprint{APS/123-QED}

\title{Heralded-Multiplexed High-Efficiency Cascaded Source of Dual-Rail Polarization-Entangled Photon Pairs using Spontaneous Parametric Down Conversion}

\author{Prajit Dhara}%
\affiliation{Wyant College of Optical Sciences, The University of Arizona, Tucson, AZ 85719}
\affiliation{NSF-ERC Center for Quantum Networks, The University of Arizona, Tucson, AZ 85719}
\author{Spencer J. Johnson}
\affiliation{Department of Physics, University of Illinois  Urbana-Champaign, Urbana, IL 61801 }
\affiliation{Illinois Quantum Information Science and Technology Center, University of Illinois  Urbana-Champaign, Urbana, IL 61801}
\author{Christos N. Gagatsos}
\affiliation{Wyant College of Optical Sciences, The University of Arizona, Tucson, AZ 85719}
\affiliation{NSF-ERC Center for Quantum Networks, The University of Arizona, Tucson, AZ 85719}
\author{Paul G. Kwiat}
\affiliation{Department of Physics, University of Illinois Urbana-Champaign, Urbana, IL 61801 }
\affiliation{Illinois Quantum Information Science and Technology Center, University of Illinois  Urbana-Champaign, Urbana, IL 61801}
\author{Saikat Guha}
\email{saikat@arizona.edu}
\affiliation{Wyant College of Optical Sciences, The University of Arizona, Tucson, AZ 85719}
\affiliation{NSF-ERC Center for Quantum Networks, The University of Arizona, Tucson, AZ 85719}

\begin{abstract}
Deterministic sources of high-fidelity entangled qubit pairs encoded in the dual-rail photonic basis, i.e., presence of a single photon in one of two orthogonal modes, are a key enabling technology of many applications of quantum information processing, including high-rate high-fidelity quantum communications over long distances. The most popular and mature sources of such photonic entanglement, e.g., those that leverage spontaneous parametric down conversion (SPDC) or spontaneous four-wave mixing (sFWM), generate an entangled (so called, continuous-variable) quantum state that contains contributions from high-order photon terms that lie outside the span of the dual-rail basis, which is detrimental to most applications. One often uses low pump power to mitigate the effects of those high-order terms. However that reduces the pair generation rate, and the source becomes inherently probabilistic. We investigate a {\em cascaded} source that performs a linear-optical entanglement swap between two SPDC sources, to generate a {\em heralded} photonic entangled state that has a higher fidelity (to the ideal Bell state) compared to a free-running SPDC source. Further, with the Bell swap providing a heralding trigger, we show how to build a multiplexed source, which despite reasonable switching losses and detector loss and noise, yields a Fidelity versus Success Probability trade-off of a high-efficiency source of high-fidelity dual-rail photonic entanglement. We find however that there is a threshold of $1.5$ dB of loss per switch, beyond which multiplexing hurts the Fidelity versus Success Probability trade-off.
\end{abstract}

\maketitle

\newpage

\section{Introduction}
Distributed high-fidelity entanglement will become a commodity as its demand stemming from a variety of promising applications increases. As the world makes progress towards realizing the vision of a {\em quantum internet}~\cite{Wehner2018-lw} to generate entanglement among many user groups at high rates, some of the biggest remaining enabling-technology challenges, are: (1) scalable sources of high-fidelity on-demand photonic entanglement, (2) high-efficiency high-bandwidth high-coherence-time universal-quantum-logic-capable quantum memories, and (3) high-efficiency converters between various qubit forms native to the leading quantum-memory contenders and optical-frequency photonic qubits. 

While, there are no viable alternatives to optical-frequency qubits for long-distance transmission, there are many ways to encode a qubit in the photon~\cite{Albert2018-tq}. Two of those most commonly studied are: (a) the Knill-Laflamme-Milburn (KLM) dual-rail photonic qubit~\cite{knill2001} where the presence of a single photon in one of two orthogonal (spatial, spectral, temporal or polarization) modes encodes the two logical quantum states of a qubit; and (b) the Gottesman-Kitaev-Preskill (GKP) encoding~\cite{Gottesman2001-yl}, which encodes the qubit in a single bosonic mode excited in one of two coherent superpositions of displaced quadrature-squeezed states that are shifted with respect to one another in the phase space. Dual-rail qubits are conceptually easy to produce and manipulate using passive linear optics, but they need a high-fidelity single-photon source  and single-photon detectors. Quantum logic on dual-rail qubits using passive linear optics and single-photon detectors is simple to build. Despite the gates being inherently probabilistic in the original KLM scheme~\cite{knill2001}, recent advancements on single-photon ancilla-assisted boosted linear-optical quantum logic~\cite{Ewert2014-nr} has ushered linear optical quantum computing using dual rail qubits into highly scalable architectures~\cite{Gimeno-Segovia2015-yr,Pant2019-ds,Bartolucci2021-ps}. Alternatively, GKP qubit is known to be the most loss-resilient photonic qubit encoding~\cite{Albert2018-tq,Noh2019-ak}, and Clifford quantum logic is deterministically implementable using squeezers and linear optics~\cite{Gottesman2001-yl}. However, not only are they hard to produce~\cite{Eaton2019-ho,Su2019-ef}, there is no known way to store GKP qubits and GKP-basis entangled states in heralded quantum memories. In this paper, we will therefore focus on the dual-rail qubit. The multiplexed heralded entanglement generation ideas we present here, however, are applicable to other photonic qubit encodings and to other (e.g., multi-qubit) entangled states.

We focus in this work on the first challenge mentioned in the first paragraph above: that of designing an on-demand photonic entanglement source that produces high-fidelity two-qubit entangled Bell states with the qubits encoded in the dual-rail photonic basis. There have been many calculations of quantum repeater protocols~\cite{Guha2015-qo,Pant2017-pc} and quantum network routing algorithms~\cite{Pant2019-ja,Nain2020-xt} which assume the availability of unit-Fidelity sources of dual-rail photonic entanglement. This results in these analyses predicting, despite inclusion of linear losses everywhere in the system, pristine dual-rail Bell states, i.e., entangled bits (ebits), be delivered to the communicating parties Alice and Bob. In reality, sources that deterministically generate dual-rail Bell states suitable for communications are quite challenging to build. Sub-unity entanglement fidelity has been incorporated in recent work on entanglement routing~\cite{Goodenough2021-qp}, but those have restricted their analyses to ideal Werner-like entangled states. Quantum dot sources~\cite{Arakawa2020-me,Uppu2020-pe} and other forms of quantum emitters~\cite{Lee2020-nk} can theoretically generate single photons and entangled photon pairs on demand, and recently these sources have achieved high (polarization) entanglement fidelity~\cite{Patel2016-lw} and over 60\% coupling efficiency into single-mode optical fiber~\cite{Yonezu2017-gv}, though not in the same experiment. Moreover, the photon frequencies from individual emitters can vary slightly, and the emitted photon frequency is usually not compatible with telecommunications hardware. The most common and reliable sources of dual-rail entanglement used in practice rely on spontaneous parametric down conversion (SPDC)~\cite{Kwiat1999-fq}, wherein single photons from a strong pump laser impinging on a carefully phase-matched (possibly periodically-poled) $\chi^{(2)}$ crystal splits into entangled photon pairs at two frequencies.  Alternatively, one can employ the process of spontaneous four-wave mixing (FWM), in which a pair of pump photons give rise to an entangled pair~\cite{armstrong1962,fejer1992}. Here we will refer to SPDC but the conclusions would be equally valid for FWM sources.

There are many variants of SPDC-based entangled photon pair generation methods. However, a detailed physical analysis of these sources has shown that the complete quantum state generated, described by two copies of the so-called two-mode squeezed vacuum (TMSV), contains contributions from vacuum and high-order, e.g., two-photon-pair terms in addition to the desired dual-rail Bell state, which can adversely affect both the distribution rates and the Fidelity of the distributed entanglement~\cite{krovi2016,kok2000}. In fact, the pump power must be carefully optimized to maximize the entanglement rate, while adhering to a desired Fidelity threshold. One common strategy is to turn down the pump power so low that the probability that the source produces two-pair (and higher-order) states becomes negligible. Of course, this entails increasing the contribution of vacuum to the emitted state and reduces the rate at which the desired Bell states are produced. The vacuum term often does not affect the usability of the source in an application, either because it gets filtered out by the `click' of a detector, e.g., in a quantum key distribution (QKD) experiment that provides a {\em post-selection trigger} to consider only those times slots that {\em had} a photon in it; or because a quantum memory provides a {\em heralding trigger} declaring that it successfully loaded a dual-rail photonic qubit into its native qubit domain (hence filtering out the vacuum). 

Other than the reduced pair-production rate of the above strategy of turning down the pump power, another inherent problem with such a `free-running' standalone SPDC-based entanglement source is that it is probabilistic, and does not have a heralding trigger. In other words, we cannot in principle know in which time slot the source actually produced an entangled photon pair, a major detriment in many applications. One method to increase the probability of emitting a single photon into a particular time slot is to use a heralded single-photon source (HSPS), \textit{e.g.}, from SPDC, combined with spatial, temporal~\cite{segovia2017,Kaneda2019-vg}, or spectral multiplexing~\cite{pseiner2021,meyer2020}. However for this to work with entangled pairs would require one of the photons to be detected immediately, undesirable for many applications. A second option is to use (four) single photons possibly from a multiplexed HSPS as inputs to a quantum circuit that probabilistically produces heralded entangled pairs~\cite{Zhang2008-jk,Stanisic2017-we,Fldzhyan2021-pw}; combined with multiplexing this could enable `entanglement on demand'.

\begin{figure}
	\centering{\includegraphics[width=\linewidth]{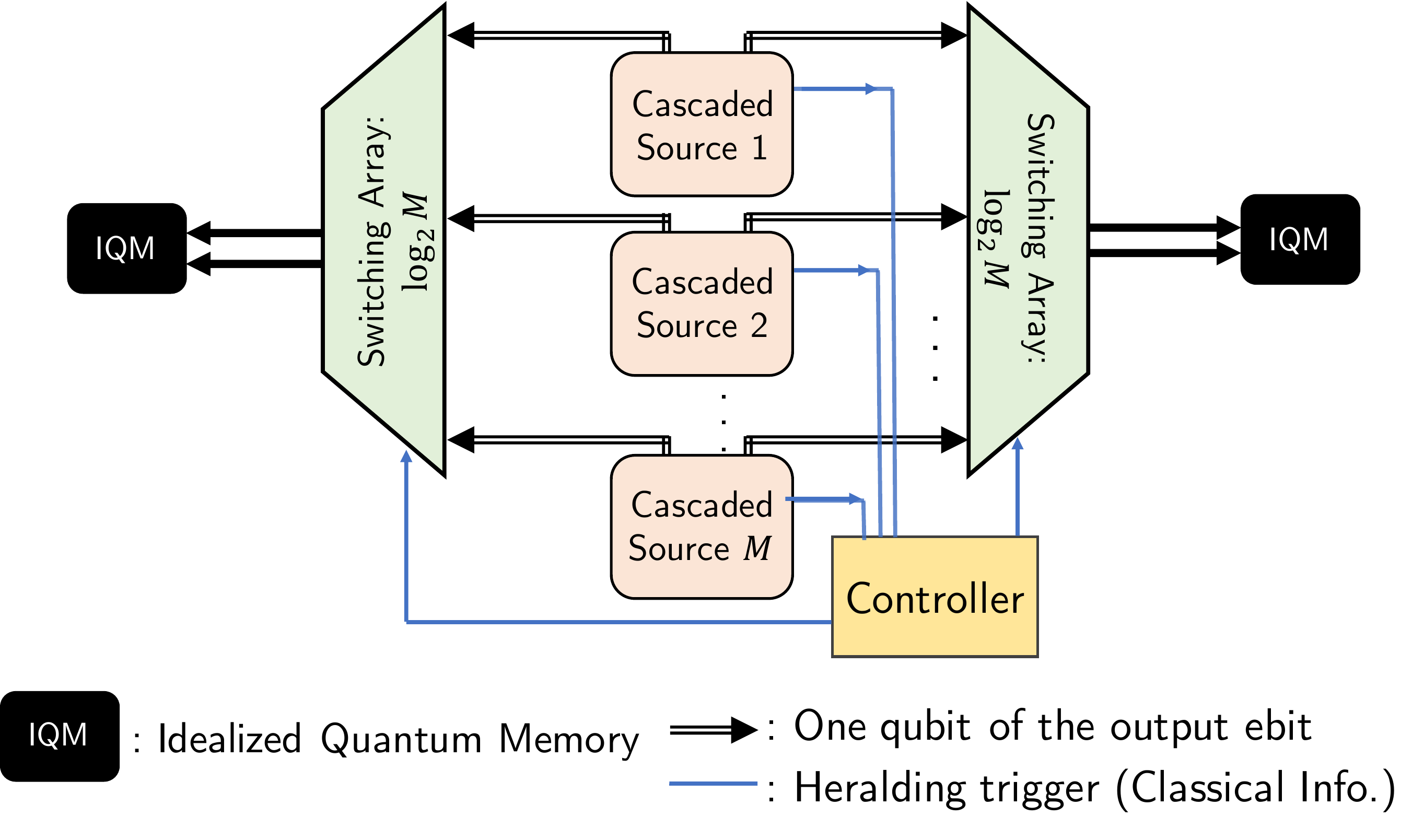} }
	\caption{Multiplexing scheme for a near-deterministic source of near-unity-Fidelity dual-rail Bell states. We multiplex $M$ cascaded sources (each with one PNR-based BSM within it), whose outputs are connected into two $M$-to-$1$ switching arrays of $ \log_2(M) $ switches, each of $\log_{10}(1/\eta_s)$ dB of loss. The switching array outputs the state of one of the successful cascaded sources in any time slot, assuming one or more succeeds in that time slot. This output photonic entangled state is then loaded into a pair of ideal heralded quantum memories (shown as black boxes marked IQM). }
	\label{fig:mux1}
\end{figure}

In this paper, we propose a source design that alleviates all of the above-listed problems, yielding a {\em high-rate}, {\em high-fidelity}, {\em near-deterministic} source of dual-rail entangled photonic qubit pairs,  at the cost of high levels of multiplexing. The concept is inspired from prior work on heralded~\cite{Kaneda2016-ps} and multiplexed~\cite{Kaneda2019-vg,hiemstra2020} SPDC based single-photon sources: we first create what we call a {\em cascaded SPDC source}, which employs two SPDC-based entanglement sources,  and performs a linear-optical Bell state measurement (BSM), commonly called an `entanglement swap', to yield an entangled state on the `outer' undetected mode pairs. This state has a much lower vacuum and high-order-photon contributions compared to a standalone SPDC source. Thereafter, we leverage the heralding trigger from the BSM to multiplex $M > 1$ such cascaded sources, using two switch-arrays (each consisting of $\log_2 M$ switches) and a controller that lets out entangled photon pairs from the `successful' source, in order to improve the pair-production rate. For the overall concept see Fig.~\ref{fig:mux1}. The switching losses, which scales up logarithmically in $M$, and photon-number-resolving (PNR) detector imperfections within the BSM, i.e., efficiency and dark counts, are incorporated into the engineering design study of the Fidelity versus entanglement generation rate of the overall heralded-multiplexed source. 


The article is organized as follows. Section~\ref{section:source_descr} discusses the full quantum-state description of an SPDC-based entanglement source, and our proposal for the cascaded source with a heralding trigger. Section~\ref{section:ideal_memory} describes an idealized model of a heralded quantum memory that we use in the remainder of our analysis. Section~\ref{section:state_analysis} presents a detailed performance evaluation of the cascaded source as a function of device impairments. Finally, we present our analysis of the multiplexed cascaded source in Section~\ref{section:mux}, including switching losses in addition to the device imperfections considered in the previous section. We evaluate a parametric trade-off of Fidelity versus Success Probability (of producing entangled pairs), while optimizing over the pump power of the individual SPDC sources and the number of cascaded sources in the multiplexed source. This trade-off shows that one can achieve a near-deterministic source of dual-rail Bell states in principle at high rates, despite reasonably non-ideal devices. Section~\ref{section:conclusion} concludes the paper with thoughts on future work and applications of this study.

\section{Cascaded SPDC Source}
\label{section:source_descr}

\subsection{Polarization-entangled SPDC source: A review}
\begin{figure}[h!]
	\centering{\includegraphics[width=\linewidth]{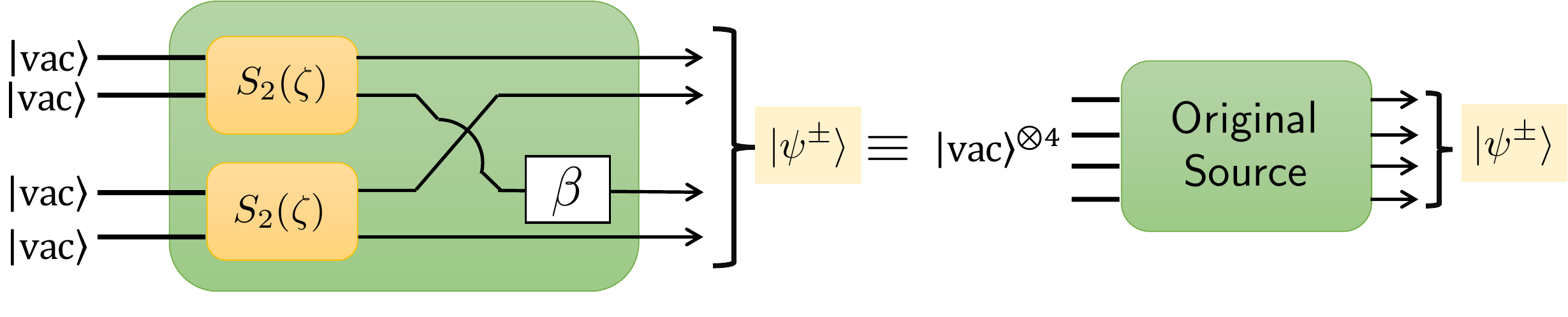}  }
	\caption{A block model of the individual SPDC-based source of dual-rail entangled qubit pairs as presented in~\cite{kok2000}. The output quantum state is given by Eq.~(\ref{eqn:srcnative}). $ S_2(\zeta) $ denotes a two-mode squeezing unitary, $|{\text{vac}}\rangle$ denotes a single-mode vacuum state, and the squeezing amplitude $ \zeta $ is related to the mean photon number per mode, $ N_s $, as $ \sinh^2 |\zeta| = N_s $. One beam from each two-mode squeezed source is swapped and a potential optical phase-shift $ \beta(=0 \text{ or }\pi) $ is introduced. Appendix \ref{appendix:source} includes a more detailed description and derivation of the quantum state description of this source.}
	\label{fig:diag_orig}
\end{figure}

A complete quantum-theoretic modeling of the polarization-dual-rail SPDC-based pulsed entanglement source was presented in~\cite{kok2000}. The physical model of this entanglement source can be seen as two copies of two-mode squeezed vacuum (TMSV) states with one mode of each TMSV swapped. See Fig.~\ref{fig:diag_orig} for a schematic representation. The output is described by four modes: two orthogonal polarization modes of each of the (spatio-temporal) modes of a pair of pulses emitted by the source. A reminder for the reader is that two orthogonal modes carry one dual-rail qubit. Hence, a two-qubit entangled Bell state requires four orthogonal modes to encode. The quantum state of this $4$-mode output is given by:
\begin{equation}
|\psi^\pm\rangle = \sum_{n=0}^\infty \sum_{k=0}^n (\pm 1)^k \sqrt{\frac{p(n)}{n+1}}\, |n-k,k;\, k,n-k\rangle,
\label{eqn:fullsourcemodel}
\end{equation}
where,
\begin{equation}
p(n) = (n+1)\frac{{N_s}^n}{(N_s + 1)^{n+2}}, 
\label{eqn:geometric_dist}
\end{equation}
with $N_s$ the mean photon number per mode. Note that, hence, the mean photon number per (dual-rail) qubit is $2N_s$. It should also be noted that this $4$-mode state is a Gaussian state, i.e., its Wigner function is an $8$-variate Gaussian function of the field quadratures of the concerned modes, because it is essentially a tensor product of two TMSV states with a pair of mode labels flipped. 

We will use the following notation for two (of the four mutually orthogonal) dual-rail two-qubit Bell states:
\begin{align}
	\ket{\Psi^\pm}\equiv\frac{\ket{1,0;0,1}\pm\ket{0,1;1,0}}{\sqrt{2}}.
	\label{eqn:target_st}
\end{align}
The ($\pm$) signs in Eqs.~\eqref{eqn:fullsourcemodel} and~\eqref{eqn:target_st} refer to the possibility of an additional $\pi$ phase that could be applied to one of the polarization modes of one output pulse, e.g., using a half wave-plate, depending upon whether the desired Bell state for the application is $|\Psi^+\rangle$ or $|\Psi^-\rangle$.

Since we will be concerned with the $N_s\ll1$ regime in this paper, we will henceforth truncate the quantum state of the source up to the photon-number (Fock) support of $2$ photon pairs~\cite{krovi2016}:
\begin{align}
	\begin{split}
	\ket{M^\pm} \!= & N_0 \left[\!\sqrt{p(0)} \ket{0,0;0,0}\! +\! \sqrt{\frac{p(1)}{2}} \left(\ket{1,0;0,1}\pm\ket{0,1;1,0} \right) \right.\\
		&\left. +  \sqrt{\frac{p(2)}{3}}\left(\ket{2,0;0,2}\pm \ket{1,1;1,1}+\ket{0,2;2,0}\right)\right],
		\label{eqn:srcnative}
	\end{split}
\end{align}
where we introduce $N_0 = {1}/{\sqrt{p(0)+p(1)+p(2)}}= {(N_s+1)^2}/{\sqrt{6 N_s^2+4 N_s+1}}$ as a normalization factor that we choose for convenience to ensure that $|\Psi^\pm\rangle$, despite the Fock truncation, is a unit-norm quantum state. In Appendix~\ref{appendix:source}, we show that for $N_s \gtrsim 0.2$ the above truncation leads to a bad approximation. Since all the results in this paper will use $N_s \ll 0.2$, this does not apply to the results reported herein. In Appendix~\ref{appendix:GBSmodel}, we show how one would do a full exact analysis of everything reported in this paper, while employing the complete Gaussian-state description of $|\psi^\pm\rangle$.


The four-mode output state $|\psi^\pm\rangle$ in~\eqref{eqn:fullsourcemodel} is a superposition of vacuum, one of the two dual-rail-basis Bell states $|M^\pm\rangle$, and additional states corresponding to $n\geq$ pairs of photons (in each of the two output pulses), with a geometrically distributed probability amplitude $p(n)$. Without the aid of auxiliary highly non-linear operations such as a quantum memory or a non-destruction measurement we discuss later in this paper, these higher-order $n$-photon-pair terms cannot be eliminated from the source output, as the very nature of the underlying TMSV model determines the proportion of these `spurious' terms. Reducing the mean photon number per mode $N_s$ by turning down the pump power reduces the proportion of two-pair terms $p(2) = 3 N_s^2 /(N_s+1)^4$, at the expense of reducing $p(1)$ as well, and hence increasing $p(0)$. In the $N_s \ll 1$ regime, the vacuum term is the dominant component. A quantum non demolition (QND) measurement that performs a {\em vacuum or not} (VON) projection~\cite{Wilde2012-iw} on the source output could aid in eliminating the vacuum component. However, the existing experimental proposals for implementing such a QND measurement involve non-linear atom-photon interactions~\cite{oi2013}, and are currently infeasible to realize efficiently on traveling modes of an optical-frequency field.

\subsection{Cascaded SPDC source using two polarization-entangled sources and a BSM}
\begin{figure}[h!]
	\centering{\includegraphics[width=\linewidth]{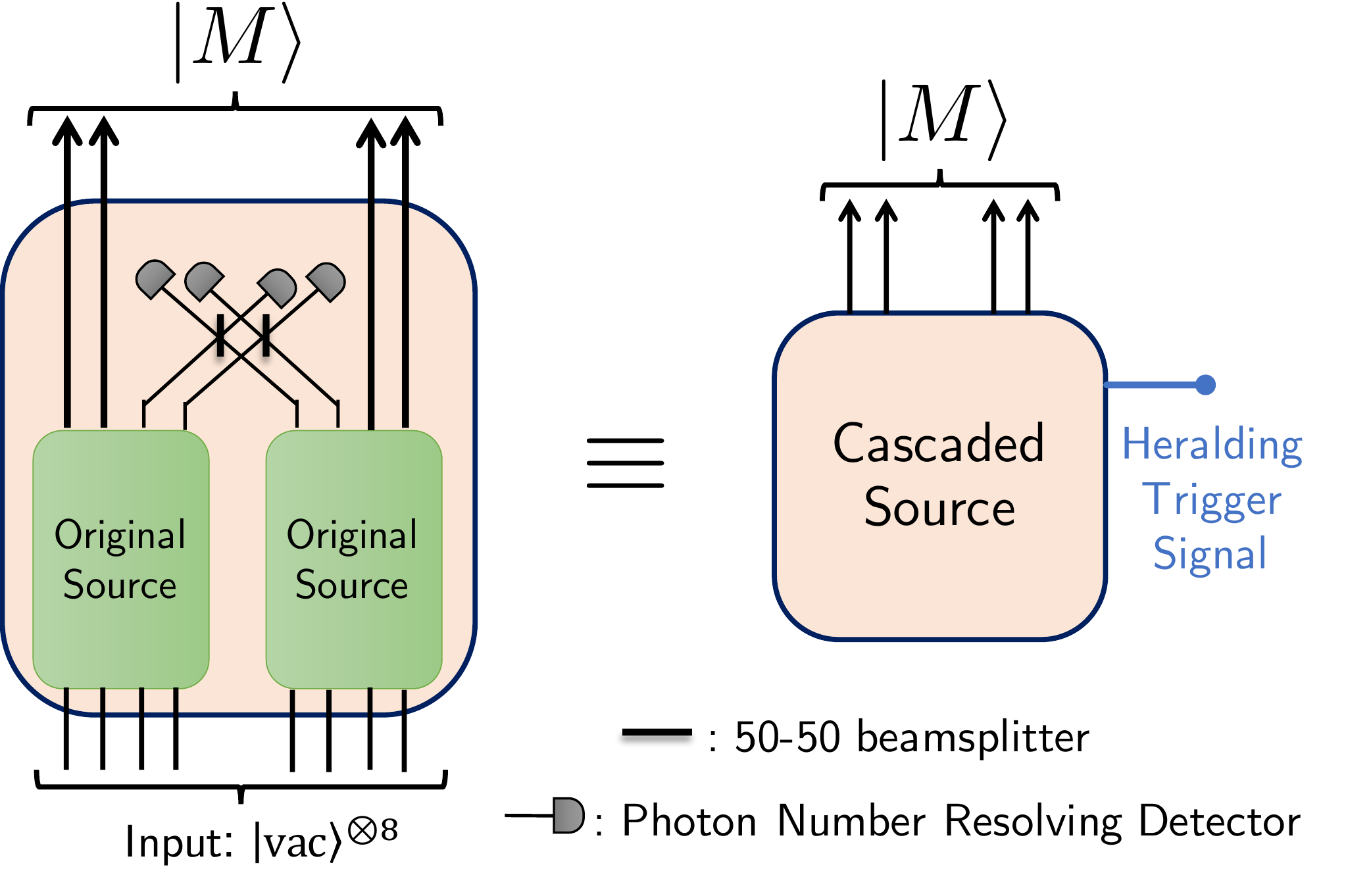}}
	\caption{Schematic of the proposed cascaded source that generates the entangled state given in Eq.~(\ref{eqn:casc_src_st}) conditioned on the linear-optical BSM in the middle producing one of the two `desirable' click patterns. This cascaded source therefore has a {\em heralding trigger}, telling us in which time slot a copy of $\ket{M}$ was produced successfully. This feature was missing in the original source. }
	\label{fig:diag_casc}
\end{figure}

We now describe our design of a {\em cascaded source} whose output state's Fidelity with the desired Bell state can be higher than that of a standalone SPDC source $\ket{\psi^\pm}$ described in the previous subsection. The proposed design is shown in Fig.~\ref{fig:diag_casc}. We take two copies of $\ket{\psi^\pm}$ (labeled `original source' in a green box), and perform a local linear optical BSM using beamsplitters and four PNR detectors. For the purposes of this section, we will assume that all the components are ideal, i.e., no coupling losses from the SPDC sources into the BSM, and ideal PNR detectors. We will relax these assumptions in the more detailed analysis in the next section. If the states fed into the BSM were ideal Bell states, i.e., with no multi-pair contributions, the resulting state of the unmeasured four outer modes (shown as black arrows in Fig.~\ref{fig:diag_casc}) would be ideal Bell states as well. However, since the outputs~\eqref{eqn:srcnative} of the original sources $\ket{\psi^\pm}$ are not ideal Bell states $ \ket{\Psi^\pm} $, despite observing a BSM `success', we might generate spurious states on those outer modes that are not Bell states.

If both the `original' sources produce the state $|\psi^+\rangle$, the heralded output state of the undetected outer modes, upon the occurrence of a \textit{desirable} click pattern, is given by:
\begin{eqnarray}
	\ket{M}&=&N'_0 \left[\frac{p(1)}{2} \left(\ket{1,0;0,1}+(-1)^{m_1}\ket{0,1;1,0}\right)+\right.\nonumber\\
	&& \hspace{-35pt}\left.(-1)^{m_2} \sqrt{\frac{p(0)p(2)}{3}} \left(\ket{0,0;1,1}+(-1)^{m_1}\ket{1,1;0,0}\right)\right],
	\label{eqn:casc_src_st}
\end{eqnarray}
where $p(0), p(1), p(2)$ are as defined in~\eqref{eqn:geometric_dist}, and the normalization constant $N_0^\prime$ is given by:
\begin{align}
	N'_0=\left(\frac{p(1)^2}{2}+ \frac{2p(0)p(2)}{3}\right)^{-1/2} .
\end{align}
By \textit{desirable}, we here signify the four click patterns (out of a possible eight) that are necessary but not sufficient to herald an entanglement swap between two dual-rail photonic modes on a linear optical BSM circuit. The reason the patterns are not sufficient is that these same patterns can also be produced by the undesirable event that both photons detected in the BSM came from only one of the SPDC sources, instead of one from each; unfortunately, the likelihood of these two processes are equal for SPDC. Therefore, to exclude the undesirable photon pair contribution from the same source, we must rely on post-selection of a photon in each outer mode- either via direct detection or via a heralded quantum memory as discussed below.

Depending upon which of the four `desirable' click patterns occur on the four PNR detectors (e.g., $0011$ implies: no-click, no-click, $1$-click, $1$-click), the values of $m_1$ and $m_2$ in the heralded state in Eq.~\eqref{eqn:casc_src_st} are given by:
\begin{table}[h]
\begin{tabular}{p{1cm}>{\bfseries}p{2.5cm}p{1cm}p{1cm}}
& {\text{Click Pattern}} & $m_1$ & $m_2$\\
\\
& \qquad 0011  & \,0 & \,0\\
& \qquad 1100  & \,0 & \,1\\
& \qquad 1001  & \,1 & \,1\\
& \qquad 0110  & \,1 & \,0\\
\end{tabular}
\end{table}

\noindent If both the `original' sources produce the state $|\Psi^-\rangle$, the heralded output state of the undetected outer modes is same as given in Eq.~\eqref{eqn:casc_src_st}, except that the values of $m_2$ in the above table are flipped.

Henceforth, we will drop the $\pm$ superscript in the state $\ket{M^\pm}$, since we will assume always to be working with the state $\ket{M^+}$. Further, we will say the source was `successful' in producing an entangled state when one of the first two desirable click patterns above (0011 or 1100) occur (i.e., $m_1 = 0$). The reason for this is that we want the output state to be (close to) the $\ket{M^+}$ state. We will use $|M\rangle$ to denote the desirable output state of the cascaded source, and not carry the $m_2$ index. This is because our results in this paper do not depend upon the value of $m_2$. Further, if the memories in which the distributed entanglement eventually gets stored have good quality native quantum logic, it is easy to apply a local single-qubit unitary operation to turn the Bell state $\ket{\Psi^-}$ into $\ket{\Psi^+}$, and vice versa. So, if one wishes to be inclusive of the output state produced to be (close to) the $\ket{\Psi^-}$ state as well, our expression for the probability of success, in Eq.~\eqref{eq:psuccess_cascaded} for instance, can be multiplied by $2$. See Appendix \ref{appendix:hybrid} for a derivation of the above results.

In what follows, we will represent the cascaded source as the orange box shown in Fig.~\ref{fig:diag_casc}. The cascaded source has a {\em heralding trigger}, telling us in which time slot a copy of $\ket{M}$ was produced successfully, a feature that was missing in the original SPDC-based entanglement source.

\section{Idealized Model of a Heralded Quantum Memory}
\label{section:ideal_memory}
Quantum memories (QMs) are an essential component of entanglement distribution protocols; especially so for building quantum repeaters for long-distance entanglement distribution, and in distilling high-Fidelity entanglement from low-Fidelity entangled qubit pairs. Memories that can efficiently load one qubit of a photonic entangled state, are necessary to store the quantum state for a time duration appropriate for the end application, or when it is ready to be interfaced to a larger quantum processor system, e.g., for performing teleported gates for distributed quantum computing. 

Although various proposals for quantum memories exist in the literature, for the purposes of the performance evaluation of the source we propose in this paper, we want to distill two important characteristics pertinent to our analysis: the QM can selectively load one dual-rail qubit (i.e., two orthogonal optical modes), and it has a heralding trigger. In other words, when the memory is successful in loading the qubit, it raises a (classical) binary-valued flag declaring {\em success} or {\em failure}.

We will consider a rather idealized model for such a memory: one that performs a vacuum-or-not (VON) measurement, in a quantum non-demolition (QND) way. The QND measurement performed by this QM on the two incident modes can be expressed by the following positive-operator-valued measure (POVM) operators:
\begin{align}
	\begin{split}
		{\hat \Pi}_0=\outprod{0,\!0};\quad {\hat \Pi}_1= {\hat I}_2-\outprod{0,\!0},
		\label{eqn:VON_POVM}
	\end{split}
\end{align}  
where ${\hat I}_2 = \sum_{m,n}|n,m\rangle \langle n,m |$ is the identity operator of the two-mode bosonic Hilbert space. If a two-mode optical quantum state $|\psi\rangle = \sum_{m,n}c_{m,n}|m,n\rangle$ is incident on this QM, with probability $p_{\text{vac}} = |c_{0,0}|^2$, the memory would raise a {\em failure} flag, and the post-measurement state will be vacuum $|\psi_{\text{vac}}\rangle = |0,0\rangle$, i.e., nothing would be loaded into the quantum memory. However, with probability $p_{\text{not-vac}} = 1 - p_{\text{vac}}$, the memory would raise the {\em success} flag, and the post-measurement state would be $|\psi_{\text{not-vac}}\rangle = N\big(|\psi\rangle - c_{0,0}|0,0\rangle\big)$, where $N = 1/\sqrt{1-|c_{0,0}|^2}$ is a normalization constant. 

An experimental proposal for this VON measurement, with a photonic-domain post-measurement state was conceived by Oi {\em et al.}\ in \cite{oi2013}, using a reversible V-STIRAP atom-photon interaction. A QND VON measurement, with the post-measurement state being stored in a spin-based qubit, is implicit in a recently-published experiment on measurement-device-independent (MDI) QKD to beat the repeater-less rate bound, using an asynchronous BSM based on a silicon-vacancy color center in a diamond nanophotonic chip~\cite{bhaskar2020,nguyen2019}.

The requirement we impose of a QM to have a heralding trigger is crucial to almost all quantum communication protocols. Practically, one way to achieve this is by using memories that entangle the incoming photonic state with the quantum state of the memory's internal qubit, for example, as in~\cite{bhaskar2020}. The heralding trigger consists of measuring the reflected photonic quantum state in the optical domain. The measurement outcome projects the quantum state of the qubit held by the QM into a local-unitary-equivalent of the photonic quantum state. 

\section{Performance Evaluation of the Cascaded SPDC Source}
\label{section:state_analysis}

In this section, we will present a detailed analysis of the cascaded source that includes coupling losses and detector non-idealities in the PNR-based BSM. We will compare the performance of the cascaded source with that of the original (SPDC-based entanglement) source.

\subsection{Fidelity, assuming ideal devices}

First, let us do a crude examination of the quality of the entangled states produced for both kinds of sources, assuming ideal devices, by looking at the proportion of the high-photon-order spurious states (multi-photon terms) to the desired Bell state. We label this metric as $ \mathcal{D}$. For the state $ \ket{\psi} $ generated by the original source shown in Eq.~(\ref{eqn:srcnative}), we get:
\begin{align}
	\mathcal{D} = \frac{p(\geq2)}{p(1)}\approx \frac{p(2)}{p(1)}=\frac{3N_s}{N_s+1}.
\end{align}
For the state $ \ket{M} $ generated by the cascaded source given in Eq.~(\ref{eqn:casc_src_st}), this proportion is given as follows:
\begin{align}
	\mathcal{D}'=\frac{4p(0)p(2)}{3 p(1)^2}=1,
\end{align}
 where, as stated above this approximate result only holds for $ N_s\lesssim 0.2 $.

\begin{figure}[h!]
	\centering
	\includegraphics[width=0.8\linewidth]{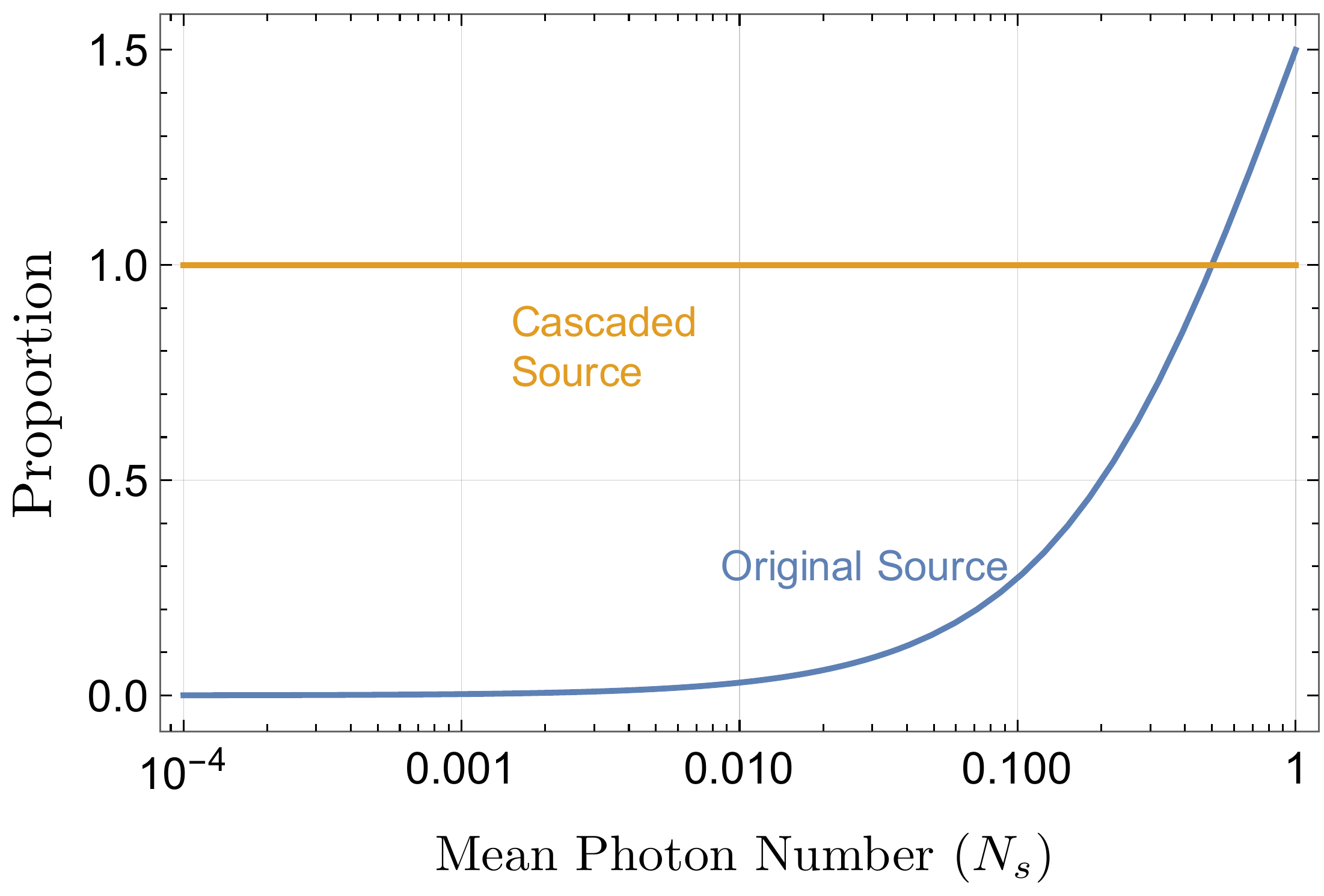}  
	\caption{Ratio of the proportion of higher photon number states to that of the $\ket{\Psi^+}$ state, for the original SPDC source (blue) and our proposed cascaded source (orange) as functions of the mean photon number per mode $(N_s)$, assuming ideal devices.}
	\label{fig:prop}
\end{figure}

We plot the above two expressions of $\mathcal{D}$ and $\mathcal{D}^\prime$ as functions of $N_s$, in Fig.~\ref{fig:prop}. As expected from the behavior of the geometric distribution, $\mathcal{D}$ increases monotonically with $N_s$. A curious observation, under the approximations we are making, is that $\mathcal{D}^\prime$ works out to exactly $1$, no matter the value of $N_s$. We recall that the output of the cascaded source $\ket{M}$ does not have any vacuum contribution, unlike $|\psi\rangle$, the output of the original source. Thus, as far as the $\mathcal{D}$-proportion metric is concerned, the only role $N_s$ (hence pump power) plays is in determining the success probability of the BSM within the cascaded source.

Next we compare the Fidelity of the generated states with the target Bell state $ \ket{\Psi^+} $. For the original source, this is given by: 
\begin{align}
 F(\ket{\psi^+},\ket{\Psi^+})= |\langle \psi^+ | \Psi^+ \rangle|^2 = \left| N_0 \sqrt{p(1)}\right|^2.
 \label{eqn:fid_orig}
\end{align} 
Similarly, the Fidelity of $ \ket{M} $ with $ \ket{\Psi^+} $ is given by:
\begin{align}
	F(\ket{M},\ket{\Psi^+})= |\langle {M} | \Psi^+ \rangle|^2=\left| N_0' \cdot\frac{p(1)}{\sqrt{2}}\right|^2 =\frac{1}{2},
	\label{eqn:fid_new}
\end{align}
 where the factor of $ 1/2 $ arises from  the unwanted cases where both detected photons came from the same source.

\begin{figure}
	\centering
	\includegraphics[width=0.9\linewidth]{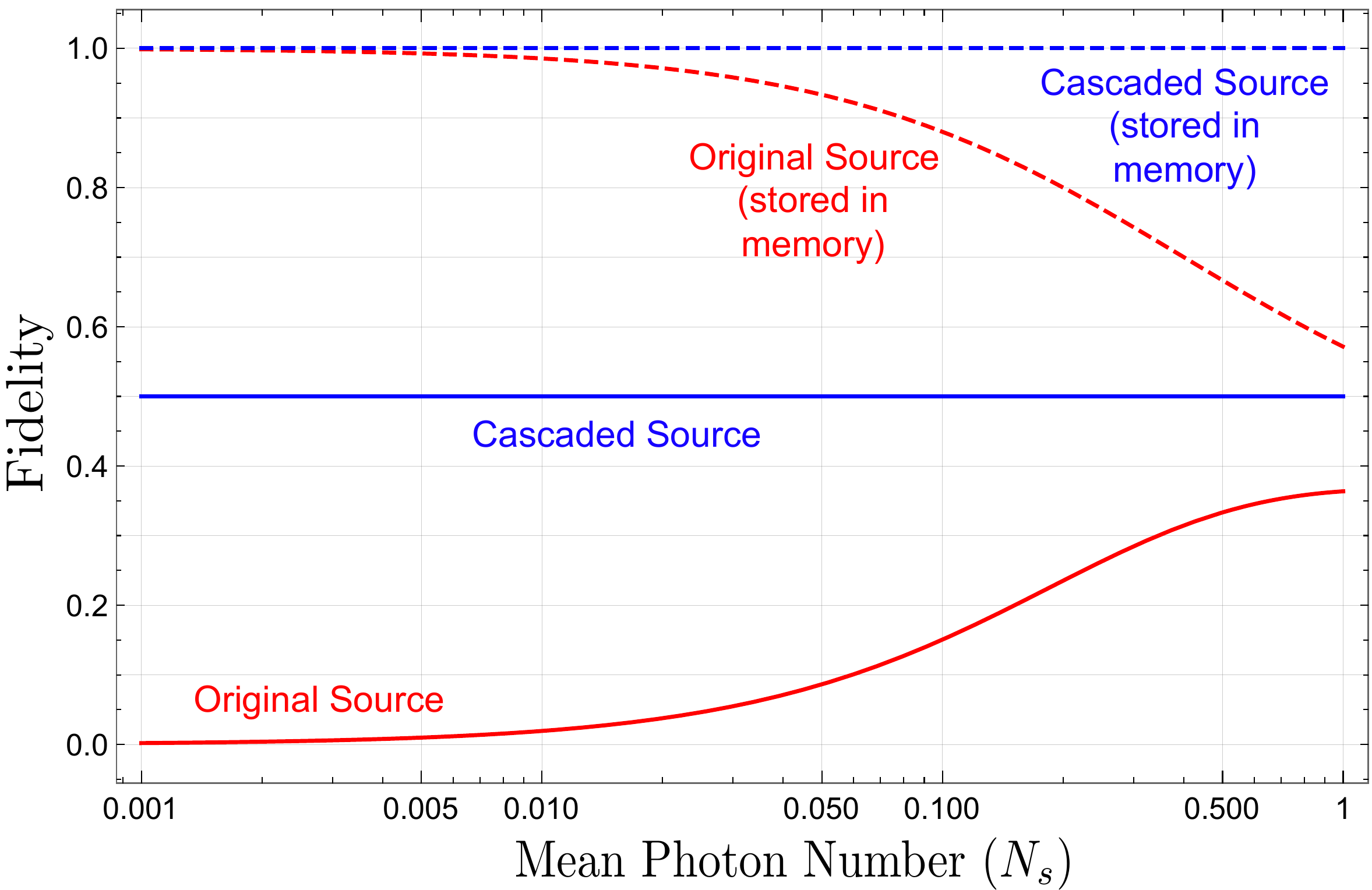}  
	\caption{Fidelity of the respective entangled states from the original SPDC source (red) and the cascaded source (blue) with the ideal Bell state $ \ket{\Psi^+} $, plotted as a function of the mean photon number per mode $(N_s)$. The corresponding state Fidelities after loading each mode pair (qubit) into an idealized quantum memory, are plotted using dashed lines.}
	\label{fig:fid}
\end{figure}

Next, let us consider the Fidelities with $\ket{\Psi^+}$ but when both mode pairs of the respective entangled states $\ket{\psi}$ (original source) or $\ket{M}$ (cascaded source) are loaded into a pair of idealized heralded quantum memories as described in Section \ref{section:ideal_memory}. It is simple to see by inspection of Eq.~\eqref{eqn:casc_src_st} that, after successful loading onto the ideal quantum memory, the cascaded source with ideal elements will yield a unit-Fidelity Bell state loaded onto the two QMs. All four of these Fidelities with $\ket{\Psi^+}$ (the original and the cascaded source, with and without a QM) are plotted as a function of $N_s$, in Fig.~\ref{fig:fid}. We see that the cascaded source, assuming ideal elements for the BSM, has a superior Fidelity compared with the original source.  Note that the preceding analysis assumes that the state generated by each `original' source is the pure state given by Eq.~\eqref{eqn:srcnative}. Realistic SPDC sources have an additional degree of freedom w.r.t. the temporal location of the emitted photons. This effect is commonly termed as a \textit{timing walk-off}, which affects the quantum description of the emitted SPDC state, which in turn sets an upper bound on the Fidelity (or equivalently, manifests as a minimum infidelity) of the emitted state from the cascaded source, in comparison to the target state, $ \ket{\Psi^+} $. We analyze this effect in detail in Appendix~\ref{appendix:timing_walkoff}.

\subsection{Probability of entangled state generation, assuming ideal devices}
\begin{figure}[h!]
	\centering
	\includegraphics[width=0.9\linewidth]{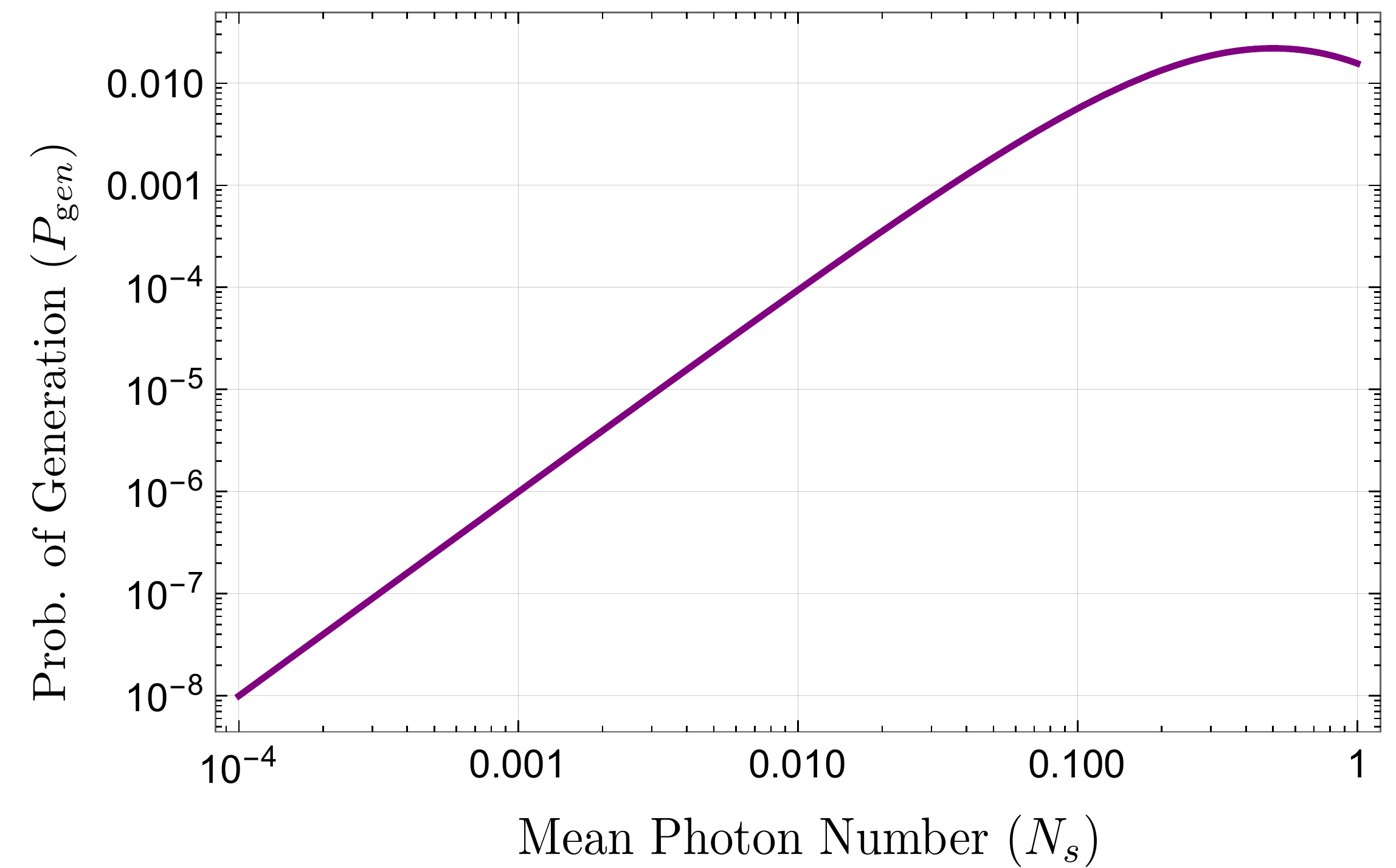}  
	\caption{Probability of generation of $ \ket{M} $ as a function of the mean photon number per mode, $N_s$, assuming ideal devices.}
	\label{fig:gen}
\end{figure}

 Since the cascaded source is a heralded-state generation scheme, there is a generation probability, $ P_{\text{gen}} $, which corresponds to the probability of the desirable click pattern and which is a function of $N_s$. Assuming ideal PNR detectors for the BSM and no other losses, this quantity is given by:
\begin{align}
	P_{\text{gen}}= \frac{1}{2} \times \left(\frac{p(1)^2}{4} + \frac{p(0)p(2)}{3}\right)=\frac{N_s^2}{(N_s+1)^6}.
	\label{eq:psuccess_cascaded}
\end{align}
 $ P_{\text{gen}} $ is plotted as a function of $ N_s $ in Fig.~\ref{fig:gen}. Here we note that the approximation made to simplify the state description is only valid up to $ N_s\approx0.2 $; above this threshold, the plots may be inaccurate (see Appendix~\ref{appendix:source}). Note that the first term in Eq.~\eqref{eq:psuccess_cascaded} describes the desired case where each source contributed one photon to the BSM, while the second term describes the undesirable case where one source produced two pairs and the other produced none.
 
\subsection{Including the effects of device non idealities}

\begin{figure}[h!]
	\centering
	\includegraphics[width=\linewidth]{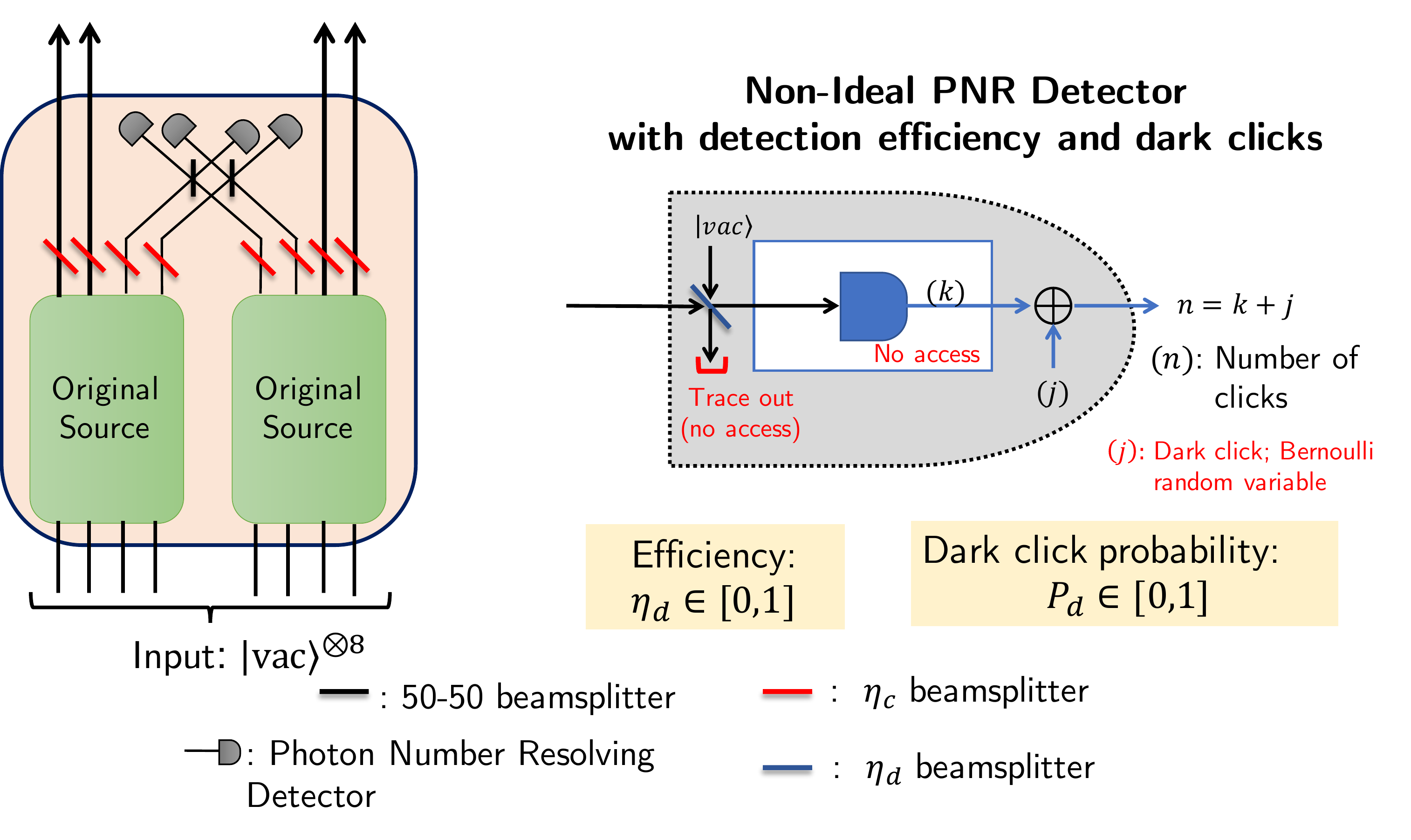}  
	\caption{Detailed schematic of the proposed cascaded source with all non idealities marked. This includes the coupling efficiency $ \eta_c $ (red beamsplitter) and the non-ideal detector- efficiency $ \eta_d$ (blue beamsplitter) and dark click probability $ P_d $ (Bernoulli random variable $ j $).}
	\label{fig:casc_source_complete}
\end{figure}

In this subsection, we will include the effects of device non-idealities into the analysis of the entangled state produced by the cascaded source. The specific non-idealities will consider in this section are as follows:

\begin{enumerate}
\item {\em Detector efficiency and dark clicks}---The leading candidates for PNR detection are the Transition Edge Sensor (TES) detectors~\cite{Arakawa2020-me,morais2020},  and Superconducting Nanowire Single Photon Detectors (SNSPDs)~\cite{Baghdadi2021-uq,cahall2017}. Each kind is influenced by multiple effects that degrade their performance~\cite{tan2016,morais2020}. In our present analysis, we will abstract off the non-ideality of a PNR detector into two parameters: a sub-unity detector efficiency $(\eta_d\leq1) $ and a non-zero dark click probability $(P_d\geq0)$ per detection gate (which will be assumed to be the length of a pulse slot for our calculations). A pictorial schematic of our model of this non-ideal PNR detector is given in Fig.~\ref{fig:casc_source_complete}. A detailed mathematical model of this two-parameter non-ideal PNR detector is discussed in Appendix~\ref{section:app_PNRD}.

\item {\em Coupling efficiency}---We will also account for losses in coupling the `inner' output modes of the SPDC sources (ones that go into the BSM) into single-mode fiber. We will label the effective efficiency of this coupling, per mode, as $\eta_c$. Our analysis of the derived formulae shows (see Appendices \ref{appendix:hybrid} and \ref{appendix:analytics}) that we can combine the two efficiency parameters into one efficiency parameter, i.e., $ \eta \equiv \eta_c \eta_d$, i.e., the output state will be identical for any given value of $\eta$, regardless of the actual values of $\eta_c$ and $\eta_d$, as long as their product equals $\eta$.
\end{enumerate}

Fig.~\ref{fig:casc_source_complete} depicts this complete model with all non-idealities accounted for.
When $P_d > 0$, the output of the cascaded source is a mixed state---a statistical mixture of pure states corresponding to various {\em true} click patterns in the detectors (several of which may not be one of the `desirable' patterns) which, with some probability, could result in the BSM using noisy detectors to conclude as a desirable pattern, and declare a success. The mixed state, derived in full detail as per the techniques in Appendix \ref{appendix:hybrid}, contains the ideal-device pure state as in Eq.\eqref{eqn:casc_src_st} along with other pure states that are generated when the apparent-desirable  click pattern actually includes one or more dark counts.

Describing the effects of detector efficiency $\eta_d$ is trickier. In general, the effect of $\eta_d < 1$ is equivalent to the detected modes being transmitted through a pure-loss bosonic channel of transmissivity $\eta_d$ prior to being incident on a unity-efficiency detector. Therefore, the pre-detection state is a mixed state, and the final density matrix for the present analysis is not compactly expressible. Appendix~\ref{appendix:hybrid} provides the detailed procedure for the calculations of this mixed state and Appendix~\ref{appendix:analytics} includes analytic formulae for the source performance metrics. Below, we present plots that shows the trend of the metrics under consideration, e.g., Fidelity and probability of success, as a function of $N_s$ at various point values of $\eta_d$ and $P_d$.  Note that, experimentally, TES detectors and SNSPDs have now demonstrated efficiencies above 98\%~\cite{fukuda2011,morais2020}. Finally, the effect of $\eta_c$ does not need to be discussed separately, since it can be subsumed into $\eta_d$, as discussed above.

\subsubsection{Analysis of the raw photonic entangled state produced by the cascaded source}

\begin{figure}[ht!]
	\centering
	\includegraphics[width=0.9\linewidth]{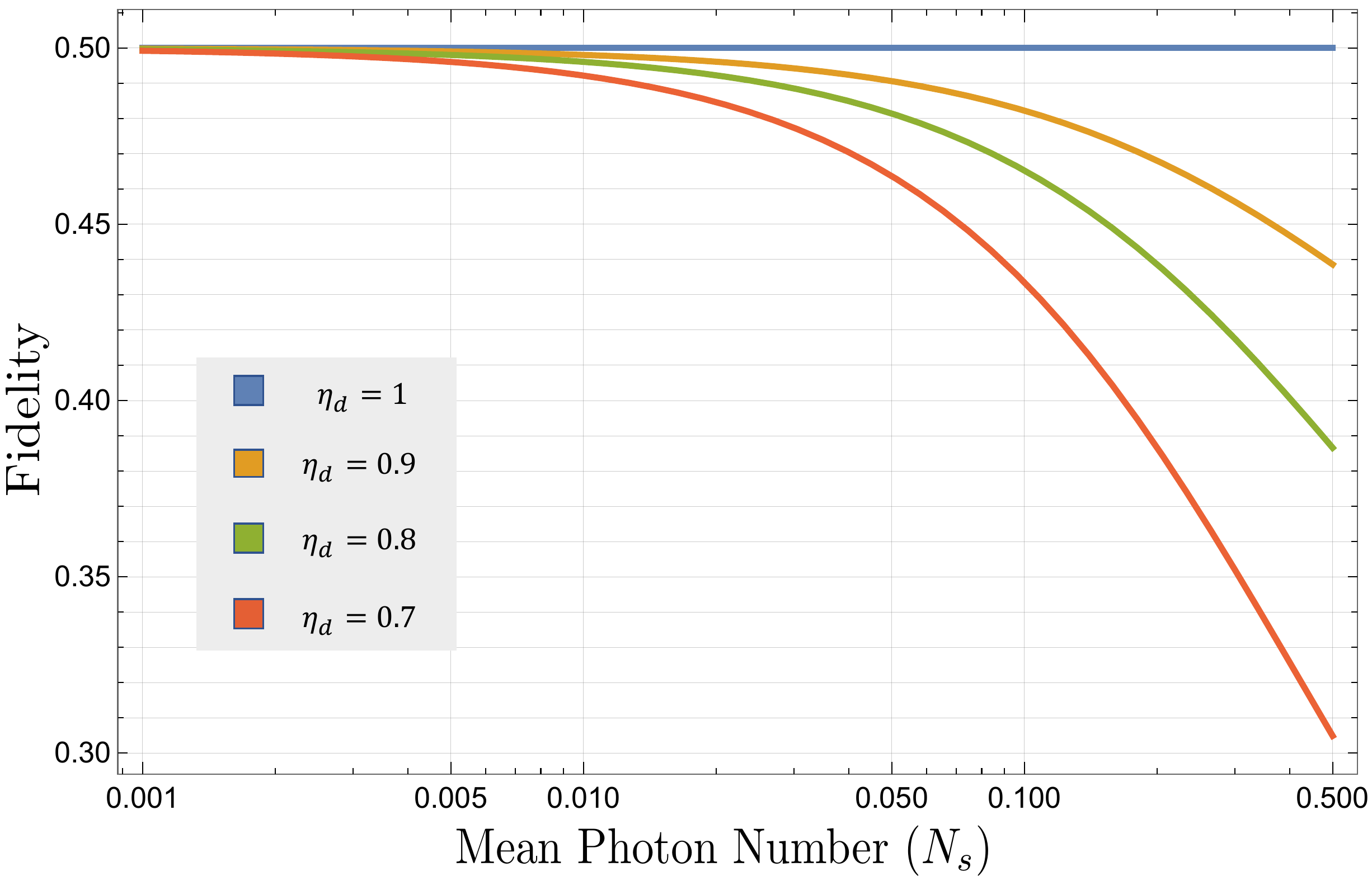}  
	\caption{Fidelity of the entangled state generated by the cascaded source with the ideal Bell state $\ket{\Psi^+}$ at various values of detector efficiency $ (\eta_d) $, plotted as a function of $N_s$. We assumed $P_d = 0$ for these plots.}
	\label{fig:fid_det}
\end{figure}
In Fig.~\ref{fig:fid_det}, we plot the Fidelity of the entangled state generated by the cascaded source with the ideal Bell state $\ket{\Psi^+}$ as a function of $N_s$, at various values of detector efficiency $\eta_d \le 1$, for $P_d = 0$. We note that the Fidelity decreases monotonically from $0.5$ (the maximum Fidelity attained by the cascaded source, as shown in Section~\ref{section:state_analysis}, Fig.~\ref{fig:fid}) with {\em increasing} $N_s$, for sub-unity efficiency, as expected. 

\begin{figure}[ht!]
	\centering
	\includegraphics[width=0.9\linewidth]{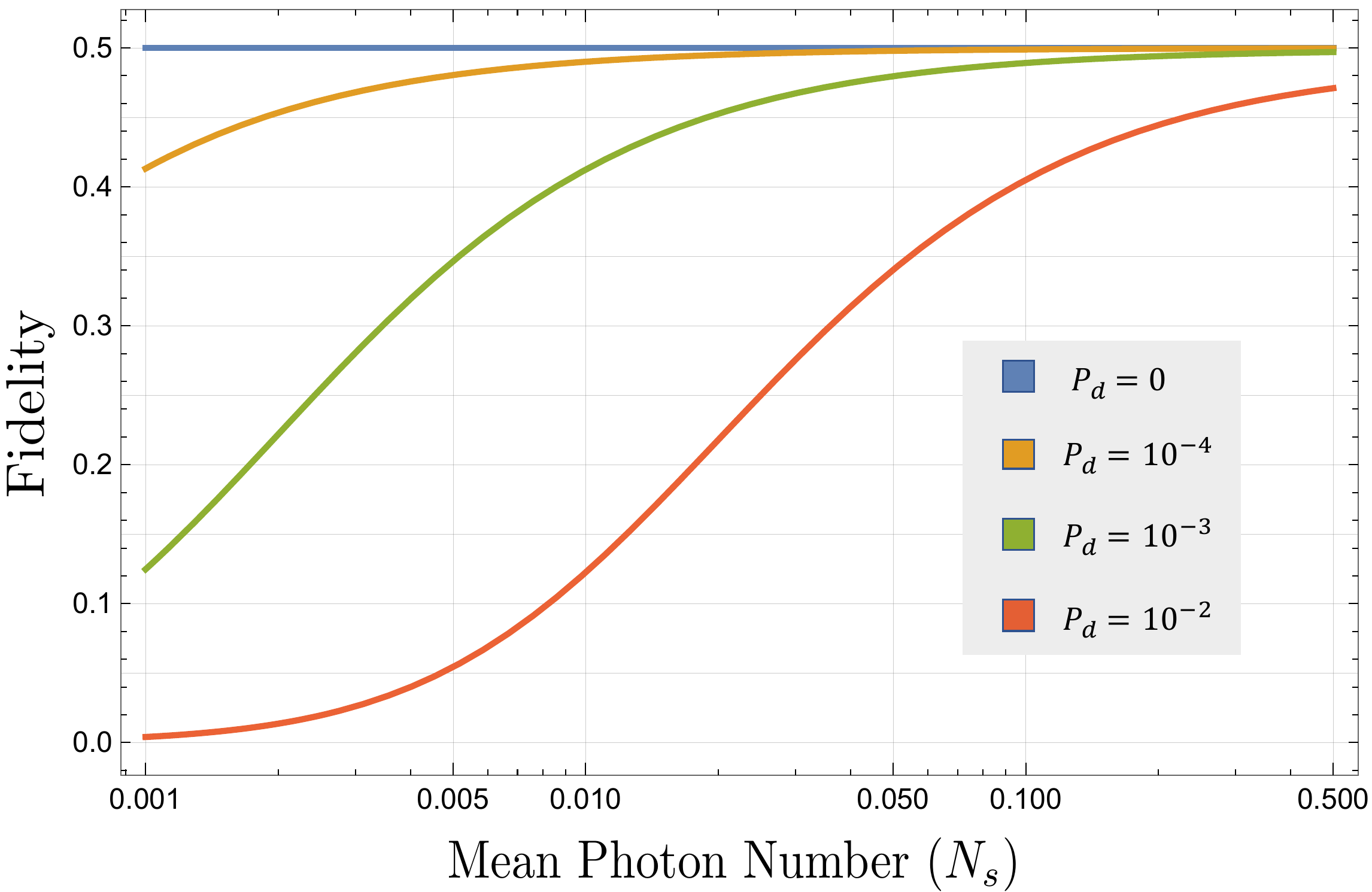}  
	\caption{Fidelity of the states from the cascaded source with the $ \ket{\Psi^+}$ Bell state for various values of detector dark click probability$ (P_d) $. We assumed $ \eta_d = 1 $ for these plots.}
	\label{fig:fid_dark}
\end{figure}
In Fig.~\ref{fig:fid_dark}, we see that as the dark click probability $P_d$ increases above zero, the Fidelity monotonically decreases with {\em decreasing} $N_s$ dropping to zero as $N_S \to 0$. This is expected, since for very low $N_s$, dark clicks account for most of the purported BSM ``success" events.

\begin{figure}[ht!]
	\centering
	\includegraphics[width=0.9\linewidth]{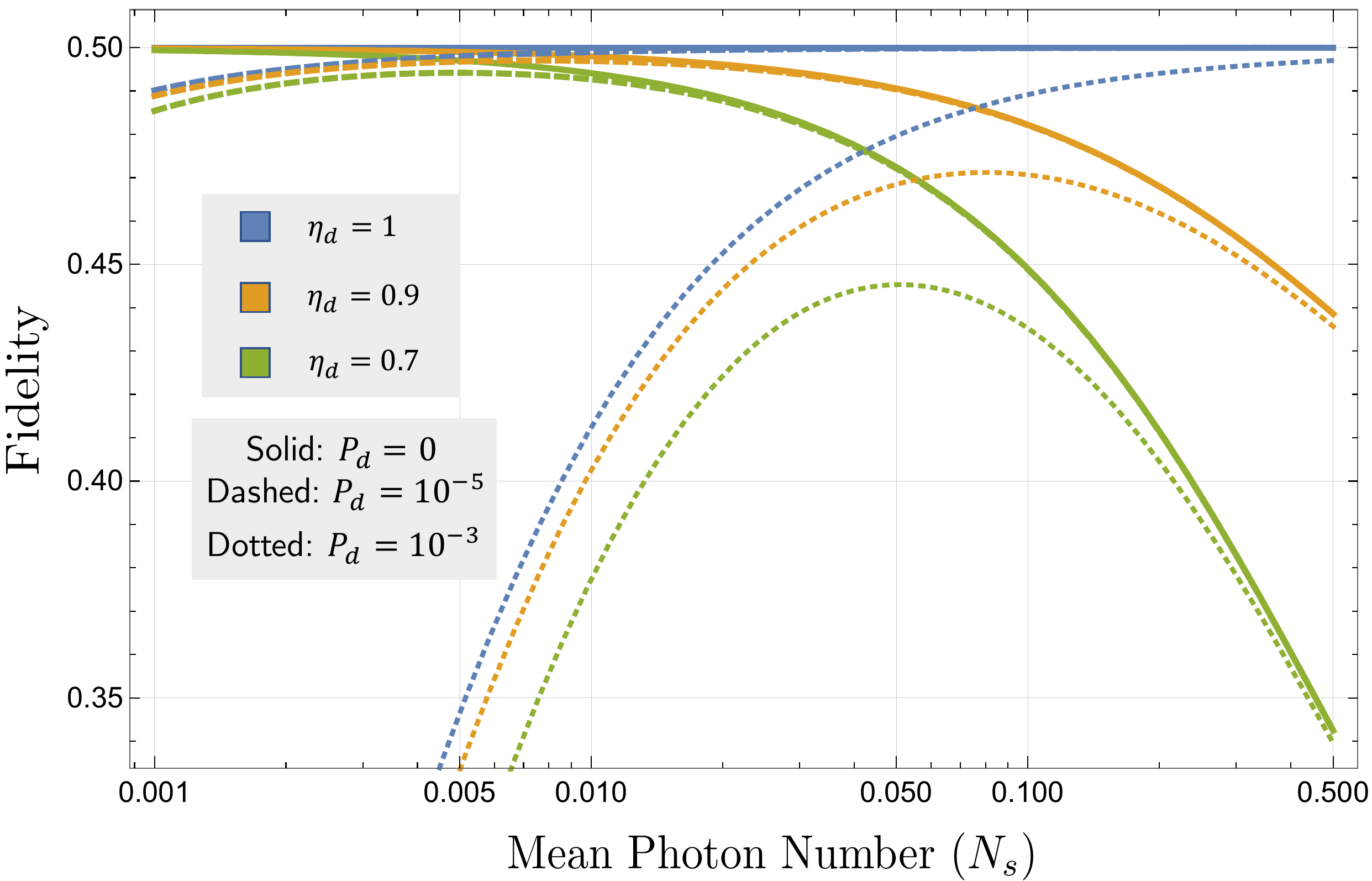}  
	\caption{Fidelity of the states from the cascaded source with the $ \ket{\Psi^+} $ Bell state for various values of detector dark click probability$ (P_d) $ and detector efficiency $ \eta_d $. The colors signify different values of $ P_d $ and the line style signifies $ \eta_d $.}
	\label{fig:fid_detectorn3}
\end{figure}
In Fig.~\ref{fig:fid_detectorn3}, we combine the effect of non-zero dark clicks ($P_d > 0$) and sub-unity detection efficiency ($\eta_d < 1$) in the PNR detectors used for the BSM. The Fidelity plots are exactly as expected---the two aforesaid forms of detector impairment pull the Fidelity down from the maximum possible value of $0.5$ at low $N_s$ and high $N_s$, respectively. 

\subsubsection{Analysis of the entangled state produced by the cascaded source, after successfully loading the qubits in a pair of ideal heralded quantum memories}
\begin{figure}[ht!]
	\centering
	\includegraphics[width=0.9\linewidth]{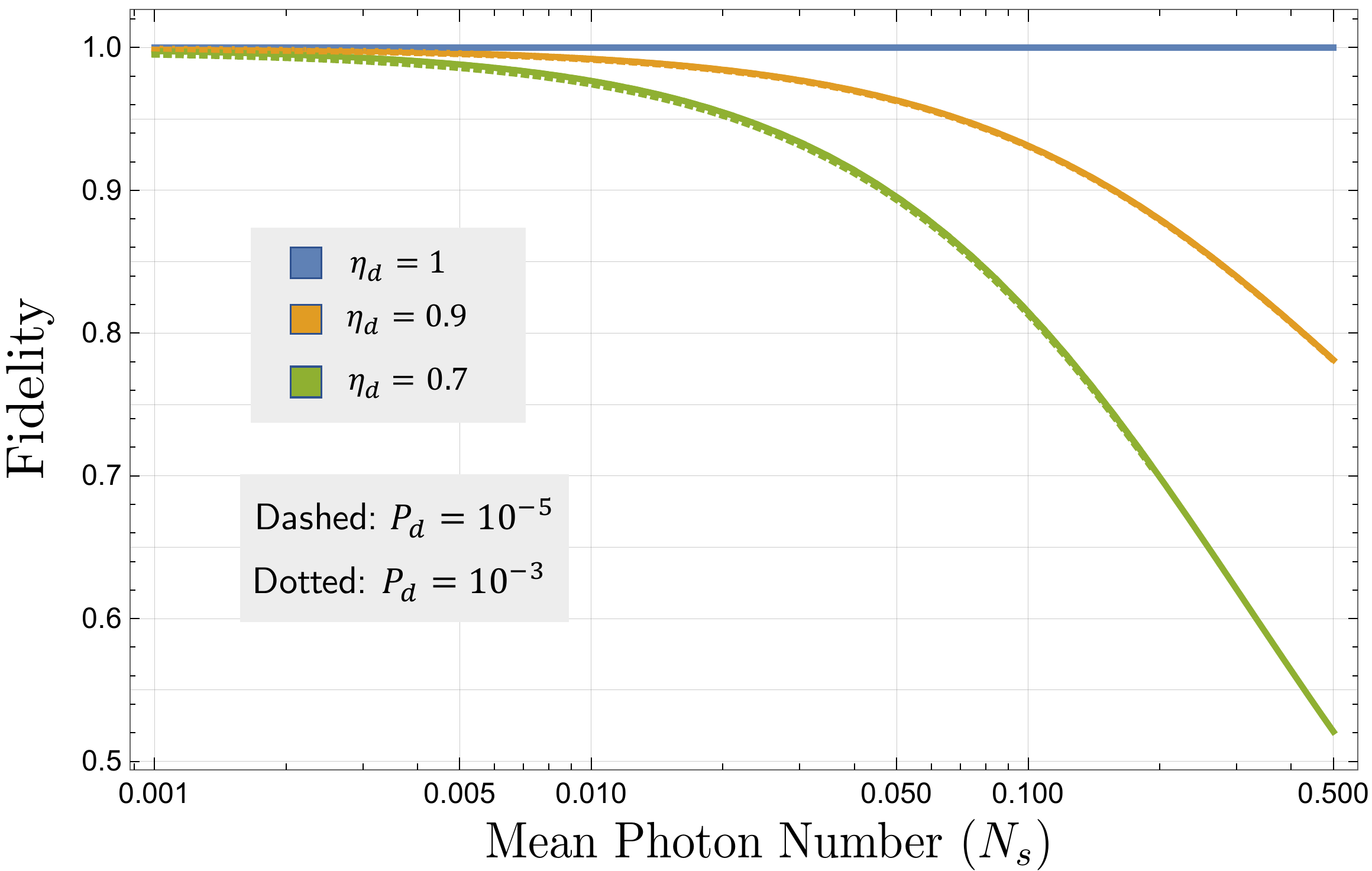}  
	\caption{Fidelity of the entangled states from the cascaded source with the $ \ket{\Psi^+} $ Bell state, after successfully loading both qubits into idealized heralded quantum memories. The colors signify different values of $ P_d $ and the line style signifies the different values of $ \eta_d $.}
	\label{fig:fid_VON_mixed}
\end{figure}
In Fig.~\ref{fig:fid_VON_mixed}, we plot the Fidelity of the entangled state with the $ \ket{\Psi^+}$ Bell state, {\em after} successfully loading the qubits into a pair of idealized heralded quantum memories. We remind the reader here of the conclusion in Section~\ref{section:state_analysis}, Fig.~\ref{fig:fid}---after successfully loading the entangled qubit pairs into idealized QMs, the Fidelity of the entangled state with the $ \ket{\Psi^+} $ Bell state, with an ideal BSM, is $1$ for the cascaded source, regardless of the value of $N_s$. The plots in Fig.~\ref{fig:fid_VON_mixed} show that the low-$N_s$ reduction of the Fidelity of the photonic entangled state produced by the cascaded source due to non-zero $P_d$, as seen in Fig.~\ref{fig:fid_dark}, is almost completely suppressed by the quantum memories. Here we have limited the discussion to the cascaded source, but the same arguments apply to the original source: The QND detection of the `outer' mode photons, implied by the ideal heralding quantum memories lifts the state fidelity at low $ N_s $ to 1 (see Fig.~\ref{fig:fid}), even in the presence of moderate BSM detector dark counts.

The aforesaid point is important, and is key to us being able to construct a near-deterministic near-unity-Fidelity source of entanglement. We do so by multiplexing several cascaded sources, as shown in the next Section.

\subsubsection{Analysis of the probabilities of success of generating the raw photonic entangled state by a cascaded source}

\begin{figure}[ht!]
	\centering
	\includegraphics[width=0.95\linewidth]{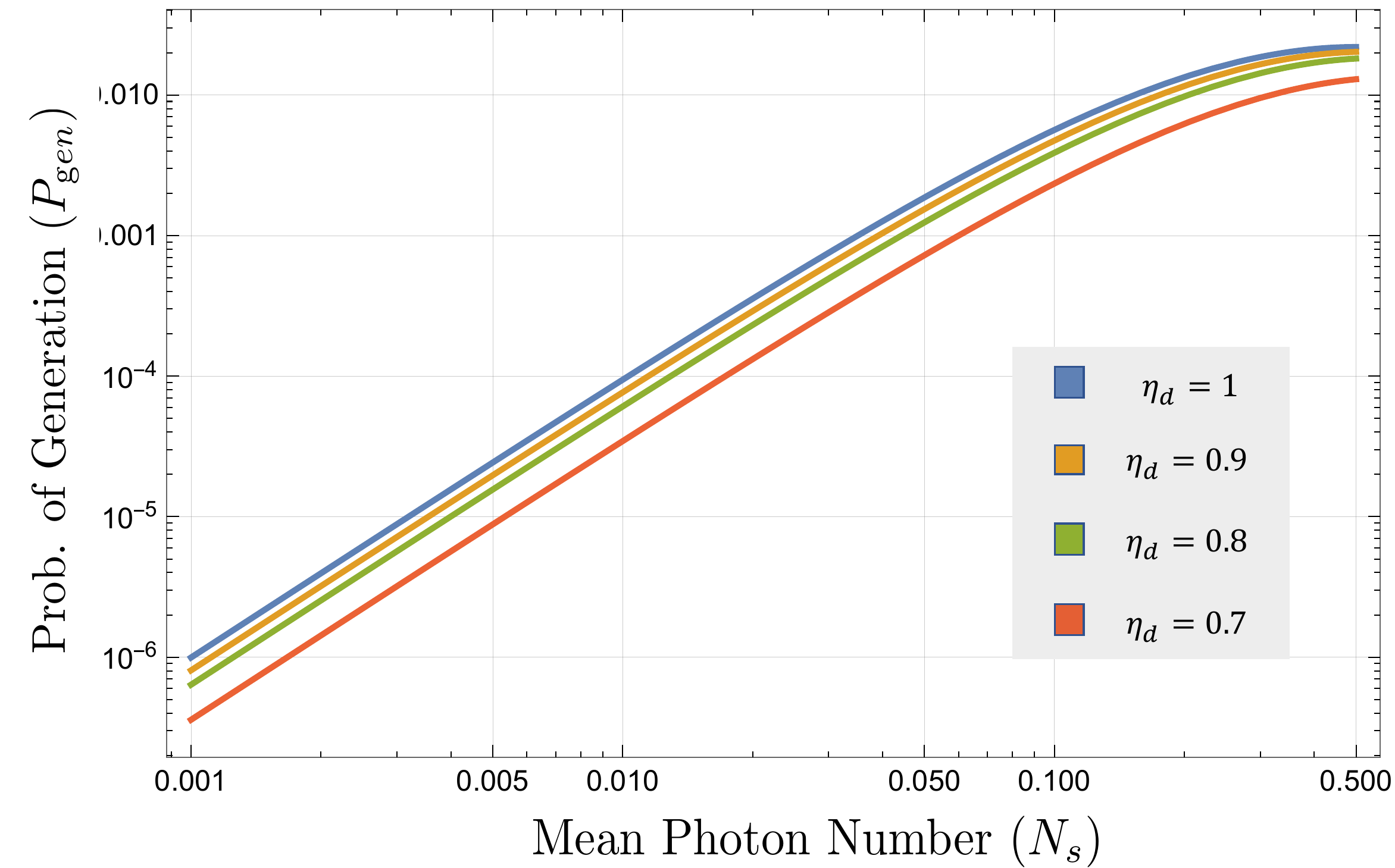}  
	\caption{Probability of generation of $ \ket{M} $ as a function of the mean photon number per mode $ (N_s) $ for various values of detector efficiency $ (\eta_d) $. We assume $ P_d = 0$.}
	\label{fig:pgen_det}
\end{figure}
In Fig.~\ref{fig:pgen_det}, we plot $P_{\rm gen}$, the probability of generation of $ \ket{M} $ by the cascaded source as a function of $N_s$, for a few different values of $\eta_d$ keeping $P_d = 0$. The entire $P_{\rm gen}$ versus $N_s$ plot shifts downwards with decreasing $\eta_d$ as compared to the ideal scenario.

\begin{figure}[ht!]
	\centering
	\includegraphics[width=\linewidth]{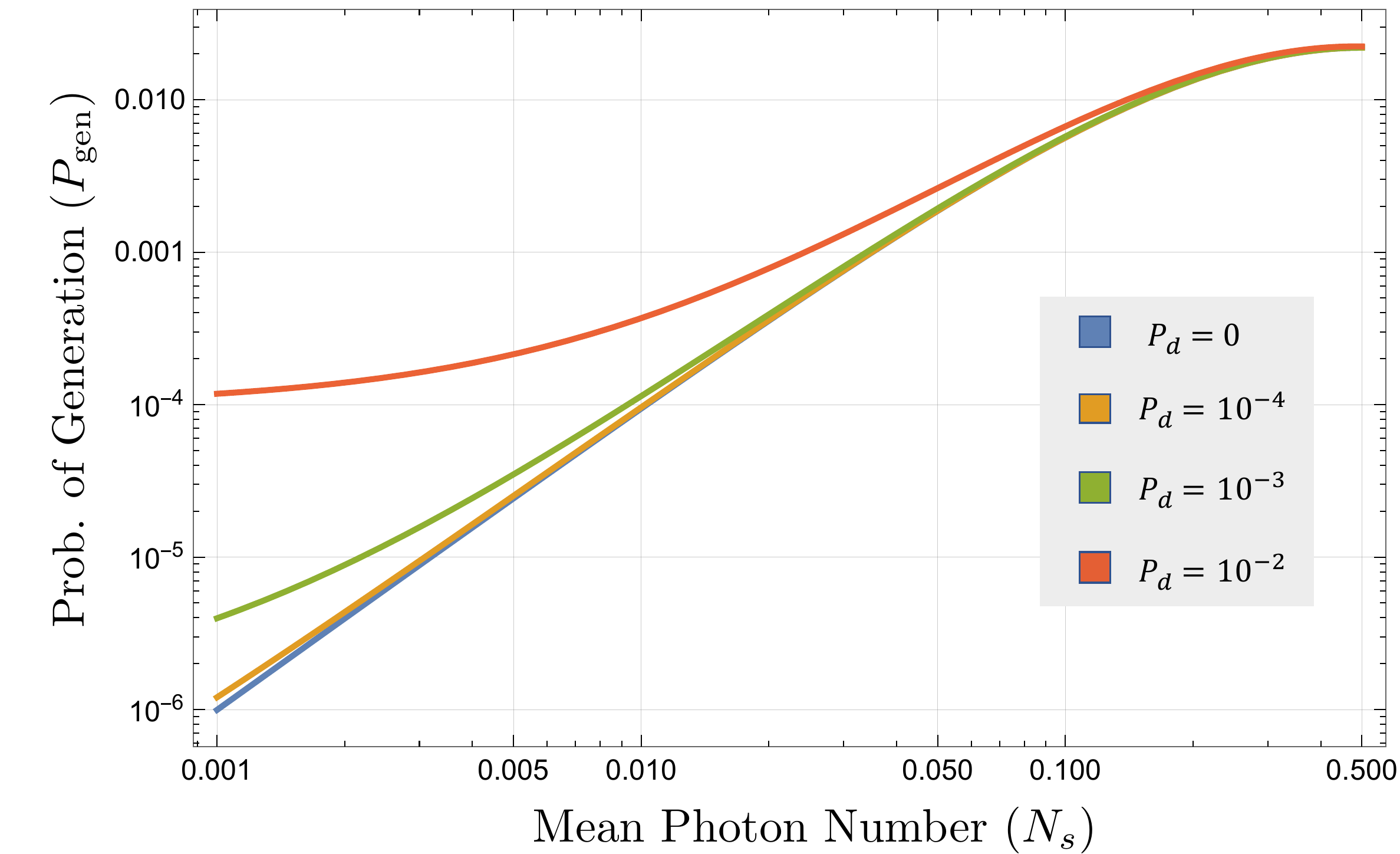}  
	\caption{Probability of generation of $ \ket{M} $ as a function of the mean photon number per mode $ (N_s) $ for various values of detector dark click probability $ (P_d) $. We assume $ \eta_d = 1 $.}
	\label{fig:pgen_dark}
\end{figure}
In Fig.~\ref{fig:pgen_dark}, we plot $P_{\rm gen}$ versus $N_s$ with $\eta_d = 1$ held fixed, but for a few non-zero values of $P_d$. Here, we see that, as $P_d$ increases, the probability of success increases, but most of those purported BSM `desirable' patterns occur due to dark-clicks, and as we already know from the low-$N_s$ regime of the Fidelity plots in Fig.~\ref{fig:fid_dark}, those spurious success events give very low Fidelity, i.e., close-to-useless output states. Interestingly, however, for higher $N_s$, the effect of non-zero dark clicks is almost completely washed away by the PNR-detection based BSM.

\begin{figure}[ht!]
	\centering
	\includegraphics[width=\linewidth]{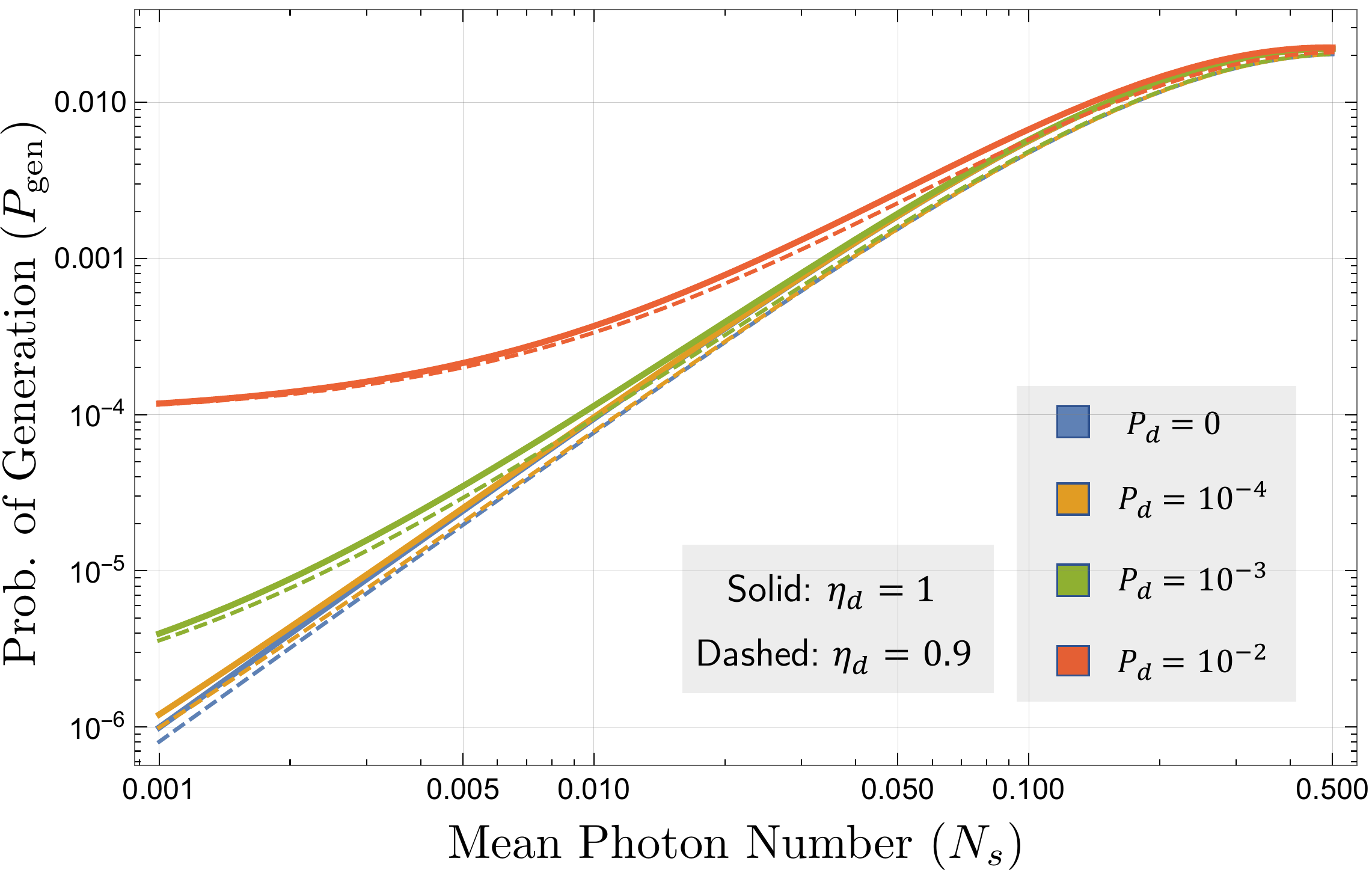}  
	\caption{Probability of generation of $ \ket{M} $ as a function of the mean photon number $ (N_s) $ for various values of detector dark click probability $ (P_d) $ and detector efficiency $ \eta_d $. The colors signify different values of $ P_d $ and the line style signifies the $ \eta_d $.}
	\label{fig:pgen_mixed}
\end{figure}
In Fig.~\ref{fig:pgen_mixed}, we plot $P_{\rm gen}$ versus $N_s$ including the effects of both detector impairments: $P_d > 0$ and $\eta_d < 1$. The results are self-explanatory, and exactly as expected by combining the qualitative effects from Fig.~\ref{fig:pgen_det} ($\eta_d < 1$, $P_d = 0$) and Fig.~\ref{fig:pgen_dark} ($\eta_d =1$, $P_d > 0$), respectively.

\section{Multiplexed Cascaded Source of On-Demand High-Fidelity Bell States}
\label{section:mux}
The greatest advantage afforded by the heralding trigger in the cascaded source is that it enables multiplexing (heralded) cascaded sources using an array of photonic switches, which releases an entangled state based on which source was successful in a given time slot. Such multiplexing has been shown to enable large enhancements in the success probability of HSPS~\cite{Kaneda2019-vg}. This construction, using a bank of $M > 1$ cascaded sources, with both output mode pairs of each of the $M$ sources fed into $ M$-to-$1$ optical switch-arrays (each built out of $\log_2M$ switches), assisted by electronic controllers, is shown in Fig.~\ref{fig:mux1}. The switching arrays output the state of one of the successful cascaded source in any time slot, assuming one or more succeeds in that time slot. If none succeed, the multiplexed source produces nothing. But the user of the source knows when such a {\em failure} event happens.

 If there were no additional losses in switching, we could generate the entangled state in Eq.~(\ref{eqn:casc_src_st}) with as high a success probability as we please, by increasing $M$ indefinitely. The probability that an ideal multiplexed cascaded source generates an entangled pair, $P_{\text{success}} = 1 - (1-P_{\text{gen}})^M$. To make the source near on-demand, we would pick $M \approx 1/ P_{\text{gen}}$, which would ensure that on average at least one of the cascaded sources in the bank would have their internal BSM declare a success.  But, this simple-minded seemingly indefinite increase of $P_{\text{success}}$ toward $1$ by increasing $M$ does not work when device non-idealities, especially the switching losses, are accounted for. Our modeling and analysis considers four device impairments: detection efficiency ($\eta_d$), coupling efficiency($\eta_c$), switching efficiency($\eta_s$) and dark-click probability($P_d$). There are two design choices: $M$ (number of cascaded sources) and $N_s$ (determines pump power). The performance of the source is quantified by the trade-off of Fidelity versus probability of success (i.e., rate of entangled pair production).



\subsection{Performance evaluation of the heralded-multiplexed source}

We consider a multiplexing scheme as described in the paragraph above (shown in Fig.~\ref{fig:mux1}), multiple cascaded sources make parallel attempts to generate the target photonic state. This output photonic entangled state is then loaded into a pair of ideal heralded quantum memories (shown as black boxes marked IQM). $M$ (number of cascaded sources) and $N_s$ (determined by pump power) are design parameters for the implementation of the heralded-multiplexed source. The device metrics in our model are quantified by: (1) coupling efficiency from the outputs of the cascaded source ($\eta_c$), (2) efficiencies of all the PNR detectors within the BSM, $ (\eta_d)$, (3) dark click probability (per qubit slot) of all the detectors in the BSM, $P_d$, and (4) switching losses per switch in the switch array, expressed as an effective transmissivity, $(\eta_s)$ (hence the overall effective transmissivity being $\eta_s^{\log_2M}$). The performance of the heralded-cascaded source is quantified by the 
probability of success $P_{\text{success}}$, of the multiplexed source producing an entangled pair of dual rail qubits, and the Fidelity of that state produced with respect to the $\ket{\Psi^+}$ Bell state.

\begin{figure}[h]
	\centering{\includegraphics[width=\linewidth]{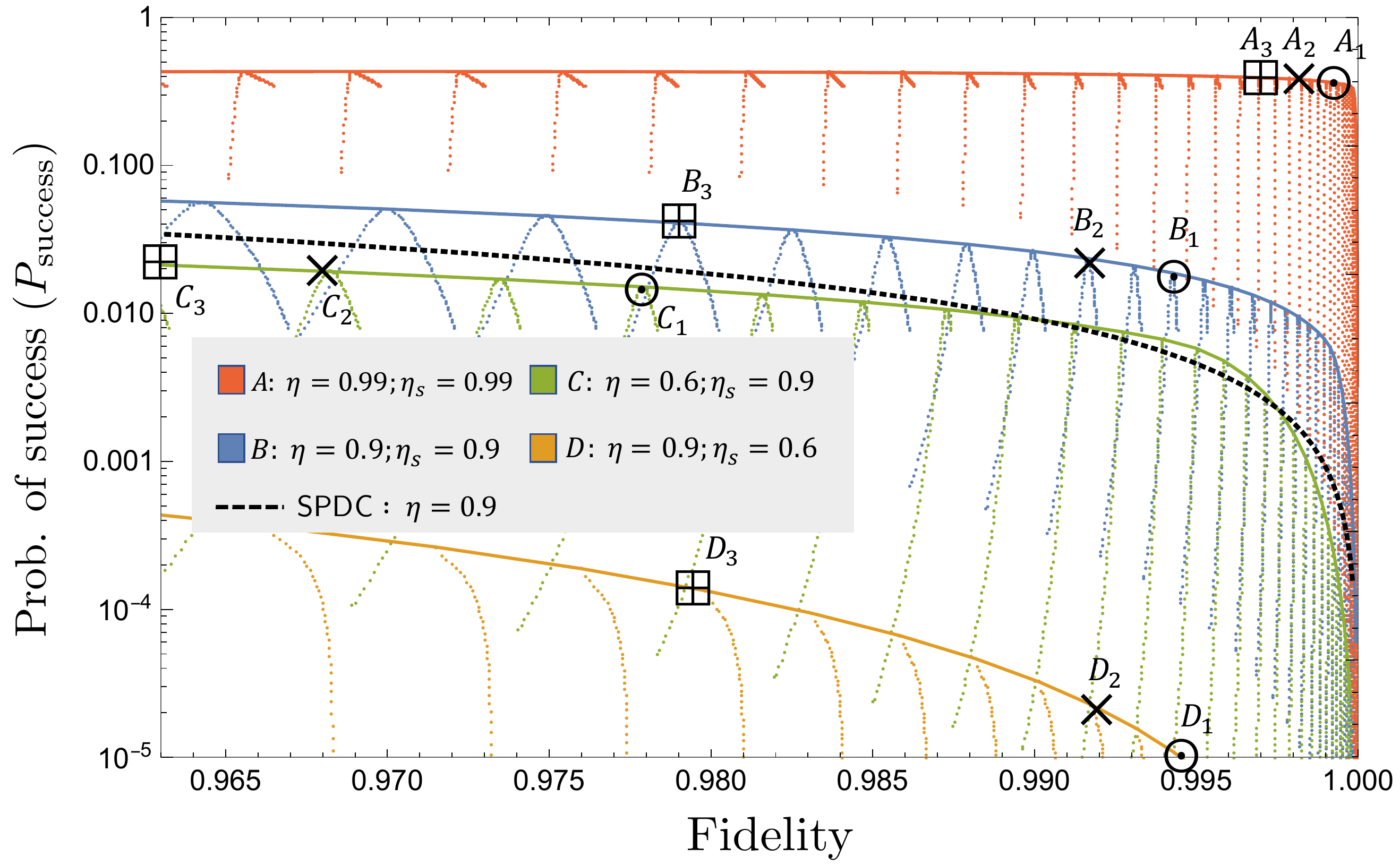} }
	\caption{Scatter plot of $ P_{\text{success}} $, the probability of successfully generating an entangled state by the heralded-multiplexed source and its Fidelity (with the ideal Bell state) for $M \in [1, 10^6]$ and $N_s \in [10^{-4}, 1]$, with $\eta = \eta_c\eta_d$ and $\eta_s$ held fixed. The solid lines are envelopes corresponding to the highest probability of success that can be achieved at a given Fidelity target. We see that for a given Fidelity target, there is an optimal value of $M$ that maximizes $ P_{\text{success}}$. The $(N_s, M)$ values corresponding to the marked points are summarized in Table~\ref{tab:valsFig15}. For comparison, the success probability of using a {\it single} SPDC source (dotted black curve) is shown; we assume its photons are also directed into the same sort of IQM (with a coupling efficiency $ \eta=0.9 $). }
	\label{fig:psucctr}
\end{figure}

\begin{table}[!htbp]
	\centering
	\begin{ruledtabular}
	\begin{tabular}{c  c c c c  c}
		\multirow{2}{*}{\textbf{Label }} &  \multirow{2}{*}{$ \eta $}&  \multirow{2}{*}{$ \eta_s $}&\multicolumn{3}{c}{\textbf{Subscript}}\\
		\midrule
		& & & 1 & 2 &3 \\
		\hline
		& \\[\dimexpr-\normalbaselineskip-0.5em]\\
		$A$& $ 0.99 $& $0.99 $ &$7.1\times 10^{4}$ &$3.2\times 10^{4}$&$5.8\times 10^{3}$\\
		$B$& $ 0.9$ & $0.9 $ &$3.2\times 10^{4}$ &$1.5\times 10^{4}$&$2.7\times 10^{3}$\\
		$C$& $ 0.6$ & $0.9 $ &$7.1\times 10^{4}$ &$3.2\times 10^{4}$&$5.8\times 10^{3}$\\
		$D$& $ 0.9$ & $0.6 $ &$1$ &$1$&$1$\\
		\hline
		& \\[\dimexpr-\normalbaselineskip-0.9em]\\
		$\mathbf{\textit{N}_s}$ \textbf{value}& & &$0.009$ & $0.015$ & $0.033$ \\
		\bottomrule
	\end{tabular}
\end{ruledtabular}
\caption{Table of $ M $ and $ N_s $  for  values for the marked points with coupling $ (\eta) $ and switching efficiencies $ (\eta_s) $ indicated, in Fig.~\ref{fig:psucctr}. Note that the $D$-points in Fig.~\ref{fig:psucctr} correspond to $\eta_s < 0.707$, and hence multiplexing ($M > 1$) does not improve $P_{\text{success}}$. This is why the optimum $M = 1$ for this row.}
\label{tab:valsFig15}. 
\end{table}

\begin{figure}
	\centering{\includegraphics[width=\linewidth]{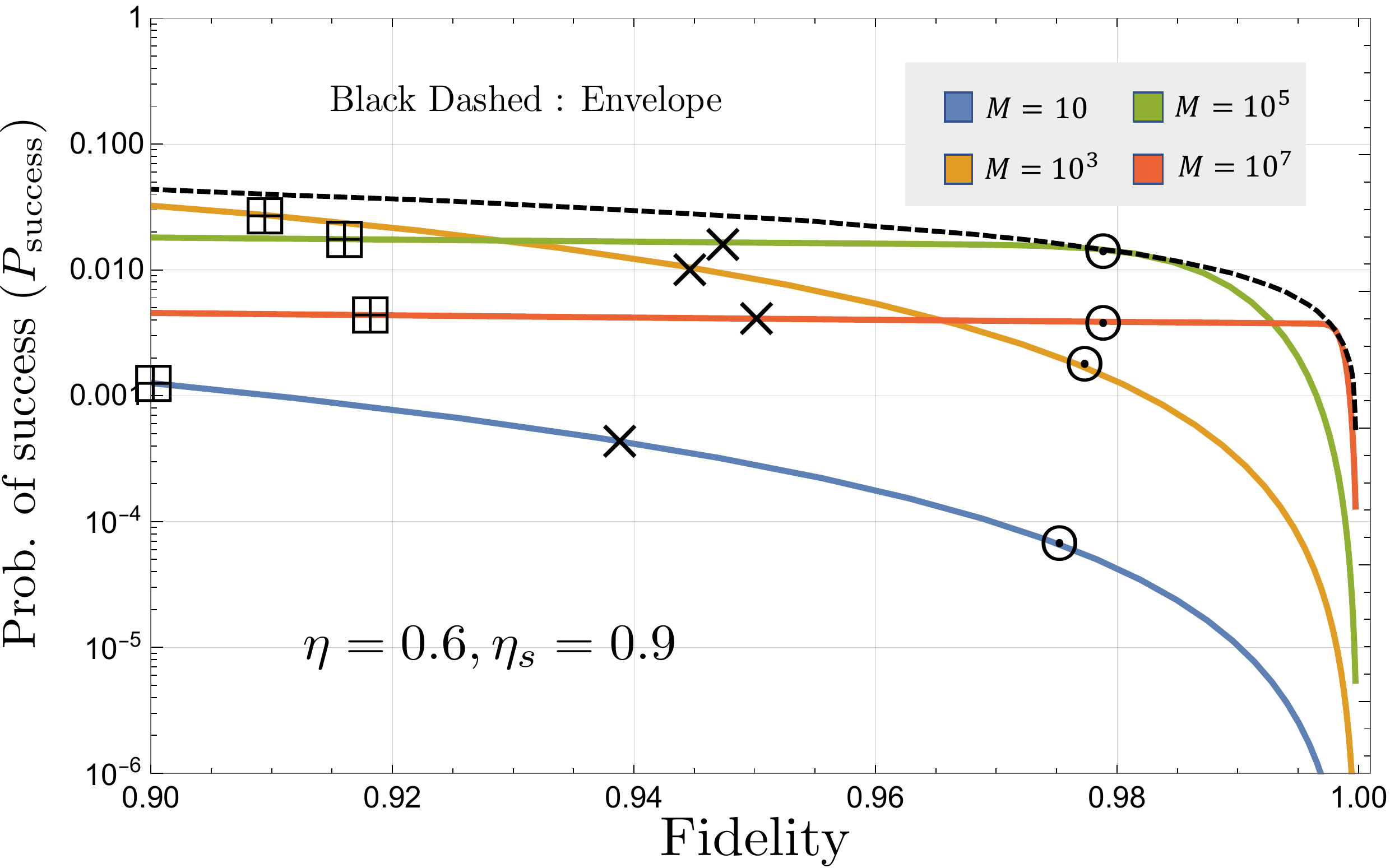} }
	\caption{Envelopes of $ P_{\text{success}} $, the probability of successfully generating an entangled state by the heralded-multiplexed source and its Fidelity (with the ideal Bell state) for varying values of $M$ and $N_s \in [10^{-4}, 1]$, with $\eta = 0.6$ and $\eta_s=0.9$. For low switching loss (i.e., higher $\eta_s$), in the high-Fidelity regime, the envelopes tend to go higher as $M$ increases. But the trend is opposite in the low-Fidelity regime. The envelope marked by the black dashed line, is identical to the green solid plot in Fig.~\ref{fig:psucctr}, i.e., the one corresponding to $\eta = 0.6$ and $\eta_s=0.9$. The marked points correspond to the following $ N_s $ requirements: $N_s=0.01\,(\odot);\; N_s=0.019,(\times);\; N_s=0.04\,(\boxplus) $. The plots here assume $P_d = 0$. }
	\label{fig:psuccvM1}
\end{figure}

\begin{figure}
	\centering{\includegraphics[width=\linewidth]{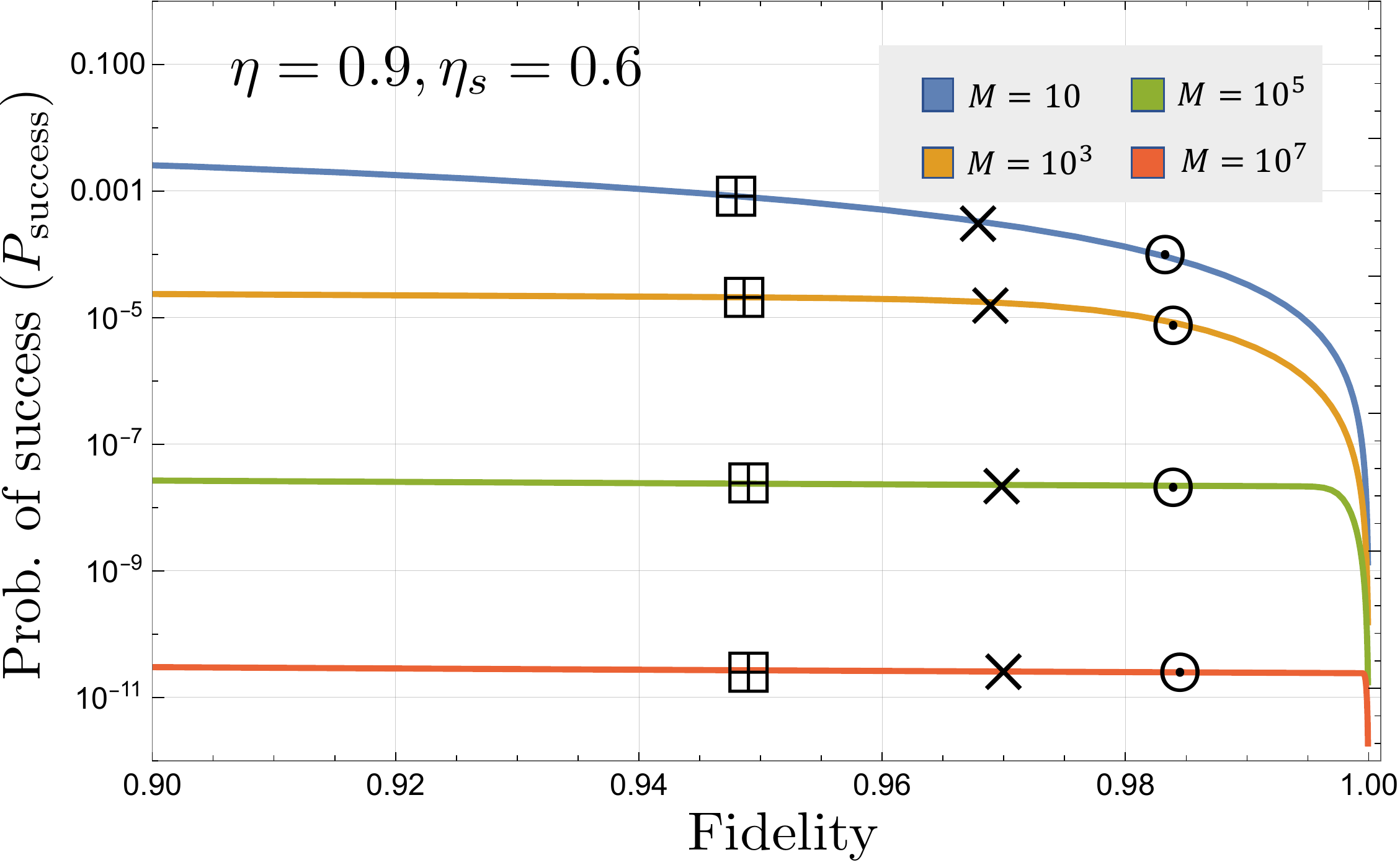} }
	\caption{Envelopes of $ P_{\text{success}} $, the probability of successfully generating an entangled state by the heralded-multiplexed source and its Fidelity (with the ideal Bell state) for varying values of $M$ and $N_s \in [10^{-4}, 1]$, with $\eta = 0.9$ and $\eta_s=0.6$. For high switching loss (i.e. lower $\eta_s$) the envelopes monotonically go lower as $M$ increases. The marked points correspond to: $ N_s $ requirements: $N_s=0.028 \,(\odot);\; N_s=0.05\,(\times);\; N_s=0.09 \,(\boxplus) $. The plots here assume $P_d = 0$. }
	\label{fig:psuccvM2}
\end{figure}

In Fig.~\ref{fig:psucctr}, we plot the trade-off between the Fidelity and $P_{\text{success}}$, for chosen values of $\eta = \eta_c \eta_d$ and $\eta_s$ in the form of a scatter plot, as $M$ and $N_s$ are both varied. Each point in this scatter plot represents a unique choice of $M$ and $N_s$, while the solid lines indicate the best trade-off the heralded-multiplexed source can achieve, given the device metrics $\eta_c$, $\eta_d$, and $\eta_s$, for $N_s \in [10^{-4}, 1]$ and $M \in [1, 10^6]$. In Figs.~\ref{fig:psuccvM1} and~\ref{fig:psuccvM2}, we plot the envelope of this aforesaid trade-off for fixed values of $M$ (while $N_s$ is varied). We note that for a given set of device metrics, there is an optimal value of $ M $ beyond which $ P_{\text{success}} $ cannot be increased any further. 

 Additionally, in Fig.~\ref{fig:psucctr} we also show for comparison the success probability of using a {\it single} SPDC source (dotted black curve), when we assume its photons are also directed into the same sort of ideal quantum memory (with a coupling efficiency $ \eta=0.9 $) that we have assumed for the multiplexed cascaded approach we introduce here. We note that performance is actually quite comparable to the much more resource-intensive approach discussed here. There are various limitations on the performance of the cascaded source, if the original SPDC sources are imperfect as discussed earlier in Section~\ref{section:state_analysis}.

\subsection{Effect of switching loss on system performance}

We note an interesting reversal in behavior when the switching loss per switch (quantified by $ \eta_s $) increases beyond a threshold value of $1.5$ dB (which corresponds to $ \eta_s = 1/\sqrt{2} \approx 0.707$). When the loss per switch is below this threshold, the envelope is seen to attain its maximal value for an optimum choice of $ M $ (see Fig.~\ref{fig:psuccvM1}). However, when the loss per switch is high, the trend reverses and increasing $ M $ is detrimental to the performance of the scheme (see Fig.~\ref{fig:psuccvM2}). This places a hard limit on the per-switch loss and the number of cascaded sources in a viable and useful implementation of the cascaded-multiplexed source. 

The intuitive reason for the aforesaid reversal in the trend is as follows. The size of the switching array scales as $\log_2(M)$. Assuming switching efficiency per switch of $\eta_s$ (i.e., $\log_{10}(1/\eta_s)$ dB of switching loss per switch), the output modes from a successful cascaded source undergo an additional loss corresponding to an effective transmission of $\eta_s^{\log_2 (M)}$. Now, unlike the case of the lossless switches, even though increasing the number of cascaded sources $M$ still increases the probability of success as $1-(1-P_{\text{gen}})^M$, it also decreases the probability that a successful output from one of the cascaded sources would be successfully loaded into the memory (because $\eta_s^{\log_2 (M)}$ decreases as $M$ increases). Given $M$ and $N_s$, the probability of success for a multiplexed source to successfully generate an entangled state is given by:
\begin{align}
	P_{\text{success}} =(1-(1-P_{\text{gen}})^M)\times(1-p_{\ket{00}})^2,
	\label{eqn:psucc}
\end{align}
where $ P_{\text{gen}}$ is the success probability of an individual cascaded source, and $p_{\ket{00}}$ is the probability the idealized quantum memory (IQM) on one side of the heralded-multiplexed source shown in Fig.~\ref{fig:mux1} {\em fails} to load the photonic qubit into the memory. This $p_{\ket{00}}$ term increases as $M$ increases due to compounding switching losses. The $M$-dependent portion of this second term $(1-p_{\ket{00}})^2$ is a multiplicative term: $\left[\eta_s^{\log_2 (M)}\right]^2$. When $P_{\rm gen}$ is small, the first term in Eq.~\eqref{eqn:psucc}, $1-(1-P_{\text{gen}})^M \approx MP_{\rm gen}$. It is simple to see that at $\eta_s = 1/\sqrt{2} \approx 0.707$, since $\left[\eta_s^{\log_2 (M)}\right]^2 = 1/M$, $P_{\text{success}}$ becomes insensitive to $M$. When $\eta_s \leq 1/\sqrt{2}$, $P_{\text{success}}$ decreases as $M$ increases, whereas for $\eta_s > 1/\sqrt{2}$, $P_{\text{success}}$ increases as $M$ increases.

\begin{figure}
	\centering{\includegraphics[width=\linewidth]{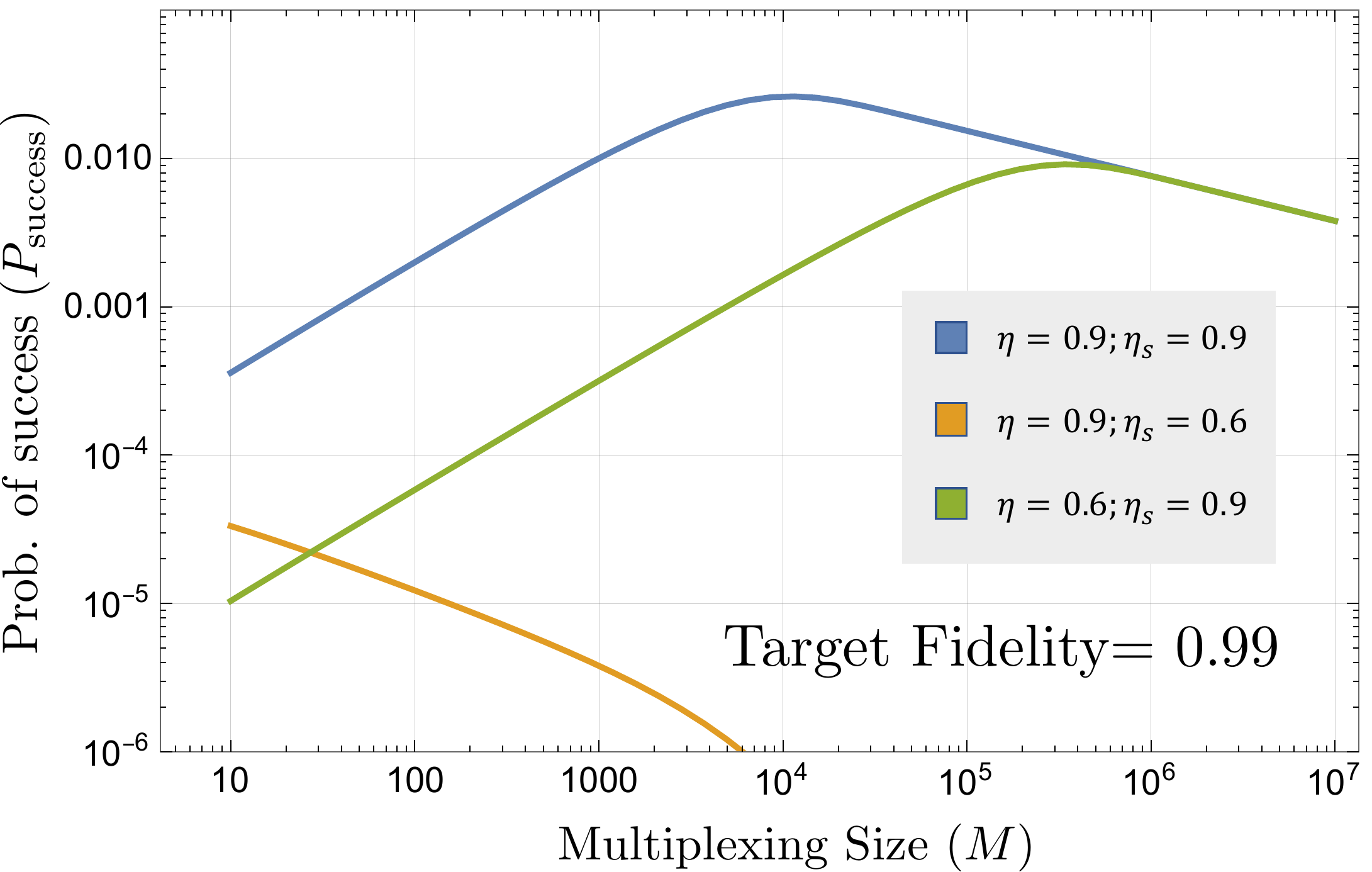} }
	\caption{$P_{\text{success}} $, as a function of the multiplexing size  $ (M) $ for a target  Fidelity of $0.99$. The pump power is optimized to achieve the given Fidelity target at the specified values of $\eta$ and $\eta_s$. For high switching loss (i.e., lower $\eta_s$; see the orange plot), we see $P_{\text{success}} $ decreasing (as opposed to increasing) as $M$ increases.}
	\label{fig:psuccvMfid}
\end{figure}
\begin{figure}
	\centering{\includegraphics[width=\linewidth]{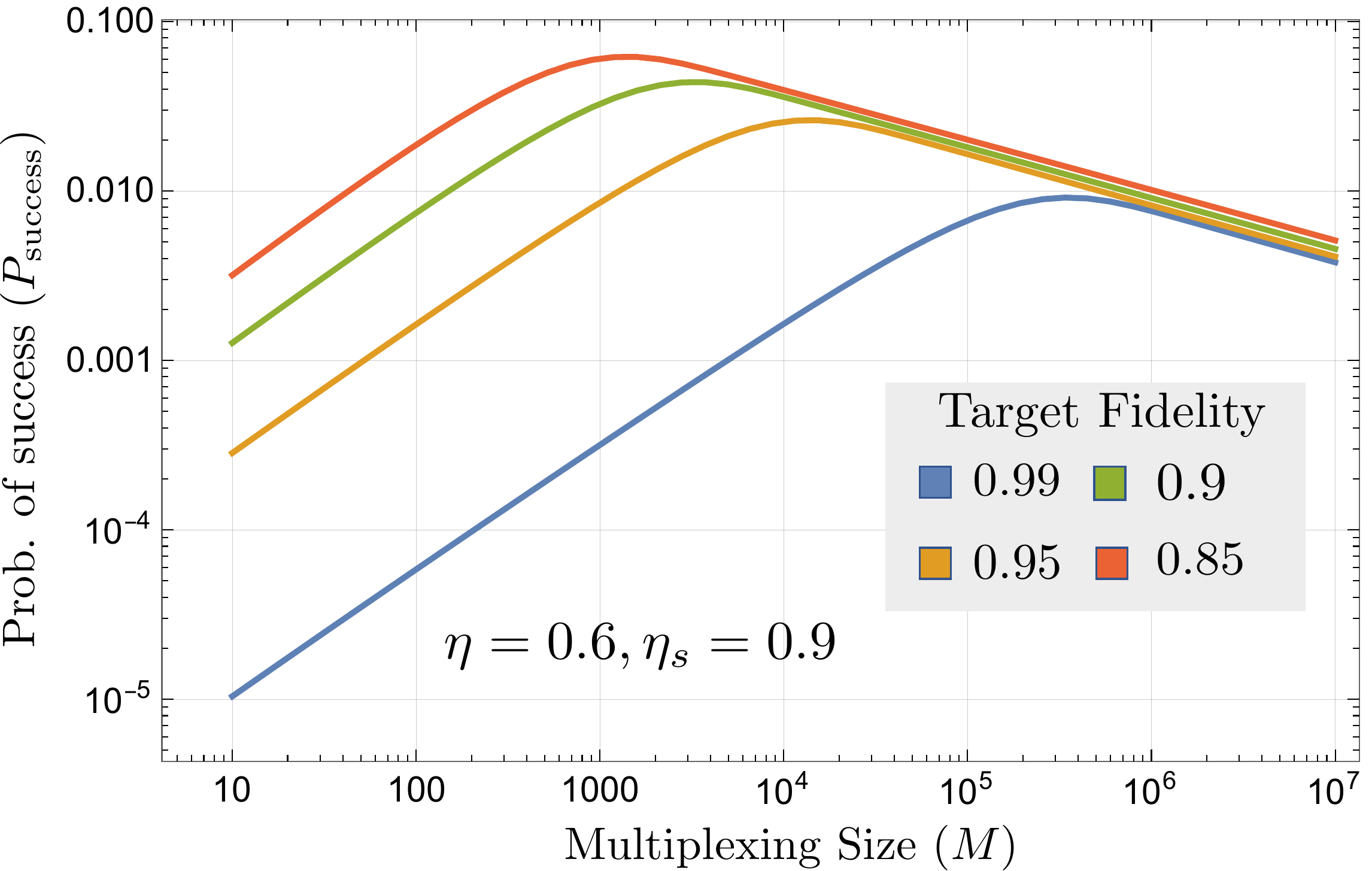} }
	\caption{$ P_{\text{success}} $, as a function of the multiplexing size  $ (M) $ for different target Fidelities. The pump power is optimized to achieve the given Fidelity target with $\eta = 0.9$ and $\eta_s=0.6$. The value of $M$ required to achieve the highest value of $ P_{\text{success}} $ increases as the target Fidelity is increased.}
	\label{fig:psuccvMsat}
\end{figure}
\begin{figure}
	\centering{\includegraphics[width=\linewidth]{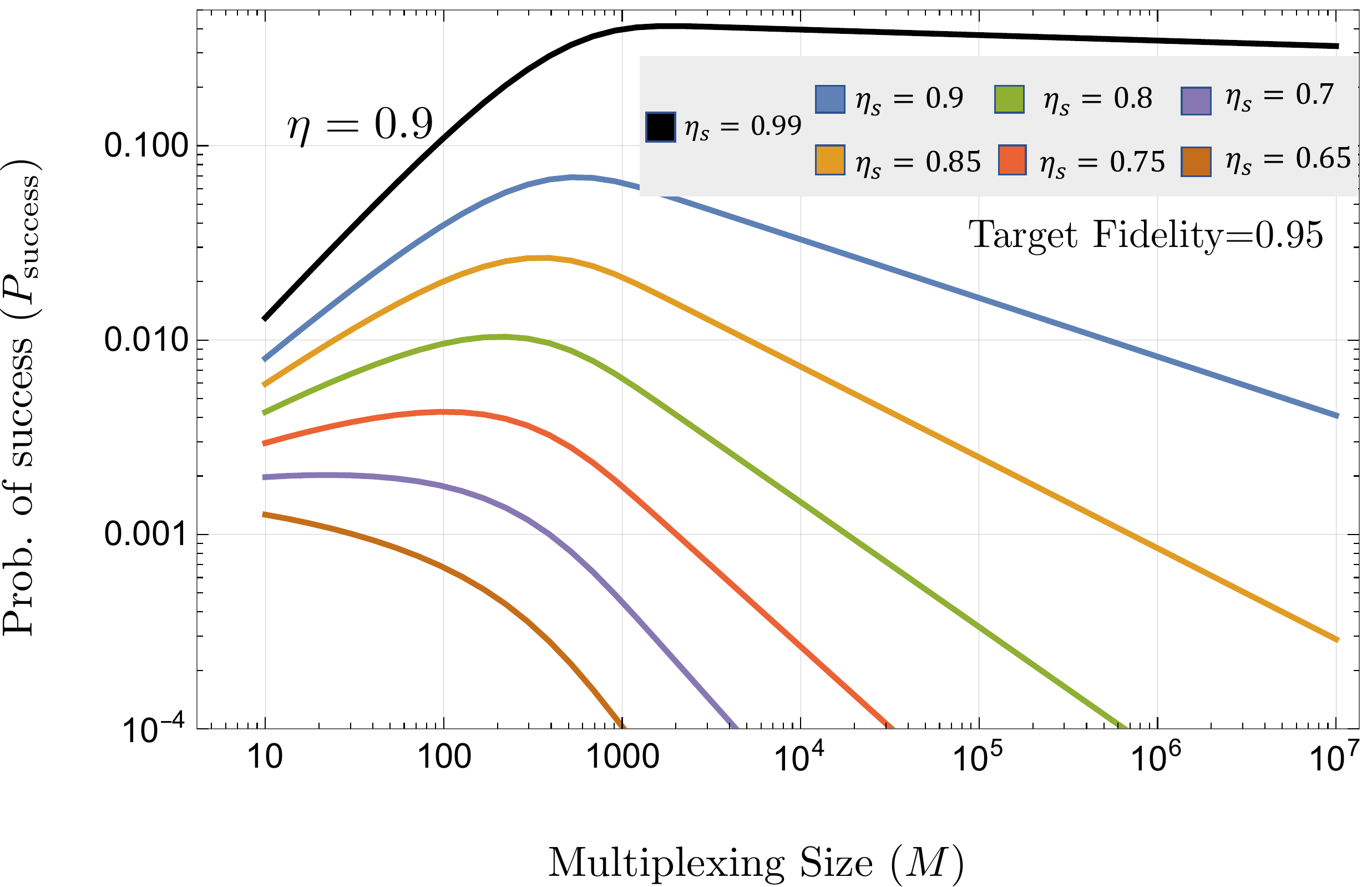} }
	\caption{$ P_{\text{success}} $, as a function of the multiplexing size  $ (M) $ for values of $ \eta_s\in(0.65,0.9)$ at a fixed $ \eta =0.9 $ and target Fidelity$=0.95 $. The pump power is optimized to achieve the given Fidelity target. We see that the value of $ M $ required to achieve the maximum value of $ P_{\text{success}} $ increases as $ \eta_s $ decreases, until at $ \eta_s \approx 0.7$ (purple line) when we see a complete turnaround and multiplexing is detrimental for the source's $ P_{\text{success}}$.}
	\label{fig:etaswturnaround}
\end{figure}
We plot $P_{\text{success}} $ as a function of $ M $ for a given Fidelity target $ (=0.99) $ in Fig.~\ref{fig:psuccvMfid}. We see that for one of the plots, for which a lower $\eta_s$ was chosen, $P_{\text{success}} $ decreases as $M$ increases, as discussed above. Further, we examine how the $ P_{\text{success}} $ as a function of $ M $ behaves in the regime of high $ \eta_s $ in Fig.~\ref{fig:psuccvMsat}. We observe that for every $(\eta, \eta_s)$ combination, where $ \eta_s>1/\sqrt{2} $, $ P_{\text{success}} $ is maximized for an optimal value of $ M $. This optimal value of $ M $ increases as we increase the target fidelity, as seen in Fig.~\ref{fig:psuccvMsat}. In the plots in Fig.~\ref{fig:etaswturnaround}, we numerically extract the per-switch loss (value of $\eta_s$) where the $ P_{\text{success}} $ versus $ M $ trend reverses, for a given value of target fidelity and $ \eta $. We find that this turnaround happens at around $\eta \approx 0.7$. As expected, and as explained in the text, this value of $\eta_s = 1/\sqrt{2}$ (corresponding to $1.5$ dB of loss per switch) where the trend reverses, is not affected by the other losses in the system (i.e., $\eta = \eta_c\eta_d$) and the Fidelity target we impose on the cascaded-multiplexed source. 

\subsection{The effect of detector dark clicks}

\begin{figure}
	\centering{\includegraphics[width=\linewidth]{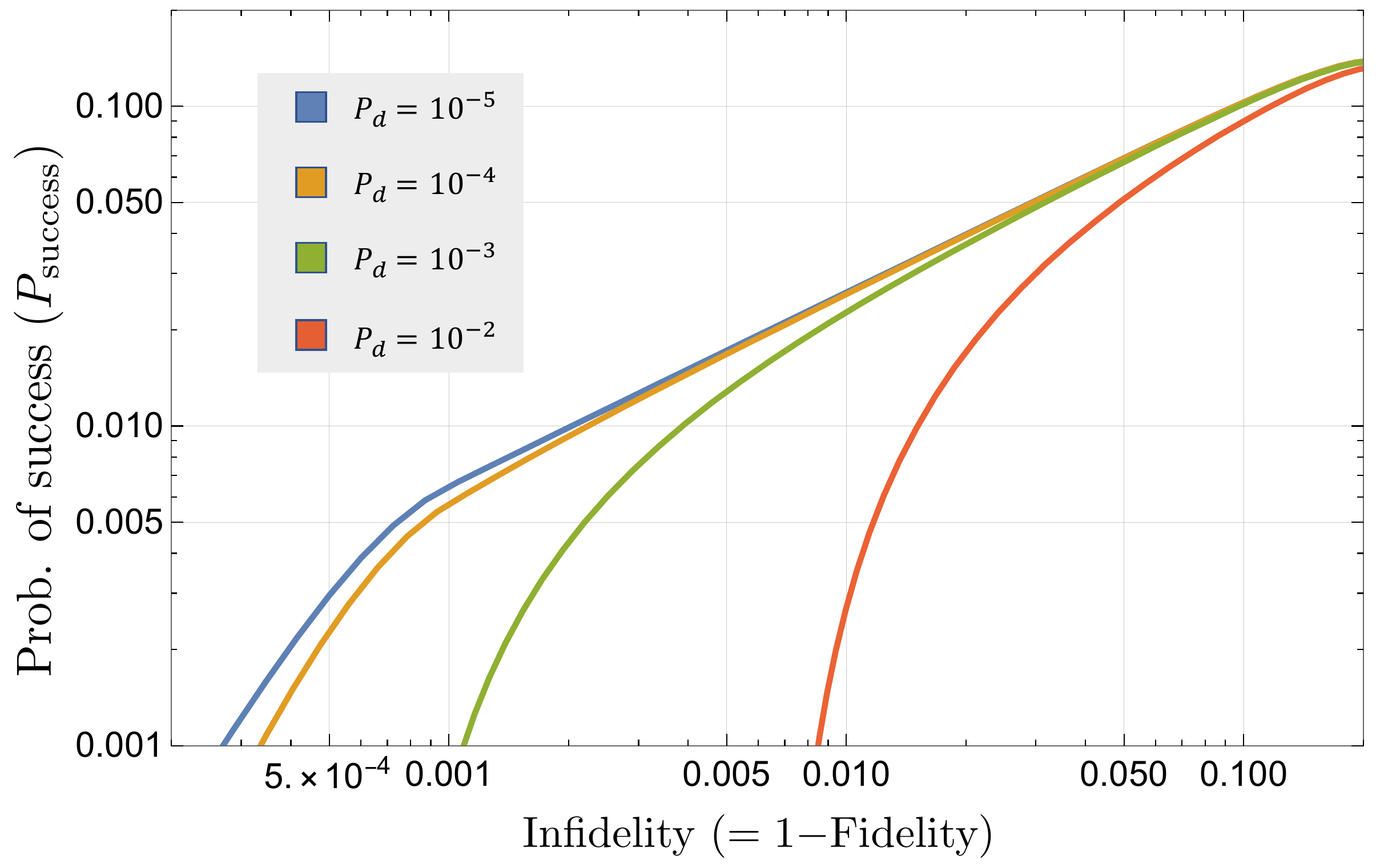} }
	\caption{ $ P_{\text{success}} $ and infidelity trade-off at various values of $ M $ and $N_s$. These plots assumed $\eta =0.9$ and $\eta_s =0.9$. The solid lines correspond to the probability of success for a given infidelity target. Note that with higher dark count probability the achievable fidelity becomes more restricted.}
	\label{fig:psucc_dark}
\end{figure}

\begin{figure}
	\centering{\includegraphics[width=\linewidth]{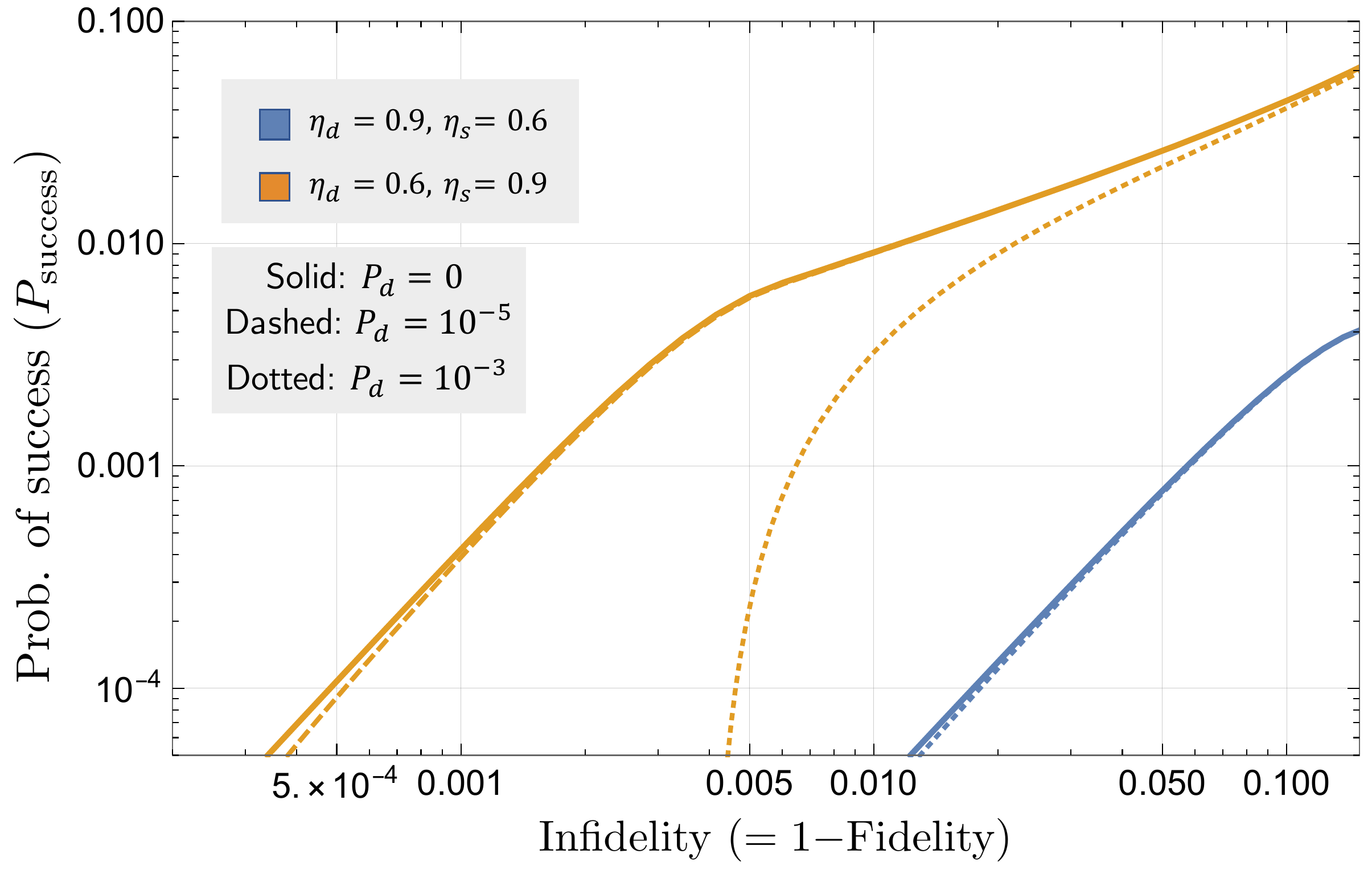} }
	\caption{ $ P_{\text{success}} $ and infidelity trade-off at various values of $ M $ and $N_s$. The values of  $\eta = 0.9$, $\eta_s = 0.6$ (blue) and $\eta = 0.6$, $\eta_s = 0.9$ (orange). We vary $ P_d $  between $ 0 $ (solid),  $ 10^{-5} $ (dashed)  and $ 10^{-3}$ (dotted). Note that with higher dark count probability the achievable fidelity becomes more restricted. The solid and dashed lines are largely indiscernible because they mostly overlap with one another. }
	\label{fig:psucc_dark_loss}
\end{figure}
\begin{figure}
	\centering{\includegraphics[width=\linewidth]{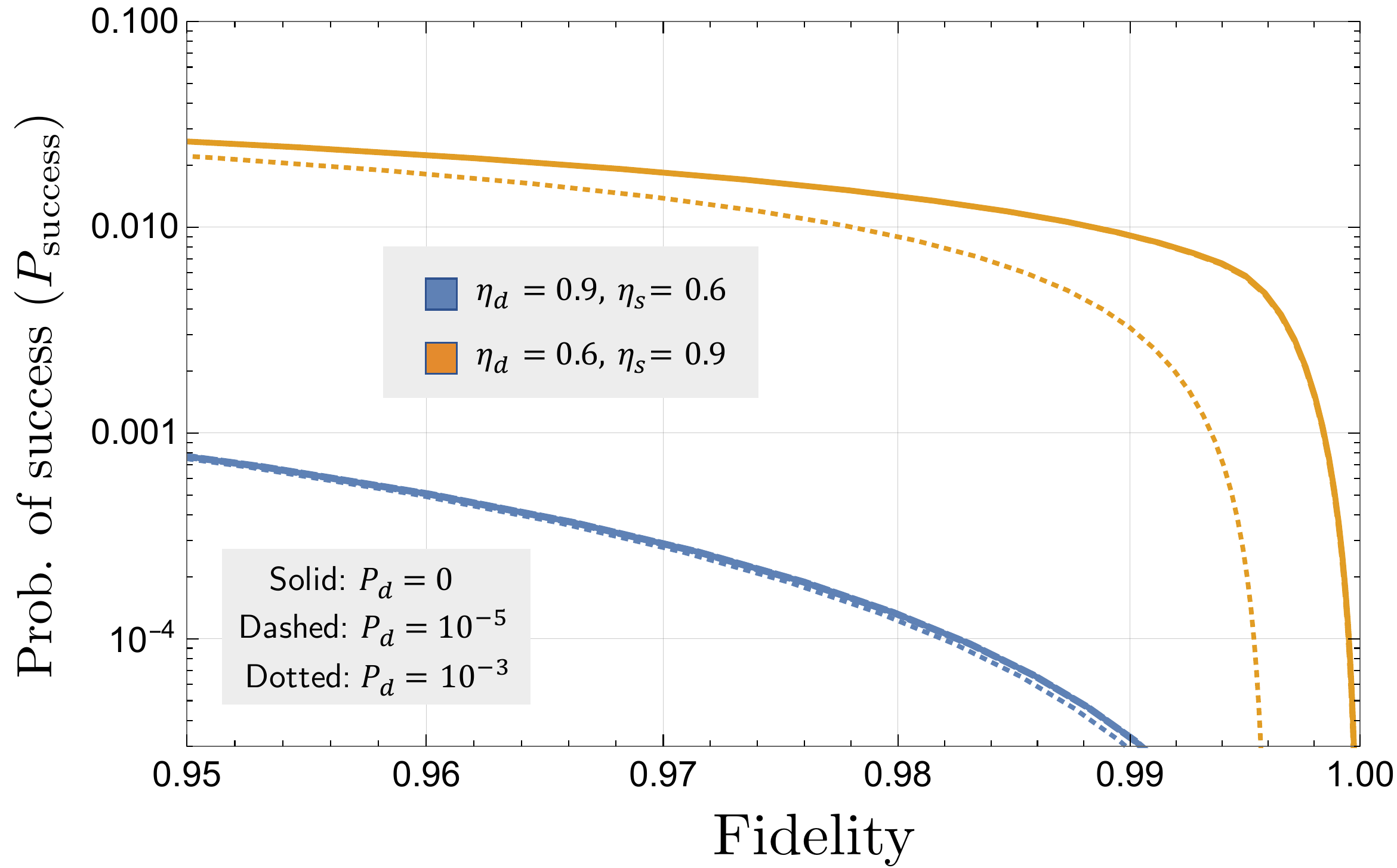} }
	\caption{This figure plots the outer envelopes of the scatter plot of $ P_{\text{success}} $ versus Fidelity when $M$ and $N_s$ are varied over: $M \in [1, 10^6]$ and $N_s \in [10^{-4}, 1]$, with $\eta = \eta_c\eta_d$ and $\eta_s$ held fixed.   The envelopes correspond to the highest probability of success that can be achieved at a given Fidelity target for an optimal value of $ M $. Note that with higher dark count probability the achievable fidelity becomes more restricted.  The solid, dashed, and dotted lines are largely indiscernible because they mostly overlap with one another.}
	\label{fig:scatter_pdark}
\end{figure}

The above analysis does not account for non-zero $P_d$. We observe that the inclusion of detector dark clicks ($P_d > 0$) only restricts the maximum achievable Fidelity. To illustrate this, we plot the $ P_{\text{success}} $ vs. infidelity ($ 1- $ Fidelity), achieved by the heralded-multiplexed source in Fig.~\ref{fig:psucc_dark}. These plots assumed $\eta = 0.9$ and $\eta_s =0.9$. In Fig.~\ref{fig:psucc_dark_loss}, we show the trade-off of $P_{\rm success}$ versus infidelity, for two sets of values of losses. These plots assumed $ M=10^6 $, $\eta = 0.9$, $\eta_s = 0.6$ (blue lines) and $\eta = 0.6$, $\eta_s = 0.9$ (orange lines), with $ P_d $ varying between $ 0 $ (solid),  $ 10^{-5}$ (dashed)  and $ 10^{-3}$ (dotted). Finally, in Fig.~\ref{fig:scatter_pdark}, we plot the $P_{\text{success}}$ versus Fidelity trade-offs as in Fig.~\ref{fig:psucctr}, but with $P_d > 0$. The main difference we note, as expected from the plots in Figs.~\ref{fig:psucc_dark} and~\ref{fig:psucc_dark_loss}, is that a non-zero $P_d$ imposes a hard upper limit on the Fidelity. However, for $P_d < 10^{-5}$, the reduction in the Fidelity cap below unity is negligible. This level of dark click probability is easily available with state-of-the-art superconducting nanowire single-photon detectors~\cite{Baghdadi2021-uq}, for a detection gate corresponding to GHz-scale repetition rates, which are readily achieved with SPDC-based entanglement sources.

\section{Discussion and Future Work}
\label{section:conclusion}

The primary pieces of intuition that drove the main results of this paper are that: (a) cascading two SPDC-based polarization-entangled sources with a linear-optical BSM built using PNR detectors in the middle produces an entangled state whose fidelity can be pushed close to unity if there were a heralded quantum memory available that can filter out the vacuum contribution. This is not possible with a stand-alone SPDC source due to the contributions from the high-order photon terms; and (b) the BSM provides a heralding trigger (again, not available in a free-running stand-alone SPDC source) which lets us multiplex many cascaded sources with a photonic switch array.  

\begin{figure}
	\centering{\includegraphics[width=\linewidth]{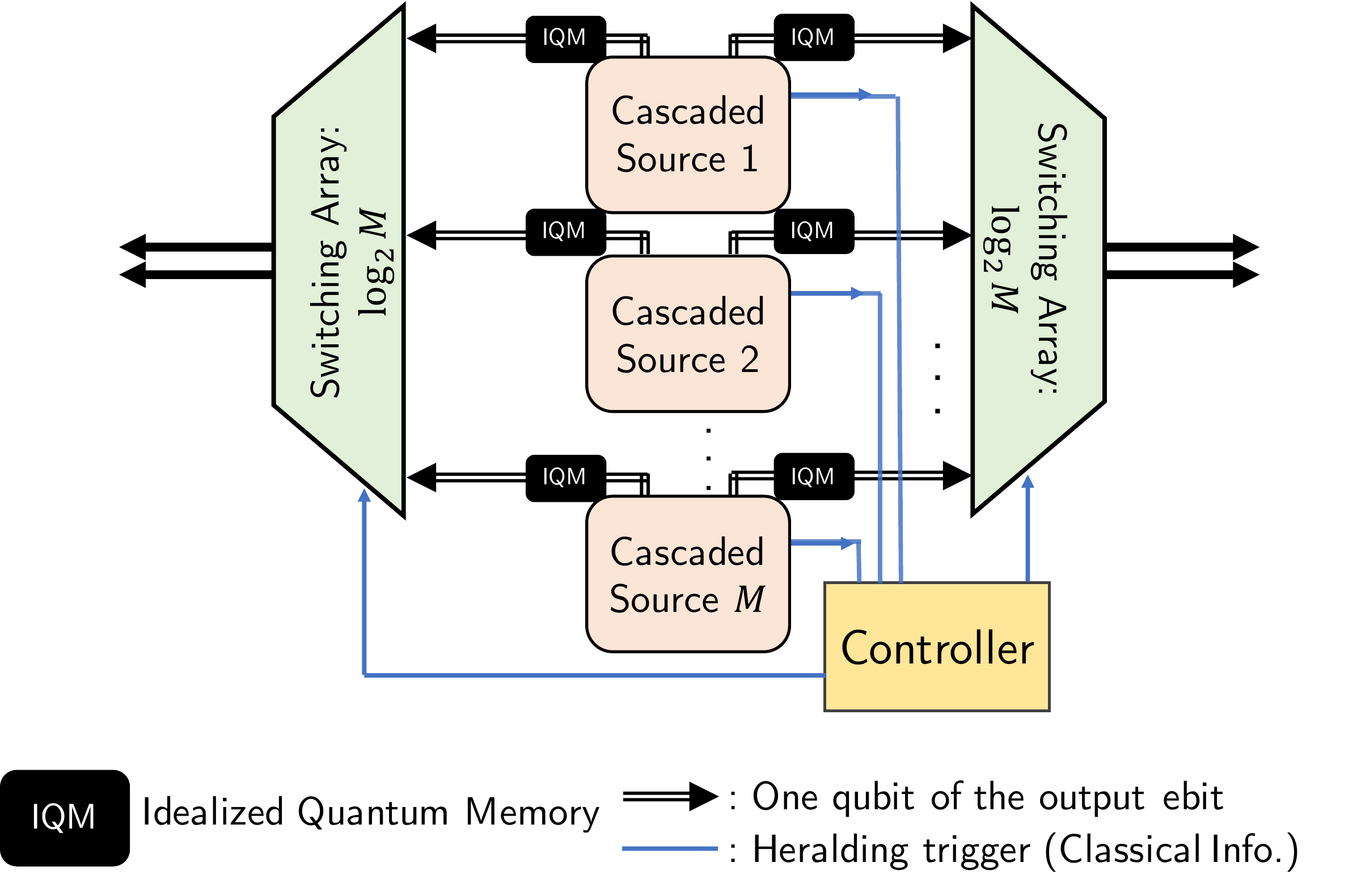} }
	\caption{Alternate multiplexing scheme for a near deterministic source of the target Bell states. Similar to the scheme in Fig.~\ref{fig:mux1} we multiplex $ M $ sources  with the VON projective measurement at each output. The heralding trigger additionally accounts for the vacuum output from the VON black box. 
	}
	\label{fig:mux2}
\end{figure} 
One limitation of the cascaded source, which the reader may have noted in Fig.~\ref{fig:fid} is that the maximum fidelity it can attain, for the raw photonic-domain entangled state it emits, is $0.5$. Obviously, the Fidelity of the stand-alone SPDC source is even worse; however, we should note (cf. Fig.~\ref{fig:psucctr}) that if we allow the standalone source access to the same ideal quantum memory, its performance becomes comparable, with many fewer required resources. So, neither of these photonic entangled sources is of use to produce high-fidelity entanglement unless a heralded quantum memory were available that can faithfully filter out the vacuum contribution, or it were used in an application where such vacuum filtering would occur naturally in a post-selected fashion as a result of photon detection, e.g., in QKD. We would like to note that it might be possible to further improve the quality of the output entangled state produced by the heralded-multiplexed source if we had an advanced version of the idealized quantum memory (IQM), wherein along with the stated characteristics of the IQM in Section~\ref{section:ideal_memory}, the IQM is additionally able to emit the stored qubit into the photonic domain, encoded in the dual-rail basis. This advanced memory would likely come with an additional efficiency cost (due to inefficiency in that storage qubit-to-photon readout process). This alternative design is depicted in Fig.~\ref{fig:mux2}. 

 The multiplexed source we analyze in this paper may find application in satellite-based entanglement distribution, quantum repeaters, resource-efficient generation of more complex multi-photon entangled states for fault-tolerant quantum computing, and quantum sensors, among others. We leave the performance analysis of this source for specific applications open for future research.

\section*{Acknowledgments}
PD, CNG and SG acknowledge the National Science Foundation (NSF) Engineering Research Center for Quantum Networks (CQN), awarded under cooperative agreement number 1941583, for supporting this research. SG additionally acknowledges support from ATA, under a NASA-funded research consulting contract. The contributions of SJ and PGK are supported in part by NASA Grant No. NNX16AM26G. The authors acknowledge useful discussions with Dr. Hari Krovi of Raytheon BBN and Dr. Babak N. Saif of GSFC, NASA. 

\nocite{*}
\bibliographystyle{unsrt}
\bibliography{bibliography}
\onecolumngrid

\appendix

\section{Detailed Source Analysis}
\label{appendix:source}
\begin{figure}[h!]
	\centering{\includegraphics[width=0.5\linewidth]{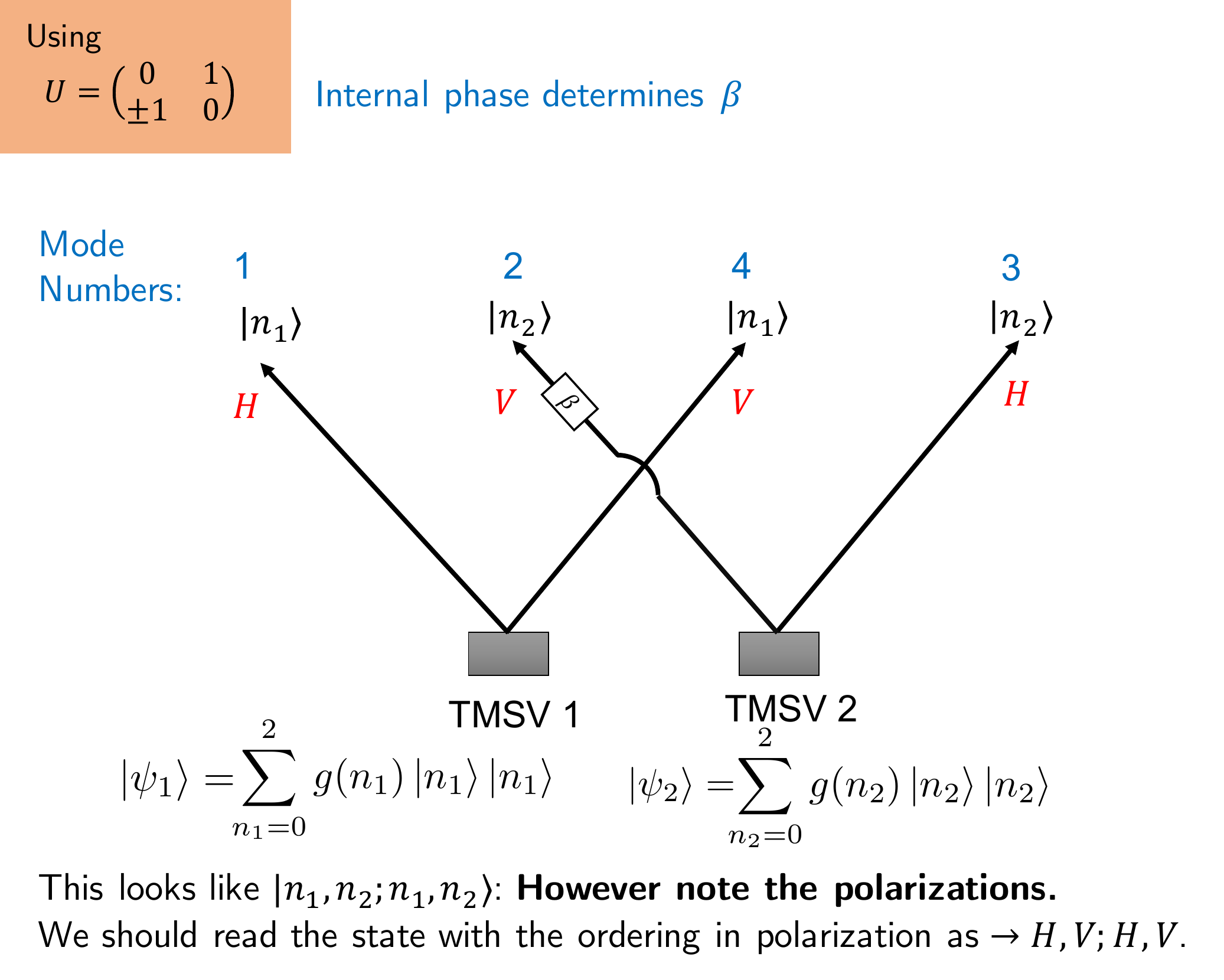}}
	\caption{Unfolded version of the source in Fig. 1 from \cite{kok2000}. The two passes through the nonlinear crystal generates two two-mode squeezed vacuum states $ (\ket{\psi_1} \text{ and } \ket{\psi_2}) $. The effect of the arm swap is explicitly shown with the choice of internal phase $ \beta $.}
	\label{fig:unfolded_source}
\end{figure}

The original source proposed in \cite{kok2000} uses simple parametric down conversion (SPDC) and additional linear optical elements to generate the state given in Eq.~\eqref{eqn:srcnative}. Analysis of the complete interaction picture of the Hamiltonian which governs the dynamics of the quantum source (i.e.,\ the weak parametric down conversion) is given in Section IA of the original work and also in \cite{krovi2016}. We present a high-level analysis of the same for our choice of notation. The source (as depicted in Fig. 1 of \cite{kok2000}) can be `unfolded' as shown in Fig.\ref{fig:unfolded_source}.

After the generation of the 2 two-mode squeezed vacuum (TMSV) states, which are given by
\begin{align}
	\ket{\psi_1}= \sum_{n_1=0}^{\infty} \sqrt{\frac{N_s^{n_1}}{(N_s+1)^{n_1+1}}} \ket{n_{1}^H}\ket{n_{1}^V}\label{eqn:SPDC1}\\
	\ket{\psi_2}= \sum_{n_2=0}^{\infty} \sqrt{\frac{N_s^{n_2}}{(N_s+1)^{n_2+1}}} \ket{n_{2}^H}\ket{n_{2}^V},
	\label{eqn:SPDC2}
\end{align}
the linear optical circuitry swaps the similarly polarized beams. The unitary $ U $ induces the swap of the photonic states in the modes labelled by $ \ket{n_1^V} $ and $ \ket{n_2^V} $ terms, which can be compactly described by the following transformation
\begin{align}
	\ket{\psi_1}\otimes \ket{\psi_2}&=\sum_{n_1=0}^{\infty}\sum_{n_2=0}^{\infty} \sqrt{\frac{N_s^{n_1+n_2}}{(N_s+1)^{n_1+n_2+2}}} \ket{n_1^H,n_1^V;n_2^V,n_2^H}\\
	&\xRightarrow{\text{Swap by }U} \sum_{n_1=0}^{\infty}\sum_{n_2=0}^{\infty} \sqrt{\frac{N_s^{n_1+n_2}}{(N_s+1)^{n_1+n_2+2}}}  (-1)^{n_2}\ket{n_1^H,n_2^V;n_1^V,n_2^H} =\ket{\psi_{src.}}.
\end{align}

We define $ g(m) \equiv \sqrt{\frac{N_s^{m}}{(N_s+1)^{m+1}}} $ and for the sake of brevity we use 
\begin{align}
	\sqrt{p(0)}= g(0)^2; \quad \sqrt{\frac{p(1)}{2}}=g(0)g(1); \quad \sqrt{\frac{p(2)}{3}}=g(0) g(2)= g(1)^2
\end{align}

\begin{figure}[h!]
	\centering{\includegraphics[width=0.65\linewidth]{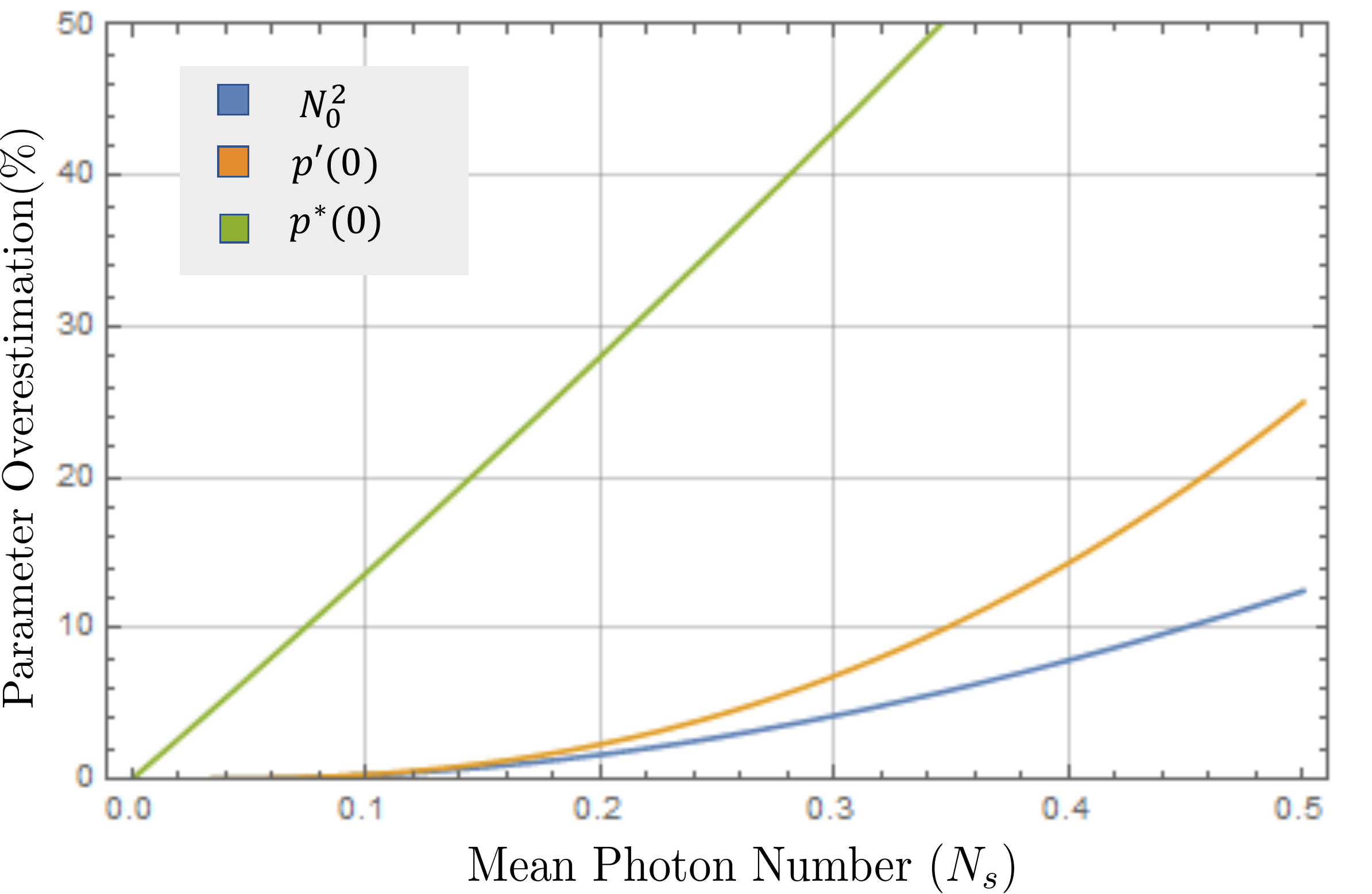}}
	\caption{Parameter overestimation caused by various normalization schemes for the state $|\psi \rangle$.}
	\label{fig:parameter_divergence}
\end{figure}

Therefore, the final  state is given by
\begin{align}
\begin{split}
	\ket{\psi^\pm} \!= & N_0 \left[\!\sqrt{p(0)} \ket{0,0;0,0}\! +\! \sqrt{\frac{p(1)}{2}} \left(\ket{1,0;0,1}\pm\ket{0,1;1,0} \right) +  \sqrt{\frac{p(2)}{3}}\left(\ket{2,0;0,2}\pm \ket{1,1;1,1}+\ket{0,2;2,0}\right)\right],
\end{split}
\end{align}
where $N_0 = 1/\sqrt{p(0) + p(1) + p(2)}$is used to normalize the state. This method preserves the probability ratio $p(1)^2/(p(0)p(2))$ even as $N_s$ increases, at the cost of overestimating each probability individually. In Fig.~\ref{fig:parameter_divergence} we plot this error, as well as the errors contributed by two additional methods of normalizing the state:
\begin{itemize}
	\item Define $p'(0) \equiv 1 - p(1) -p(2);$
	\item Define $p^*(2) \equiv 1 - p(0) - p(1)$. 
\end{itemize} 

The $p'$ normalization preserves the 1- and 2-photon pair probabilities of $|\psi\rangle$, but overestimates vacuum contributions. Similarly, $p^*$ normalization allows for accurate representation of the vacuum and 1-photon pair terms, but all multi-pair events are treated as having 2 pairs. These alternate normalization schemes have some advantages since they do not overestimate the 1-photon pair probability. However, this comes at the cost of faster divergence and less convenience when performing analytic calculations.

\section{Analysis of SPDC Timing Walk-off  }
\label{appendix:timing_walkoff}

In our present analysis, we assume that the simple parametric down conversion (SPDC) process which generates the states given by Eq.~(\ref{eqn:SPDC1}) and (\ref{eqn:SPDC2}) is devoid of any imperfections that may influence the output state. However, in practical experimental implementations of SPDC, there is an additional degree of freedom w.r.t. the temporal location of the emitted photons, which may affect the overall description of the output two-mode squeezed vacuum state. A complete and rigorous analysis of this effect is beyond the scope of this work. However, as the dual-rail Bell state is the target state for the present article, we examine the single-pair emission terms in the complete photonic state emitted by the SPDC (i.e.\, $ n_1=1$ in Eq.~\ref{eqn:SPDC1}; $ n_2=1$ in Eq.~\ref{eqn:SPDC2}). 

It is well understood that timing walk-off between the emitted photons induces a partial decoherence in the output state~\cite{}. In the basis of the single-pair emission term, the density matrix of the SPDC source resembles
\begin{align}
	\hat{\rho}_{\text{single}}=\frac{1}{2}
	\begin{pmatrix}
		0 & 0 & 0 & 0 \\
		0 & 1 & 1-\epsilon & 0 \\
		0 & 1-\epsilon & 1 & 0 \\
		0 & 0 & 0 & 0\\
	\end{pmatrix}.
\end{align}

\begin{figure}[h!]
	\centering{\includegraphics[width=0.7\linewidth]{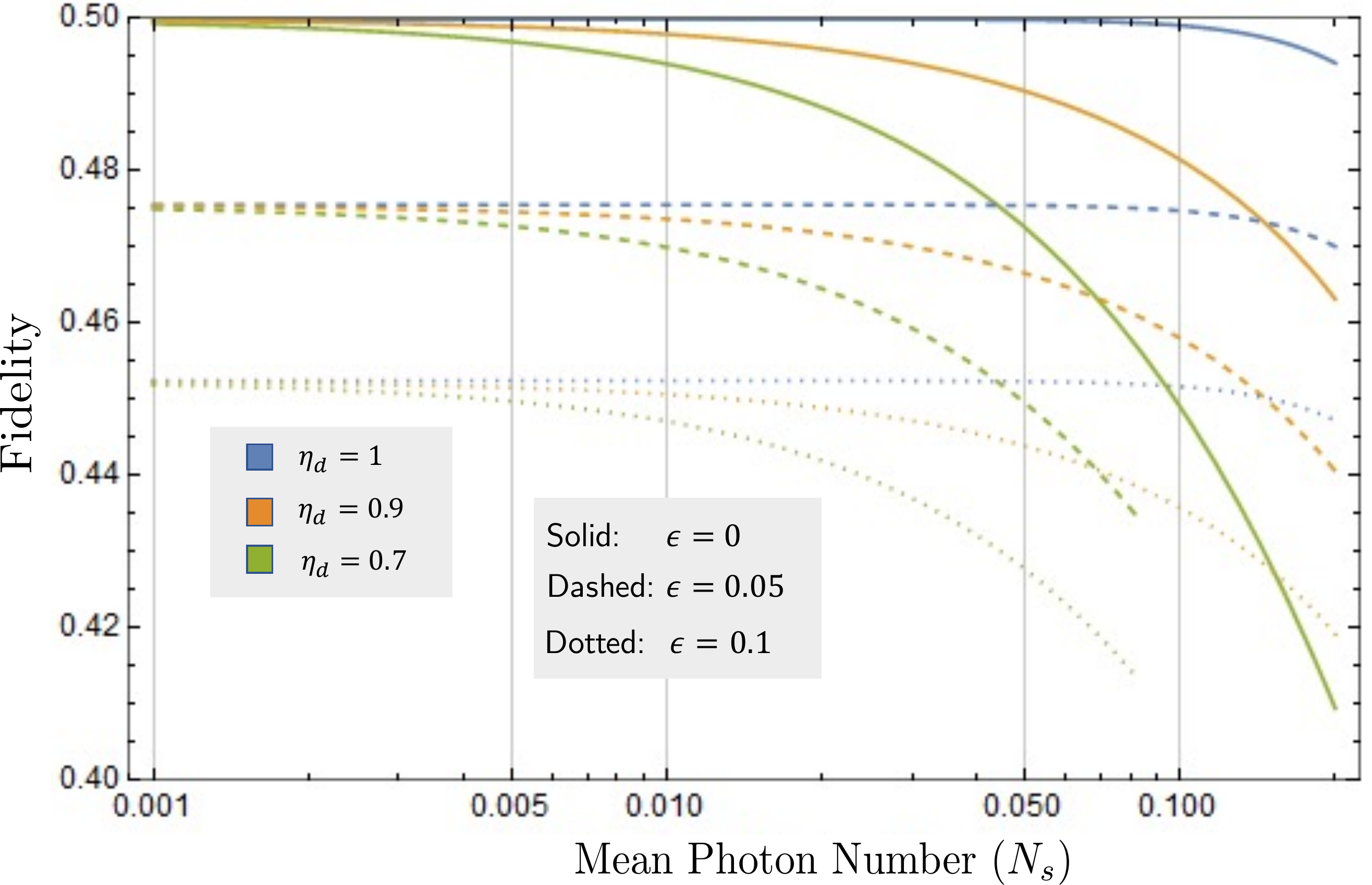}}
	\caption{Fidelity of the entangled state generated by the cascaded source with the ideal Bell state $\ket{\Psi^+}$ at various values of state decoherence $ (\epsilon) $ and detector efficiency $ (\eta_d) $, plotted as a function of $N_s$. We assumed $P_d = 0$ for these plots.}
	\label{fig:fid_decoh1}
\end{figure}

\begin{figure}[h!]
	\centering{\includegraphics[width=0.7\linewidth]{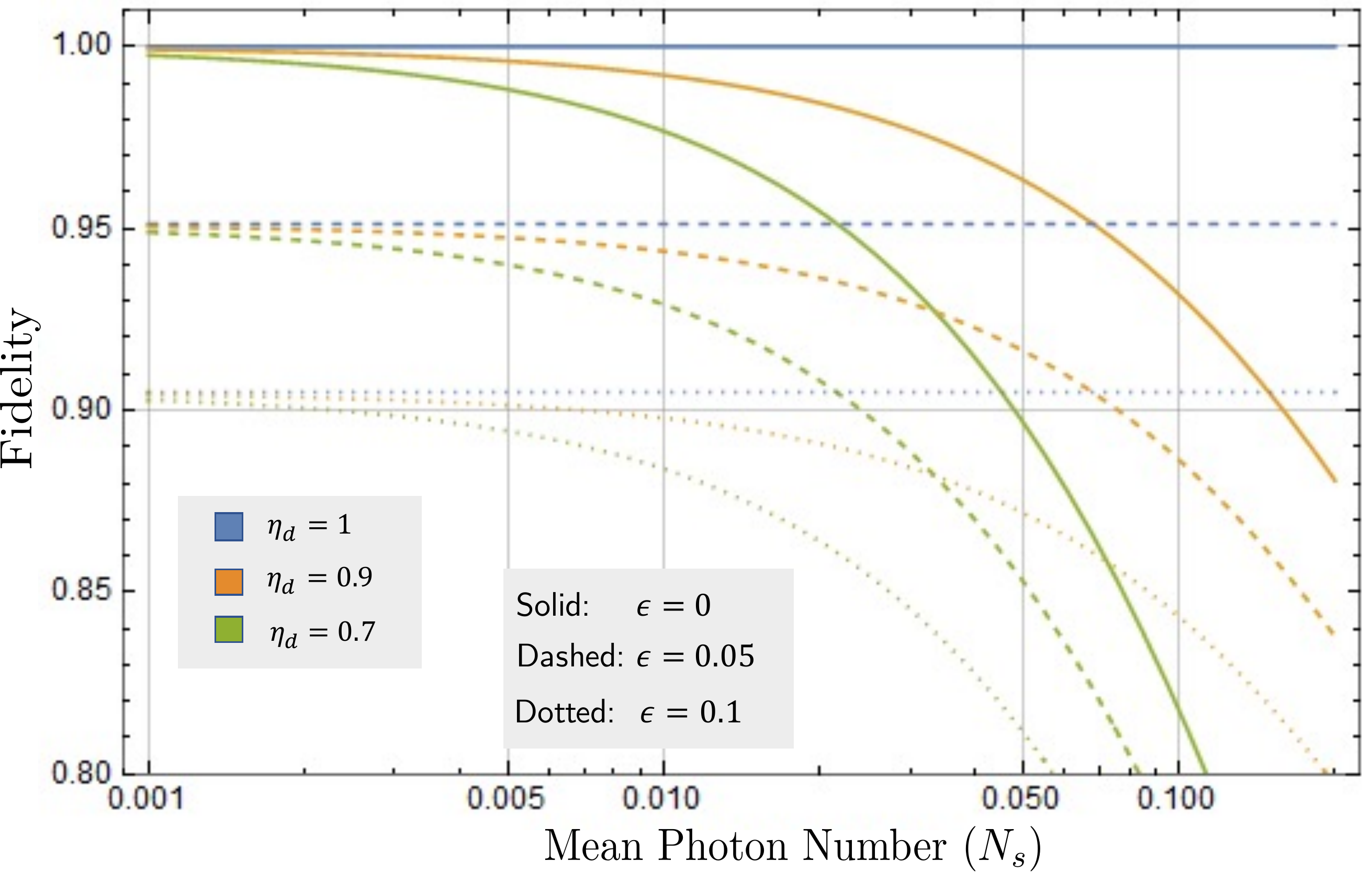}}
	\caption{Fidelity of the entangled state generated by the cascaded source with the ideal Bell state $\ket{\Psi^+}$ after successfully loading into idealized heralded quantum memories. The different lines signify various values of state decoherence $ (\epsilon) $ and detector efficiency $ (\eta_d) $, plotted as a function of $N_s$. We assumed $P_d = 0$ for these plots.}
	\label{fig:fid_decoh2}
\end{figure}

The parameter $ \epsilon $ accounts for the decoherence of the state, where $ \epsilon=0 $ denotes the absence of any decoherence due to the timing walk-off effect. We observe that the introduction of decoherence in the state description limits the maximum fidelity achievable by the state. This can be explained by the presence of the non-zero decoherence parameter (in the cross terms of the density matrix), which decreases the overlap with the target $ \ket{\Psi^+} $ state. This is depicted in Fig.~\ref{fig:fid_decoh1}, where we plot the Fidelity of the entangled state from a \textit{cascaded} source w.r.t. the ideal Bell state $ \ket{\Psi^+} $ as a function of the mean photon number $ (N_s) $, when the underlying SPDC source state has decoherence values $ (\epsilon) $  as marked in the legend. In Fig.~\ref{fig:fid_decoh2} we plot the state Fidelity as a function of $ N_s $, after the output photonic state has been loaded into the ideal quantum memory (IQM).

Based on the results of the figure, we can roughly derive the following empirical relation between the maximum achievable fidelity (both pre- and post-loading into the IQM) as follows
\begin{align}
	\max_{\{N_s,\eta_d\}} F(\hat{\sigma},\ket{\Psi^+})= 0.5(1-\epsilon)\\
	\max_{\{N_s,\eta_d\}} F^{\text{VON}}(\hat{\sigma},\ket{\Psi^+})= (1-\epsilon),
\end{align}
where $ \hat{\sigma} $ denotes the density operator of the output state (we omit the complete description for brevity). Therefore, it is clear that given the value of $ \epsilon $ for the underlying SPDC sources, the maximum fidelity target for the cascaded source (and in extension the multiplexed source from Section~\ref{section:mux}) is limited to $ 1-\epsilon $.

\section{Hybrid Fock-Coherent System Modeling}
	\label{appendix:hybrid}
	Although complete system modeling in the Fock basis is exact and complete for truncated basis states, it poses a variety of computational difficulties. Inclusion of component efficiency involves the consideration of additional environment modes that need to be traced out. Treating the pure loss effects in the Kraus operator formalism is one way to circumvent this difficulty. The Fock-basis representation is not well suited to this treatment. In our calculations we adopt a hybrid Fock-Coherent approach. 
	
 In this approach, the action of the 50-50 beamsplitter and efficiencies $ \eta_c,\eta_d,\eta_d $ are treated in the coherent basis, and then projected onto the Fock basis to generate the complete density matrix description. We highlight the key steps to this approach in the subsequent paragraphs. The complete density matrix description is omitted for brevity in this Appendix. 
	
 We identify that the original state from Eq.~\eqref{eqn:srcnative} is comprised of a pair of two-mode squeezed vacuum states. Hence, it is conceptually much simpler to treat the whole link as two different TMSVs that are connected as shown in Fig.~\ref{fig:unfolded_source}. The whole state is a tensor product of two such setups, with the mode labels suitably rearranged as described in Appendix \ref{appendix:source}.
 
 One may think of this as two TMSV states with their mode label/ordering changed. Given a single TMSV state $ \ket{\psi}=\sum_m c(m) \ket{m;m} $, the corresponding density matrix  is
 	\begin{align}
 		\rho=  \sum_{m,m'=0}^{n,n'} c_{m} c^*_{m'}  \ket{m;m}\bra{m';m'}.
 	\end{align}
We now take two such sources, changing the labels of their Fock state to keep the indices distinct; our final state is then represented as
		\begin{align}
			\rho_{A} \otimes\rho_{B} =& \sum_{n_A,n'_A,n_B,n'_B} c_{A,m_A} c^*_{A,m'_A} \times c_{B,m_B} c^*_{B,m'_B} \sum_{\text{all m and m'}}\ket{m_A,m_A,m_B,m_B}\bra{m'_A,m'_A,m'_B,m'_B}\\
			=&\frac{1}{\pi^4}\int \sum_{\text{all sums}} C_{m_A,m_B,m'_A,m'_B} \ket{m_A}\ket{\alpha_1,\alpha_2}\braket{\alpha_1,\alpha_2|m_A,m_B}\ket{m_B} \times \bra{m'_A}\braket{m'_A,m'_B|\gamma_1,\gamma_2} \bra{\gamma_1,\gamma_2}\bra{m'_B}	\\
			=&\frac{1}{\pi^4}\int \sum_{\text{all sums}} C_{m_A,m_B,m'_A,m'_B} \ket{m_A}\ket{\alpha_1,\alpha_2}  \ket{m_B} \bra{m'_A} \bra{\gamma_1,\gamma_2} \bra{m'_B} \exp{(-(|\alpha_1|^2+|\alpha_2|^2+|\gamma_1|^2+|\gamma_2|^2)/2)} \nonumber\\
			&\times \frac{\alpha^{m_A}_1}{\sqrt{m_A!}}	\frac{\alpha^{m_B}_2}{\sqrt{m_B!}} \frac{\gamma^{m'_A}_1}{\sqrt{m'_A!}} \frac{\gamma^{m'_B}_2}{\sqrt{m'_B!}}.
		\end{align}
		
		We can make the following changes to the coherent basis vectors (ignoring Fock basis vectors for brevity; same process for the corresponding bras) after each step:
	{\small
		\begin{align}
		&\ket{\alpha_1,\alpha_2}  \xRightarrow{\eta_c} \ket{\alpha_1 \sqrt{\eta_c},\alpha_2\sqrt{\eta_c}} \xRightarrow{ 50:50 \text{BS}} \Ket{\frac{\alpha_1 \sqrt{\eta_c}+\alpha_2 \sqrt{\eta_c}}{\sqrt{2}},\frac{-\alpha_1 \sqrt{\eta_A}+\alpha_2 \sqrt{\eta_c}}{\sqrt{2}}} \xRightarrow{\eta_d} \Ket{\sqrt{\eta_d}\frac{\alpha_1 \sqrt{\eta_c}+\alpha_2 \sqrt{\eta_c}}{\sqrt{2}}, \sqrt{\eta_d}\frac{-\alpha_1 \sqrt{\eta_A}+\alpha_2 \sqrt{\eta_c}}{\sqrt{2}}}.
	\end{align}
	}%

		The complete action can be thought to be
		\begin{align}
			(\rho_A \otimes \rho_B)^{\otimes 2}=\bar{\rho}\xRightarrow{\text{loss+BSM}} \tilde{\rho}\xRightarrow{\text{det.}} \tilde{\sigma}_i
		\end{align}
		where $ \tilde{\sigma}_i $ is the output state conditioned on a certain click pattern as given in the main text. Note that $ \tilde{\sigma}_i $ is expressed in a mixed basis (part Fock and part Coherent). In order to get back to the Fock-basis density matrix, we adopt the techniques developed in \cite{gagatsos2019}
		to perform a Fock projection on the coherent basis description of the quantum state of a bosonic mode. 
		
		Performing the preceding mathematical operations on the final density matrix lets us evaluate the various metrics of interest:
		\begin{enumerate}
			\item Calculating $ \Tr(\tilde{\sigma}_i) $ yields the generation probability ($ P_{\text{gen}} $).
			\item The Fidelity can be determined as follows:
			\begin{align}
				F= \frac{\braket{\Psi^+|\tilde{\sigma}_i|\Psi^+}}{\Tr(\tilde{\sigma}_i)}.
			\end{align}
		\end{enumerate}

\section{Modeling a Practical Photon-Number Resolving Detector}
\label{section:app_PNRD}
Photon-number resolving (PNR) detectors \cite{tan2016,morais2020,cahall2017} are key components of the linear optical Bell state measurement(BSM) circuit, heralding the outcome of an entanglement swap attempt. When swapping entanglement between two ideal dual-rail basis Bell pairs, only specific click patterns across the multiple detectors indicates a possible `success'. In our proposal for the improved source, this `success' information heralds the generation of the state given in Eq.(\ref{eqn:casc_src_st}).

\begin{figure}[h!]
	\centering{\includegraphics[width=0.7\linewidth]{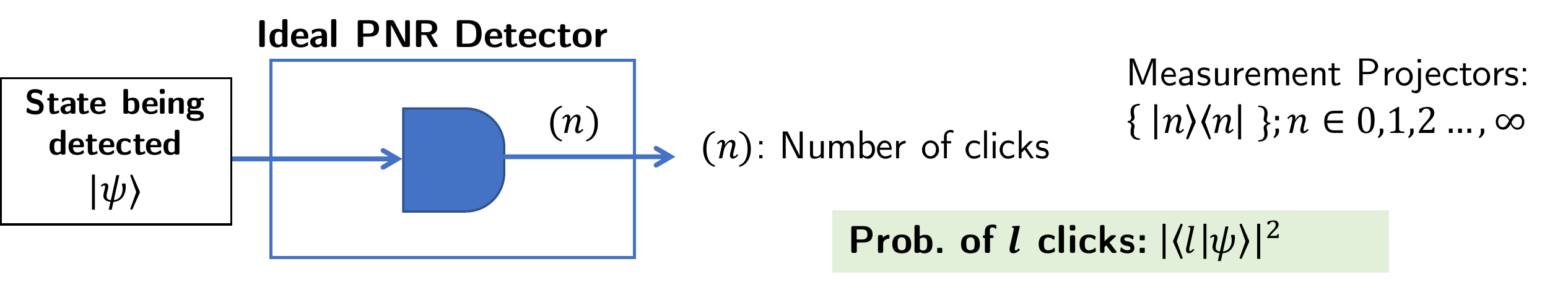}  }
	\caption{Model of the ideal photon number resolving (PNR) detector. The ideal PNR generates clicks depending upon the photon statistics of the input state. }
	\label{fig:ideal_PNR}
\end{figure}

Theoretically, the ideal PNR measurement is a projective measurement given by the action of POVM elements $ {\Pi_n=\ket{n}\!\!\bra{n}} , n\in \mathbb{Z}^+$, on the input state, say $ \ket{\psi} $. A measurement result of $ k $ clicks, would collapse the input state onto one of the Fock basis elements, in this case $ \ket{k} $. One can use ideal PNRDs repeatedly to get the photon statistics of the input state and `reconstruct' the state (this is the whole focus of quantum state tomography). This simple model is shown in Fig.\ref{fig:ideal_PNR}.

Since we consider the use of imperfect PNRDs, there are multiple factors that must now be taken into account. Presently, we only focus on the effects that affect the quantum state being detected. Factors that influence the physical operability of the detector are not treated by this model; timing jitter, post-detection dead time, after-pulsing and count saturation are not considered here. We assume that the PNR model is synchronized with the rest of the circuit and the pulse profile, bandwidth and frequency are optimized a priori. The PNRD has a detection efficiency $( \eta_d) $ and dark click probability (uncorrelated to the input state) of $ P_d $. Detector efficiency can be interpreted as the input state being transmitted through a bosonic pure loss channel of transmissivity $ \eta_d $ before the actual detection happens, which yields an (inaccessible) outcome $ k $. The detector dark clicks can be treated as a Bernoulli random variable with the probability of success $ P_d $. The outcome of this  random variable  is convoluted with the actual output from the ideal PNR measurement after loss. Thus an observation of $ l $ where $ l\geq1 $ clicks may signify one of two cases: 
\begin{itemize}
	\item There were $ l $ clicks and no dark clicks.
	\item There were $ l-1 $ clicks and a single dark click.
\end{itemize} 
Thus, the probability of $ l $ clicks in the detector is given by 
\begin{align}
	P[l]=P[k=l-1]\times P_d + P[k=l] \times (1-P_d).
\end{align}

\begin{figure}[h!]
	\centering{\includegraphics[width=\linewidth]{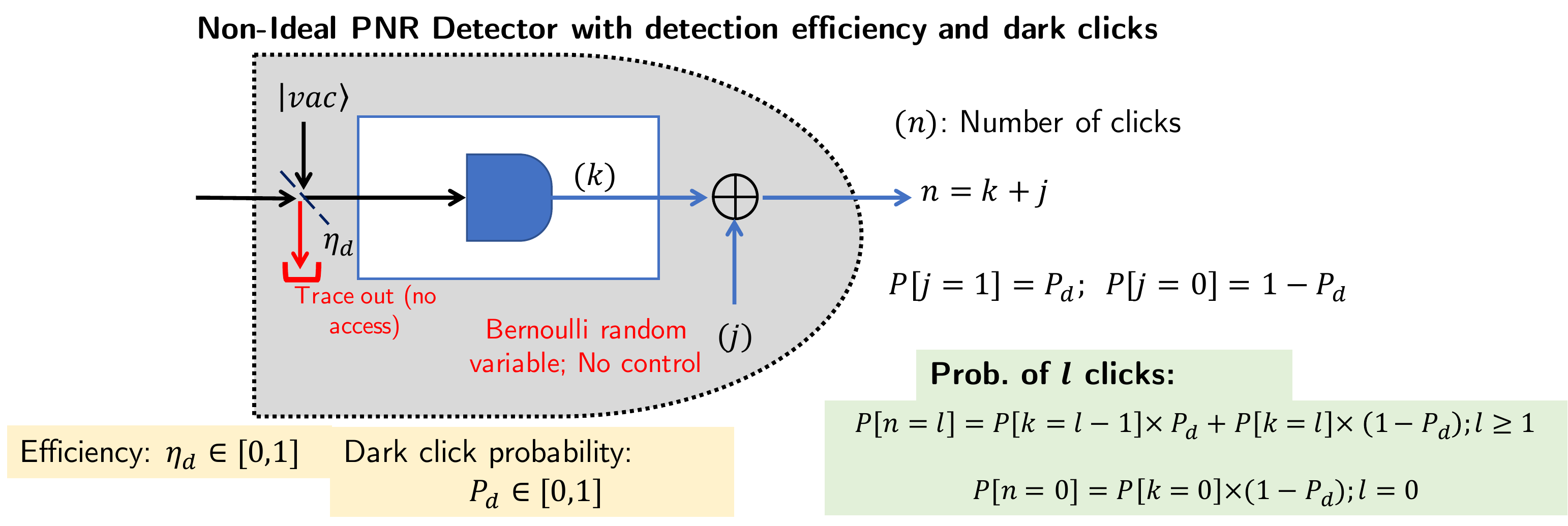}  }
	\caption{Model of the non-ideal photon number resolving (PNR) detector.The non-ideal PNR detector has two parameters that affect the operation, namely the detector efficiency ($ \eta_d $) and dark click probability $ (P_d) $. }
	\label{fig:ni_PNR}
\end{figure}

The complete model of the non-ideal PNR measurement is depicted in Fig.~\ref{fig:ni_PNR}.

\noindent Thus the observation of a correct click pattern indicates successful entanglement swap only for a small subset of cases. For example, an observed pattern of $ \vec{n}_{\text{observed}}=(1,1,0,0) $ may correspond to any of the following cases:
\begin{align}
	\begin{split}
		P^{\text{observed}}_{(1,1,0,0)} = &(1-P_d)^4 \times P_{(1,1,0,0)} \\+& P_d (1-P_d)^3 \times[P_{(1,0,0,0)}+P_{(0,1,0,0)}] \\+& P^2_d (1-P_d)^2 \times P_{(0,0,0,0)},
	\end{split}
	\label{eqn:dark1}
\end{align}
where $ P_{(\vec{n})} $ is the probability that in reality the detection pattern was $ (\vec{n})$ prior to the dark clicks.

\noindent Thus, for the chosen click pattern we have the mixed state
\begin{align}
	\begin{split}
		\sigma= &(1-P_d)^4 \outprod{\phi_0} + P_d (1-P_d)^3  \sum_{k_1=1}^2\outprod{\phi^{k_1}_1} \\
		& + P^2_d (1-P_d)^2 \outprod{\phi_2},
	\end{split}
	\label{eqn:dark2}
\end{align}
where $ \ket{\phi_k} $ denotes all the states generated when we have $ k $ photons less than the ideal detection pattern, and $ k $ dark clicks. For example, in the proposed cascaded source, if we consider only detector dark clicks (no detector loss, i.e.\, $ \eta_d=1$) in the present model, we would  the following mixed state at the output
\begin{align}
	\bar{\rho}&=  (1-P_d)^4  \hat{\rho}_1 + (1-P_d)^3 P_d  \, \hat{\rho}_2+ (1-P_d)^2 P^2_d \, \hat{\rho}_3,
\end{align}
with
\begin{align}
	\hat{\rho}_1= & \, c_1 \left(\ket{1,0;0,1}+\ket{0,1;1,0}\right) \times \left(\bra{1,0;0,1}+\bra{0,1;1,0}\right)\\
	\hat{\rho}_2= & \, c_2	\left({\ket{1,0;0,0}+\ket{0,0;1,0}}+{\ket{0,0;0,1}+\ket{0,1;0,0}}\right)\nonumber\\ &\times 	\left({\bra{1,0;0,0}+\bra{0,0;1,0}}+{\bra{0,0;0,1}+\bra{0,1;0,0}}\right)\\
	\hat{\rho}_3= & \, c_3 \outprod{0,0;0,0},
\end{align}
where
\begin{align}
	c_1 = \frac{p(1)^2}{16}; \quad c_2 = \frac{p(0)p(1)}{4}; \quad c_3 = p(0)^2.
\end{align}

\section{`Vacuum or Not' Quantum Non Demolition Measurement}
\label{section:app_VON}
One essential processing tool required to attain a high Fidelity for the generated quantum state w.r.t. the target entangled state is a `vacuum or not' (VON) quantum non-demolition measurement. Such a measurement is a theoretical tool essential to filter out the vacuum component of the quantum state; which is a major part component that drives down the state fidelity. Given an $ N $-mode quantum state $ \ket{\psi} $, the VON measurement can be ideally modeled as a black box producing one of two outcomes:
	\begin{itemize}
	\item[$ - $] the $ N $-mode vacuum state: $ \ket{\psi_0}=\ket{0}^{\otimes N} $ with a probability of $ p_{\ket{0}^{\otimes N}}$;
	\item[$ - $] the vacuum-subtracted quantum state from $ \ket{\psi} $, $\ket{\psi_1} =\frac{\ket{\psi}-\sqrt{p_{\ket{0}^{\otimes N}}} \, \ket{0}^{\otimes N}}{\sqrt{1-|p_{\ket{0}^{\otimes N}}|^2}} $  with a probability of $ 1-p_{\ket{0}^{\otimes N}} $.
	\end{itemize}
The probability of the vacuum outcome is given by $ |p_{\ket{0}^{\otimes N}}|^2 =\left|\Braket{00\ldots0|\psi}\right|^2$. It must, however be noted that while a photonics-based implementation of the VON measurement is still an open problem, there are preliminary proposals to implement the same in atomic systems coupled to optical cavities~\cite{oi2013}. 
	
	\begin{figure}[h!]
		\centering
		\includegraphics[width=0.7\linewidth]{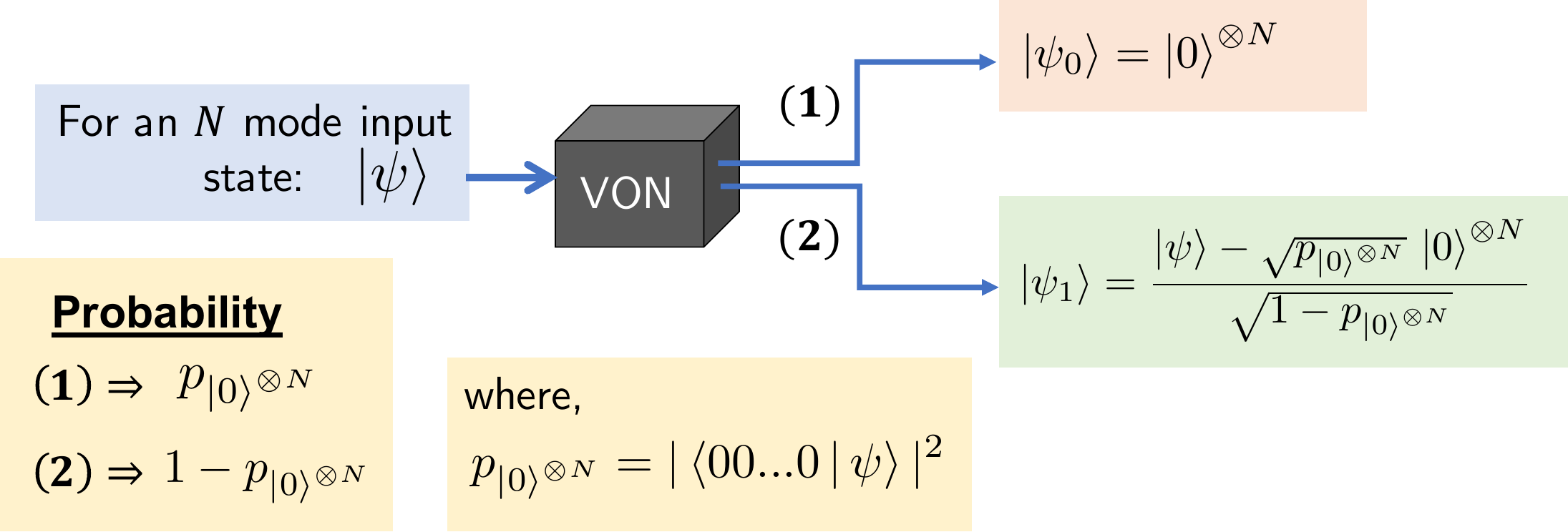}  
		\caption{Representation of the `vacuum or not' quantum non-demolition measurement. Given an input $ N $-mode quantum state, with a probability $ p_{\ket{0}^{\otimes N}} $, the VON black box separates the $ N $-mode vacuum, and with the complementary probability it separates the vacuum state from the input state. }
		\label{fig:von_qnd}
	\end{figure}

\section{Analytic Expressions of Fidelity and $ P_{\rm gen} $}
\label{appendix:analytics}

The most general formula for Fidelity and probability of generation $ P_{\text{gen}} $ of the output quantum state from the cascaded source (considering inefficiencies in coupling and non ideal detectors) is given by
\begin{align}
	F(\ket{M},\ket{\Psi^+})= \frac{\mathcal{A}_1}{\mathcal{B}_1}\\
	P_{\text{gen}}=\mathcal{B}_1,
\end{align}

where 
\begin{align}
\mathcal{A}_1=	\frac{2 N_s^2 \left(\left(5 P_d^2-4 P_d+1\right) \eta _c^2 \eta _d^2+2 (1-3 P_d) P_d \eta _c \eta _d+2P_d^2\right)}{\left(N_s+1\right)^6}
\end{align}

\begin{align}
\begin{split}
	\mathcal{B}_1=\frac{4}{\left(N_s+1\right)^8} \times \Biggl[&3 \eta _c^4 \eta _d^4 \left(P_d \left(11 P_d-10\right)+2\right) N_s^4-2 \eta _c^3 \eta _d^3 \left(P_d \left(17 P_d-13\right)+2\right) N_s^3 \left(4 N_s+1\right)\\
	&+\eta_c^2 \eta_d^2 N_s^2 \left(P_d^2 \left(N_s \left(211 N_s+118\right)+21\right)-2 P_d \left(N_s \left(61 N_s+34\right)+6\right)+11 N_s^2+6 N_s+1\right)\\
	&-2 \eta _c \eta
	_d P_d \left(3 P_d-1\right) N_s \left(4 N_s+1\right) \left(6 N_s^2+4 N_s+1\right)+P_d^2 \left(6 N_s^2+4 N_s+1\right)^2 \Biggr].
\end{split}
\end{align}

\noindent The Fidelity and probability of generating the output quantum state after the IQM's VON filtering (considering inefficiencies in coupling and non-ideal detectors) is given by
\begin{align}
	F^{\text{VON}}(\ket{M},\ket{\Psi^+})= \frac{\mathcal{A}_2}{\mathcal{B}_2}\\
	P^{\text{VON}}_{\text{gen}}=\mathcal{B}_2,
\end{align}

where 
\begin{align}
	\mathcal{A}_2=\frac{ \left(\eta _c^2 \eta _d^2+P_d^2 \left(-3 \eta _c \eta _d+\eta _c \eta _d \left(5 \eta _c \eta _d-3\right)+2\right)+\eta _d P_d \left(2
		\eta _c-4 \eta _c^2 \eta _d\right)\right)}{2 \left(N_s+1\right)^6}
\end{align}

\begin{align}
\begin{split}
		\mathcal{B}_2= \frac{1}{2(1+N_s)^8}  \Biggl[&P_d^2 \biggl(17 \eta _c^2 \eta _d^2-24 \eta _c \eta _d+N_s^2 \left(66 \eta _c^4 \eta _d^4-272 \eta _c^3 \eta _d^3+397 \eta _c^2 \eta _d^2-240 \eta _c \eta _d+50\right)\\
		& -2N_s \left(34 \eta _c^3 \eta _d^3-93 \eta _c^2 \eta _d^2+78 \eta _c \eta _d-20\right)+8\biggr)+2 \eta _c \eta _d P_d \biggl(-5 \eta _c \eta _d+N_s^2 \bigl(-30 \eta _c^3
		\eta _d^3+104 \eta _c^2 \eta _d^2\\
		&-115 \eta _c \eta _d+40\bigr) +N_s \left(26 \eta _c^2 \eta _d^2-54 \eta _c \eta _d+26\right)+4\biggr)+\eta _c^2 \eta _d^2 \bigl(N_s^2
		\left(12 \eta _c^2 \eta _d^2-32 \eta _c \eta _d+21\right)\\
		&+N_s \left(10-8 \eta _c \eta _d\right)+1\bigr)  \Biggr].
\end{split}
\end{align}

\section{Gaussian Modeling of the Cascaded Source}
\label{appendix:GBSmodel}

\begin{figure}[h!]
	\centering
	\includegraphics[width=0.9\linewidth]{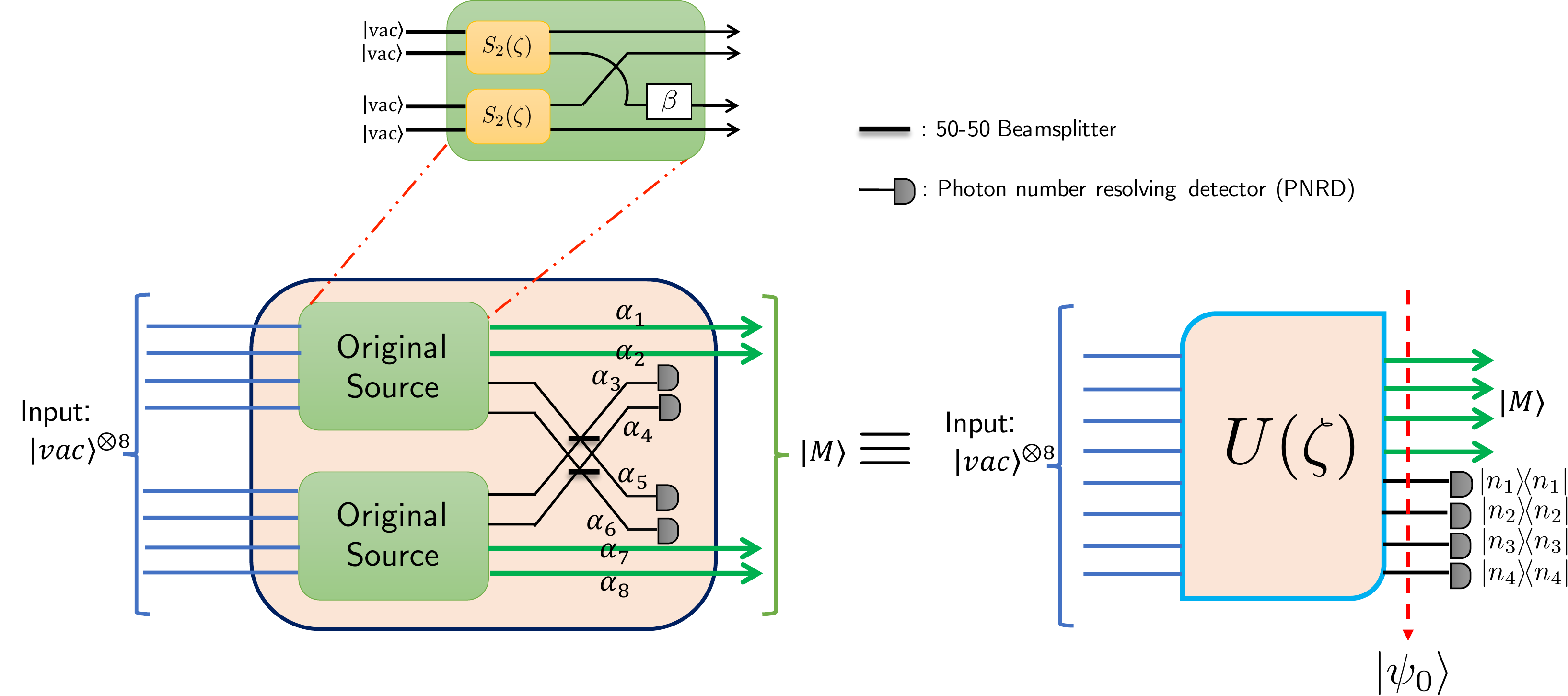}  
	\caption{ The complete abstraction of the \textit{cascaded} source as a Gaussian Boson sampling circuit setup for state preparation. We start with eight input vacuum modes, perform a unitary  operation $ U(\zeta) $(which includes the unitary for the two-mode squeezers and intermediate beamsplitters) parameterized by $ \zeta $. This corresponds to the mean photon number per mode$ (N_s) $ of the underlying two-mode squeezed states.}
	\label{fig:GBS_full}
\end{figure}

Let us begin with the Gaussian pure state after the unitary $ U(\zeta) $ as $ \ket{\psi_0} $(marked by the red dashed line in Fig.~\ref{fig:GBS_full}). Since the internal beamsplitters of the circuit are balanced, we can commute all our coupling losses to manifest just before detection. This is an operational trick to keep the state description rid of difficulties with swapping mixed states. This is justified because coupling losses are uniform and the beamsplitters are balanced. )

We can write the $ K$-function of the state $ \ket{\psi_0} $ as $ K(\vec{\alpha}) $ \cite{gagatsos2019}  as 
\begin{align}
	\ket{\psi_0}=\frac{1}{(2\pi)^8}\int d^{16}\alpha \;K(\vec{\alpha}) \ket{\vec{\alpha}},
\end{align}
where, $ \ket{\vec{\alpha}}=\Ket{\alpha_1,\ldots,\alpha_8}$ is the coherent basis vector and $ \alpha_i=(q_{\alpha_i}+i p_{\alpha_i})/\sqrt{2} $. Here $ q_{\alpha_i}, p_{\alpha_i} $ are the quadrature variables of the $ i^{th} $ mode as marked in Fig.~\ref{fig:GBS_full}.

Similarly the density operator description of the state $ \ket{\psi_0} $ can be expressed in terms of the $ K$-function as,
\begin{align}
	\hat{\rho}_0= \outprod{\psi_0}= \frac{1}{(2\pi)^{16}}\int d^{16}\alpha \, d^{16}\beta\;K(\vec{\alpha}) K^*(\vec{\beta})  \ket{\vec{\alpha}}\!\!\bra{\vec{\beta}}.
\end{align}
We adopt a Kraus operator-based approach (similar to Appendix~\ref{appendix:hybrid}) to account for the pure loss (due to coupling and detection efficiency). The Kraus operators for the action of a channel of transmissivity $ \eta $ are given by
  \begin{align}
  A_k=\sqrt{\frac{(1-\eta)}{k!}}\eta^{\hat{n}/2}\hat{a}^k .
  \end{align}  

\noindent The action of these operators on a general coherent basis term $ \proj{\gamma}{\delta} $, where $ \gamma,\delta\in\mathbb{C} $ is given by 
\begin{align}
	\sum_{k=0}^{\infty} \hat{A}_k \proj{\gamma}{\delta} \hat{A}^{\dagger}_k &= \sum_k \frac{(1-\eta)^k}{k!} (\sqrt{\eta})^{\hat{n}} \hat{a}^k\proj{\gamma}{\delta}\hat{a}^{\dagger k} (\sqrt{\eta})^{\hat{n}}\\
	&= \sum_k \frac{(1-\eta)^k}{k!} (\gamma\delta^*)^k \exp\left(-\frac{(|\gamma|^2 +|\delta|^2)(1-\eta)}{2}\right) \proj{\gamma\sqrt{\eta}}{\delta\sqrt{\eta}}\\
	&= \exp\left((\gamma\delta^*)(1-\eta)-\frac{(|\gamma|^2 +|\beta|^2)(1-\eta)}{2}\right) \proj{\gamma\sqrt{\eta}}{\delta\sqrt{\eta}}.
\end{align}

\noindent Hence, we observe that the basis elements are modified as
\begin{align}
	&\ket{\alpha_1;\,  \alpha_2 ;\,\alpha_3;\, \alpha_4 ;\,\alpha_5;\, \alpha_6;\, \alpha_7;\, \alpha_8}\!\! \bra{\beta_1;\,  \beta_2 ;\,\beta_3;\, \beta_4 ;\,\beta_5;\, \beta_6;\, \beta_7;\, \beta_8}\\
	\Rightarrow& g(\vec{\alpha},\vec{\beta},\vec{\eta})\times \ket{\alpha_1\sqrt{\eta_c};\,  \alpha_2\sqrt{\eta_c} ;\,\alpha_3 \sqrt{\eta_c \eta_d};\, \alpha_4 \sqrt{\eta_c \eta_d};\,\alpha_5\sqrt{\eta_c \eta_d};\, \alpha_6\sqrt{\eta_c \eta_d};\, \alpha_7\sqrt{\eta_c} ;\, \alpha_8\sqrt{\eta_c} } \nonumber\\
	&\times \bra{\beta_1\sqrt{\eta_c};\,  \beta_2\sqrt{\eta_c} ;\,\beta_3 \sqrt{\eta_c \eta_d};\, \beta_4 \sqrt{\eta_c \eta_d};\,\beta_5\sqrt{\eta_c \eta_d};\, \beta_6\sqrt{\eta_c \eta_d};\, \beta_7\sqrt{\eta_c} ;\, \beta_8\sqrt{\eta_c} }.
\end{align}
The function $ g(\vec{\alpha},\vec{\beta},\vec{\eta}) $  accounts for the mixed nature of the final state after loss, and is compactly expressed as 
\begin{align}
	g(\vec{\alpha},\vec{\beta},\vec{\eta})=	\prod_{i=1}^{8}  \exp\left((\alpha_i \beta_i^*)(1-\eta_i)-\frac{(|\alpha_i|^2 +|\beta_i|^2)(1-\eta_i)}{2}\right),
\end{align}
where
\begin{align}
	\vec{\alpha}&=(\alpha_1,\,  \alpha_2 ,\,\alpha_3;\, \alpha_4 ,\,\alpha_5,\, \alpha_6,\, \alpha_7,\, \alpha_8);\\
	\vec{\beta}&=(\beta_1,\,  \beta_2 ,\,\beta_3,\, \beta_4 ,\,\beta_5,\, \beta_6,\, \beta_7,\, \beta_8)\\
	\vec{\eta}&=(\sqrt{\eta_c},\sqrt{\eta_c},\sqrt{\eta_c \eta_d},\sqrt{\eta_c \eta_d},\sqrt{\eta_c \eta_d},\sqrt{\eta_c\eta_d},\sqrt{\eta_c },\sqrt{\eta_c}).
\end{align}

\noindent Hence, we can write down the density matrix of the state after loss as 
\begin{align}
	\hat{\rho}_1=\frac{1}{(2\pi)^{16}}\int d^{16}\alpha \, d^{16}\beta\;K(\vec{\alpha})\; K^*(\vec{\beta})\; g(\vec{\alpha},\vec{\beta},\vec{\eta}) \ket{\vec{\alpha}\vec{\eta}}\!\!\bra{\vec{\beta} \vec{\eta}}.
\end{align}
In the general approach to analyze GBS circuits, we perform photon-number projection on the modes to be detected \cite{gagatsos2019,Su2019-ef}, which yields a non-Gaussian state (pure if there are no losses in any of the modes; mixed otherwise) that can be characterized. In the current analysis,
\begin{enumerate}
	\item we have a specific requirement for the quantum state in the undetected modes, i.e., $ \ket{\Psi^+}$ ;
	\item we know the `correct' detection patterns from our preliminary analysis.
\end{enumerate}
With this knowledge, we may subsume the photon-number detection step into our Fidelity calculation. Since we know that the target Bell state exists in the undetected `outer' modes (along with spurious terms), given the measurement outcomes from the PNRDs (say we get $ n_1, n_2,n_3 $ and $ n_4 $ clicks respectively), we effectively know the 8-mode state that would yield a Bell state should the intermediate 4 modes be detected.  Therefore, using this insight, our simplified technique for calculating the Fidelity is equivalent to evaluating the following overlap:
\begin{align}
	F(\ket{\xi},\hat{\rho}_1)=\Braket{\xi|\hat{\rho}_1|\xi},
	\label{eqn:GBS_fid}
\end{align}
where
\begin{align}
	\ket{\xi}=\frac{1}{\sqrt{2}} \left(\ket{1,0,n_1,n_2,n_3,n_4,0,1}+(-1)^{m_1} \ket{0,1,n_1,n_2,n_3,n_4,1,0}\right)
\end{align}
and $n_1,n_2,n_3,n_4  $ can be  any of the patterns from the table that determines the internal phase terms (refer Section~\ref{section:source_descr}).

The probability $ P_{\text{gen}} $ to generate the state from Eq.~\ref{eqn:casc_src_st} is given by 
\begin{align}
	\text{Tr}(\Pi_e \hat{\rho}_1),
\end{align} 
where
\begin{align}
	\Pi_e=\mathds{1}^{\otimes 2} \otimes\outprod{n_1,n_2,n_3,n_4}\otimes \mathds{1}^{\otimes 2}.
\end{align}

\noindent This simplifies to
\begin{align}
	\text{Tr}	(\Pi_e \hat{\rho}_1)=\frac{1}{(2\pi)^{16}}\int& d^{16}\alpha \, d^{16}\beta\;K(\vec{\alpha})\; K^*(\vec{\beta})\; g(\vec{\alpha},\vec{\beta},\vec{\eta}) \nonumber \\
	\times& \braket{\beta_1\sqrt{\eta_c}|\alpha_1 \sqrt{\eta_c}} \braket{\beta_2\sqrt{\eta_c}|\alpha_2 \sqrt{\eta_c}} \braket{\beta_7\sqrt{\eta_c}|\alpha_7 \sqrt{\eta_c}}\braket{\beta_8\sqrt{\eta_c}|\alpha_8 \sqrt{\eta_c}}\nonumber\\
	\times & \braket{n_1|\alpha_3 \sqrt{\eta_c \eta_d}} \braket{n_2|\alpha_4 \sqrt{\eta_c \eta_d}}\braket{n_3|\alpha_5 \sqrt{\eta_c \eta_d}} \braket{n_4|\alpha_6 \sqrt{\eta_c \eta_d}} \nonumber\\
	\times & \braket{\beta_3\sqrt{\eta_c \eta_d}|n_1} \braket{\beta_4\sqrt{\eta_c \eta_d}|n_2} \braket{\beta_5\sqrt{\eta_c \eta_d}|n_3} \braket{\beta_6\sqrt{\eta_c \eta_d}|n_4}.
	\label{eqn:GBS_trace}
\end{align}

\noindent We identify two distinct types of overlap terms in the integral, that simplify as 
\begin{align}
	\braket{\beta_i \eta_i|\alpha_i \eta_i}&=\exp\left[-\frac{\eta_i^2}{2}\left(|\alpha_i|^2+ |\beta_i|^2- 2 \alpha_i\beta_i^*\right) \right]\\
	\braket{\beta_i\eta_i|n_j}\!\!\braket{n_j|\alpha_i\eta_i}&=\eta_i^{n_j}\frac{(\alpha_i\beta_i^*)^{n_j}}{n_j !} \exp\left[-\frac{\eta_i^2}{2}\left(|\alpha_i|^2+ |\beta_i|^2\right)\right].
\end{align}

In the present manuscript, we have sufficient evidence from the hybrid analysis in Appendix~\ref{appendix:hybrid} to claim that the cascaded source can be efficiently operated for $ N_s<0.2 $. The reader may solve the detailed integrals in Eq.~(\ref{eqn:GBS_fid}) and (\ref{eqn:GBS_trace}) if they wish to analyze the performance of the source in the regime of $ N_s>0.2 $.

\end{document}